%% file: arXiv-linear-multi-winner.tex
\documentclass[format=acmsmall, review=false]{acmart} 
\usepackage{booktabs} 
\PassOptionsToPackage{numbers}{natbib}

\input{macro.tex}

\begin{document}
\title[A Linear Theory of Multi-Winner Voting]{A Linear Theory of Multi-Winner Voting} 
 \author{
Lirong Xia}  
\affiliation{
\institution{Rutgers University-New Brunswick and DIMACS} 
\country{USA} 
}
\email{lirong.xia@rutgers.edu}



\newpage

\begin{abstract} 


We introduces a general linear framework that unifies the study of multi-winner voting rules and proportionality axioms, demonstrating that many prominent multi-winner voting rules—including Thiele methods, their sequential variants, and approval-based committee scoring rules—are linear. Similarly, key proportionality axioms such as Justified Representation (JR), Extended JR (EJR), and their strengthened variants (PJR+, EJR+), along with core stability, can fit within this linear structure as well.

Leveraging PAC learning theory, we establish general and novel upper bounds on the sample complexity of learning linear mappings. Our approach yields near-optimal guarantees for diverse classes of rules, including Thiele methods and  ordered weighted average rules, and can be applied to analyze the sample complexity of learning proportionality axioms such as approximate core stability. Furthermore, the linear structure allows us to leverage prior work to extend our analysis beyond worst-case scenarios to study the likelihood of various properties of linear rules and  axioms. We introduce a broad class of distributions that extend  Impartial Culture for approval preferences, and show that under these distributions, with high probability, any  Thiele method is resolute,  $\core$ is non-empty, and any Thiele method satisfies $\core$, among other observations on the likelihood of commonly-studied properties in social choice.

We believe that this linear theory offers a new perspective and powerful  new tools for designing and analyzing multi-winner rules in modern social choice applications. 
 
\end{abstract}
\maketitle

\newpage

\tableofcontents

\newpage
\setcounter{page}{1}
 \section{Introduction}

Multi-winner voting studies how to select a set of $k$ alternatives fairly and efficiently based on agents' preferences. This domain of social choice has gained significant attention recently, driven by practical applications such as parliament selection, shortlisting, recommender systems,  participatory budgeting, etc.~\citep{Lackner2023:Multi-Winner}.


Traditionally, the design and analysis of multi-winner rules have followed the classical axiomatic approach in social choice theory~\citep{Plott76:Axiomatic}. This approach mirrors "first principles thinking," where researchers propose desiderata (called {\em axioms}), design voting rules, and evaluate these rules based on their satisfaction of the axioms. The study of multi-winner rules traces back to the pioneering work of Thiele and \phrag{} in the 19th century. Following the seminal work of~\citet{Aziz2017:Justified}, the field has seen substantial progress in the past decade: researchers have proposed and extensively studied new proportionality axioms (e.g.,~\citep{Aziz2017:Justified,Sanchez-Fernandez2017:Proportional,Brill2023:Robust}, designed and analyzed novel rules (e.g.,~\citep{Aziz2018:On-the-Complexity,Peters2020:Proportionality,Brill2023:Robust}), and implemented some of these rules in real-world scenarios~\citep{Peters2020:Proportionality}.


However, as with classical social choice theory, the existence of an ``optimal'' multi-winner voting rule that performs well across all scenarios remains unlikely, evidenced by the ``{\em widespread presence of paradoxes and impossibility results}"~\citep{Sen1999:The-Possibility}. This creates a practical challenge: decision makers must formulate desiderata, choose or design a voting rule, and compare different rules based on their satisfaction of these desiderata. But how can this be accomplished when the decision maker lacks expertise in social choice theory, as is often the case?


The conventional solution is to consult an expert. While this may work well for well-studied domains like political elections, it faces severe limitations in novel applications. Experts may need significant time to formulate and propose desirable axioms (a process still ongoing for multi-winner voting), and their definitions might be subjective due to lack of information or understanding of the application and the stakeholders. Additionally, designing new rules  and analyzing their axiomatic satisfaction can be time-consuming  and mathematically challenging.


A promising direction lies in leveraging AI and Machine Learning (ML) to  design and analyze voting rules~\citep{Xia2022:Group}. AI and ML could   learn axioms from experts or stakeholders, design  voting rules through expert consultation~\citep{Procaccia09:Learnability,Caragiannis2022:The-Complexity} while maximizing axiom satisfaction~\citep{Xia13:Designing,Mohsin2022:Learning}, and with the help of computer-aided tools, verify axiom satisfaction or even prove impossibility theorems~\citep{Tang2009:Computer-aided,Geist2011:Automated,Geist17:Computer-aided,Brandl2018:Proving}. Recent advances in foundation models have demonstrated remarkable reasoning capabilities and human-level intelligence, potentially enhancing all aspects of collective decision-making. This potential is already being explored, as evidenced by a recently developed framework that uses generative AI to make collective decisions with guaranteed proportionality~\citep{Fish2024:Generative}. While these approaches effectively complement  the classical axiomatic approach, the theoretical foundations for learning axioms and rules remain underdeveloped.


To further enhance AI and ML-powered  design and analysis of voting, we need well-defined classes of axioms and voting rules. Defining these classes presents significant challenges: they must be general enough to allow for exploration and discovery, yet remain explainable and constrained enough to permit good theoretical guarantees. Ideally, we should also be able to prove general results for axioms and rules within these classes. Previous work has aimed to address some of these goals, including the study of positional scoring rules for learning single-winner rules~\citep{Procaccia09:Learnability}, approval-based committee scoring rules~\citep{Lackner2021:Consistent} for learning multi-winner rules~\citep{Caragiannis2022:The-Complexity}, and generalized (decision) scoring rules for analyzing likelihood of manipulation~\citep{Xia08:Generalized,Xia15:Generalized}. We are unaware of a previous work that aims at learning social choice axioms. Is it too ambitious to achieve all the goals at once?

\subsection{Our contributions}

We propose a linear framework for mappings from agents' preferences in a general preference space $\prefspace$ to a general decision space $\decspace$, aiming to address three fundamental goals simultaneously: learning axioms, learning rules, and analyzing properties of rules and axioms. Our contributions are threefold:

\myparagraph{Contribution 1: Modeling.}   Let $\prefspace^* = \cup_{n=1}^\infty (\prefspace)^n$ represent the set of all possible collections of agents' preferences. We say that a mapping $f:\prefspace^*\ra \decspace$ is {\em linear}, if there exists a finite set $\vec H = (\vec h_1,\ldots, \vec h_K)\in \mathbb R^{|\prefspace|}$ of $K\in\mathbb N$ hyperplanes in  $R^{|\prefspace|}$ such that the decision  depends solely on the relative position of the profile's histogram to each   hyperplane  (i.e., on the hyperplane, on the positive side, or on the negative side).  We show that many existing approval-based committee (ABC) rules, proportionality axioms, and properties about them are linear. Specifically: 
\begin{itemize}
\item {\bf For rules (Section~\ref{sec:linear-rules})}, we introduce  {\em generalized approval-based committee scoring (GABCS) rules}, extending ABCS rules~\citep{Lackner2021:Consistent}, which already cover many existing ABC rules  such as Thiele methods. We prove that both GABCS rules and their sequential variants  are linear (Thm.~\ref{thm:linear-rule} and~\ref{thm:seq-combination});
\item {\bf For axioms (Section~\ref{sec:linear-axiom})}, we introduce   {\em group-satisfaction-threshold (GST) proportionality axioms}, a broad class that includes many existing proportionality axioms (Thm.~\ref{thm:GST}). We prove that all  axioms in this class are linear (Thm.~\ref{thm:linear-axiom}); 
\item {\bf For properties (Section~\ref{sec:linear-operations})}, we establish that various operations on linear mappings preserve linearity (Thm.~\ref{thm:preserve}). These  operations correspond to important properties  in social choice theory, such as whether a  rule is resolute (the winning $k$-committee is unique), whether an axiom is satisfied at a profile, whether an axiom is stronger than another axiom,   whether a   rule satisfies an   axiom, etc. (Table~\ref{tab:properties}).
\end{itemize}

\myparagraph{Contribution 2: Learning.}  We analyze the {\em sample complexity} of learning  linear mappings in the PAC framework by bounding their {\em Natarajan dimensions}, which determine  the  number of samples needed for learning with high accuracy and confidence.   While general linear mappings require substantial samples  due to their expressiveness (Thm.~\ref{thm:ndim-linear-mappings}), we show that compact parameterization can significantly reduce the sample complexity. We introduce two compactly parameterized classes: 
\begin{itemize}
\item [1.] {\em Parameterized maximizers} (Def.~\ref{dfn:parameterized-maximzers}): We show that many popular classes of multi-winner rules belong to this category (Prop.~\ref{prop:gmp}), leading to precise Natarajan dimension bounds in Table~\ref{tab:ndim}.
\item [2.] {\em Parameterized hyperplanes} (Def.~\ref{dfn:parameterized-lm}): While GST axioms require doubly exponentially many samples to learn due to their expressiveness (Thm.~\ref{thm:ndim-pa} and~\ref{thm:ndim-spa}), we show that approximate $\core$~\citep{Peters2020:Proportionality} and the approximate version of any proportionality axiom mentioned in this paper, including   approximate  EJR~\citep{Halpern2023:Representation}, are parameterized hyperplanes requiring  only polynomially many samples to learn (Thm~\ref{thm:ndim-linear-feature}--\ref{thm:ndim-aGST}, Coro.~\ref{coro:ndim-axioms}). 
\end{itemize} 
\renewcommand{\arraystretch}{1.2}
\begin{table}[htp]
\begin{tabular}{|@{}c@{}|c|  c|c|}
\hline 
$\prefspace$&  Class &   Lower bound (Thm.~\ref{thm:ndim-lower}) & Upper bound (Prop.~\ref{prop:gmp} \& Thm.~\ref{thm:ndim-parameterized-maximizer})\\
\hline
\multirow{3}{*}{ $2^\ma$} & Thiele~\citep{Thiele1985:Om-Flerfoldsvalg} &     $k-2$&  $k-2$\\
 & ABCS~\citep{Lackner2021:Consistent} &       $\Omega(|\calX_{m,k}|)$~\citep{Caragiannis2022:The-Complexity} & $|\calX_{m,k}|-1$\\
& $\GABCS$ (Def.~\ref{dfn:gabcs})  &     $(2^m-2)\left ({m\choose k }-1\right)$ & $2^m\times {m\choose k}-1$\\
\hline 
 \multirow{2}{*}{   $\listset m$ }&  Committee scoring rules~\citep{Elkind2017:Properties}&   ${m\choose k} -2$ & ${m\choose k} -1$\\
& Ordinal OWA~\citep{Skowron2016:Finding} &     $k-2$ & $k-1$\\
\hline 
$\mathbb R^m$ &   OWA~\citep{Skowron2016:Finding} &   $k-2$ & $k-1$\\
\hline
\end{tabular}
\caption{Natarajan dimension of  multi-winner rules. $\calX_{m,k}$ is a set of indices to the  parameters of ABCS~\citep{Lackner2021:Consistent}. \label{tab:ndim}}
\end{table} 
\renewcommand{\arraystretch}{1}
Our results extend beyond multi-winner voting to other preference and decision spaces. For instance, 
our result on parameterized maximizers (Thm.~\ref{thm:ndim-parameterized-maximizer})  yields asymptotically tight bounds for positional scoring rules   and  simple rank scoring functions (SRSFs)~\citep{Conitzer09:Preference} (Prop.~\ref{prop:gmp} and Thm.~\ref{thm:ndim-parameterized-maximizer}).



\myparagraph{Contribution 3: Likelihood Analysis.} The linear structures of voting rules, axioms, and properties (Table~\ref{tab:properties}) enable the application of the {\em polyhedral approach}~\citep{Xia2021:How-Likely} for precise likelihood estimation. We demonstrate this approach for multi-winner voting under   {\em independent approval distributions} (Def.~\ref{dfn:iad}), where each alternative $i$ is independently approved by each agent with probability $p_i$, generalizing {\em Impartial Culture}  distributions~\citep{Bredereck2019:An-Experimental,Elkind2023:Justifying}. 
\begin{itemize}
\item {\bf For rules (Section~\ref{sec:likelihood-rules}),} we prove that any  Thiele method is resolute  with high probability  (Thm.~\ref{thm:Thiele}). We also prove a dichotomy theorem on the comparison of the winners under two different Thiele methods, showing that the probability for one to be contained in the other is either $1-\exp(-\Omega(n))$ or $\Theta(1)\wedge (1-\Theta(1))$;  similar observations hold for the two sets being equal, and for the two sets to overlap with each other (Thm.~\ref{thm:Thiele-refinement}). 
\item {\bf For axioms (Section~\ref{sec:likelihood-axioms}),} we prove that  $\core$ is non-empty  with high probability (Thm.~\ref{thm:core-iad})---as a corollary of this result and another result on $\ejrp$ satisfaction (Thm.~\ref{thm:ejrp-iad}), all $k$-committees are in $\core$ and satisfy $\ejrp$ with high probability under Impartial Culture  (Coro.~\ref{coro:core-ic}). We also prove that there exists an independent approval distribution under which $\core$ is strictly stronger than $\jr$ with high probability (Prop.~\ref{prop:jr-core}). 
\item {\bf For properties (Section~\ref{sec:likelihood-axiomatic-satisfaction}),}  we prove that any Thiele method satisfies $\core$ with high probability (Thm.~\ref{thm:Thiele-core}).  We also prove a dichotomy theorem on the probability for Thiele methods to satisfy GST axioms, showing that either there exists a GST axioms that is satisfied by any given Thiele method with probability $1-\exp(-\Omega(n))$, or any given Thiele method fails any given GST axiom with probability  $\Theta(1)\wedge (1-\Theta(1))$ (Thm.~\ref{thm:Thiele-GST}). 
\end{itemize}

\subsection{Related work and discussions}


\myparagraph{Multi-winner voting.} As discussed previously,  an extensive literature exists on the design and analysis of multi-winner voting rules, comprehensively surveyed in~\citep{Lackner2023:Multi-Winner}. Our paper introduces a unified linear framework (Section~\ref{sec:modeling}) and develops general bounds on sample complexity (Section~\ref{sec:learning}) while outlining procedures for likelihood analysis (Section~\ref{sec:analysis}). The newly proposed generalized ABCS rules (Def.~\ref{dfn:gabcs}) and GST axioms (Def.~\ref{dfn:GST-axiom}) may be of independent interest. While we demonstrate our framework's utility primarily through multi-winner voting, our theoretical foundations naturally extend to general preference and decision spaces. A comprehensive investigation of this linear theory's applications to other rules, axioms, and other social choice settings presents a promising direction for future research.

\myparagraph{Linear mappings. } Our linear mappings are mathematically equivalent to {\em generalized decision scoring rules}~\citep{Xia15:Generalized} and {\em perceptron trees}~\citep{Utgoff1988:Perceptron}. However, we introduce the new terminology ``linear'' because the existing names can be misleading. ``Generalized decision scoring rule'' suggests a focus on voting rules alone, whereas our linear mappings encompass rules, axioms, and properties. Similarly, "perceptron trees" implies a tree representation of the $g$ function. Moreover, neither term captures the application of the polyhedral approach   in Section~\ref{sec:analysis} to analyze voting rules, axioms, and properties, which fundamentally relies on the linearity of the separating hyperplanes.

\citep{Xia08:Generalized,Xia15:Generalized}  primarily focused on single-winner rules and likelihood of manipulability, examining only two multi-winner rules (Chamberlin-Courant and Monroe's rule). Our approach is substantially broader. Notably, our conceptualization of proportionality axioms as linear mappings opens new avenues for learning and analysis beyond worst-case scenarios. Our work provides novel sample complexity bounds for linear mappings and parameterized classes, contributing to the machine learning literature. The application of perceptron tree learning algorithms to axiom and rule learning remains a  promising area for future investigation.

\myparagraph{Learning voting rules and axioms.} Two previous works have examined the sample complexity of learning voting rules in the PAC framework.  \citet{Procaccia09:Learnability} formulated the learning problem and focused on learning resolute positional scoring rules with  certain lexicographic tie-breaking. Our theory extends to irresolute positional scoring rules, yielding comparable upper bounds (Prop.~\ref{prop:gmp} and Thm.~\ref{thm:ndim-parameterized-maximizer}).   \citet{Caragiannis2022:The-Complexity} established that the Natarajan dimension of ABCS rules is  $\Theta(|\calX_{m,k}|)$. By proving that ABCS rules are a parameterized maximizer  with $|\calX_{m,k}|-1$ parameters (Prop.~\ref{prop:gmp}), we  obtain a stronger, non-asymptotic $|\calX_{m,k}|-1$ upper bound through a simplified, unified approach. 

 
 
 \citet{Halpern2023:Representation} formulated opinion aggregation as an ABC voting problem, introduced approximate EJR, and developed preference query-based algorithms for axiom satisfaction. \citet{Fish2024:Generative} proposed a generative AI-powered framework for social choice and explored BJR satisfaction through LLM queries. Other related work explores PAC learning  of concepts in economics and game theory~\citep{Jha2020:A-Learning,Zhang2019:A-PAC-Framework}. None of these work aimed at learning social choice axioms. Our work addresses the sample complexity of learning both voting rules and axioms within a unified  framework.

\myparagraph{Likelihood analysis in social choice.}  While extensive literature exists on likelihood analysis of axiomatic satisfaction  for single-winner voting~\citep{Diss2021:Evaluating}, little was done for multi-winner voting. \citet{Elkind2023:Justifying} proved that under the  $p$-Impartial Culture model for approval preferences, any $s$-committee with $s\ge k/2$  satisfies JR with high probability, supporting  \citet{Bredereck2019:An-Experimental}'s empirical finding on the large number of $k$-committees satisfying JR. Our Corollary~\ref{coro:core-ic} strengthens the $s=k$ case of this result by proving that $\core$ (which is stronger than $\jr$) and $\ejrp$ (which is neither stronger nor weaker than $\core$)  contain all $k$-committees with high probability. 

Additionally, while $\core$'s universal non-emptiness remains an open question (with various approximations~\cite{Fain2018:Fair,Peters2020:Proportionality,Jiang2020:Approximately,Munagala2022:Approximate} and proven non-emptiness for $k\le 8$~\citep{Peters2025:The-Core}), our Theorem~\ref{thm:core-iad} establishes $\core$'s non-emptiness with high probability under  independent approval distributions, which are broader than  Impartial Culture.  More generally, we demonstrated the power of  the  polyhedra approach~\citep{Xia2020:The-Smoothed,Xia2021:How-Likely} in analyzing probability of other properties in Section~\ref{sec:analysis} Thm.~\ref{thm:ejrp-iad} to~\ref{thm:Thiele-GST}. We believe that this general approach together with the linear theory help us go beyond worst-case analysis in multi-winner voting.

\section{Preliminaries}
{\bf General voting setting.} Let $\prefspace$ denote agents' preference space and let $\decspace$ denote the collective decision space.  Each agent uses an element $R\in \prefspace$ to represent his or her preferences. Let $\prefspace^* = \bigcup_{n=1}^\infty \prefspace^n$ denote the set of all collections of preferences for arbitrary number of agents. A voting rule $r:\prefspace^*\ra \decspace$ maps agents' preferences, called a {\em profile}, to a collective decision. When $|\prefspace|$ is finite, for any profile $P\in \prefspace^*$,  let $\hist(P)\in\mathbb Z_{\ge 0}^{|\prefspace|}$ denote the {\em histogram} of $P$, which consists of the numbers of occurrences of all $|\prefspace|$ types of preferences in $P$.
 
\myparagraph{\bf Multi-winner voting.} Let  $\ma\triangleq [m]= \{1,\ldots,m\}$ denote the set of  $m\in\mathbb N$ alternatives. Let $\committee{k}$ denote the set of all {\em $k$-committees} of $\ma$, i.e., all subsets of $\ma$ with $k\le m$ alternatives. In multi-winner voting, $\decspace = 2^{\committee k}\setminus \{\emptyset\}$ and there are two commonly studied preference space. When $\prefspace = 2^\ma$, a profile $P\in (2^\ma)^n$ is called an {\em approval profile} and this case is commonly known as {\em approval-based committee (ABC)} voting. When $\prefspace = \listset m$, which consists of all linear orders over $\ma$, we say that agents have {\em rank preferences}. We call a voting rule $r$ a {\em multi-winner rule}, if $\decspace = 2^\ma\setminus\{\emptyset\}$, and we call a multi-winner rule an ABC rule  if  $\decspace = 2^\ma\setminus\{\emptyset\}$ and $\prefspace = 2^\ma$.
 

\myparagraph{ABC rules.} A {\em Thiele method} is an ABC rule that is specified by a score vector $\vec s = (s_0,s_1,\ldots,s_k)\in   \mathbb R^{k+1}$, often in non-decreasing order. We assume that the components in $\vec s$ are nonidentical. The rule is denoted by $\Thiele_{\vec s}$. For any approval profile $P$ and any $k$-committee $W$, define the Thiele score of $W$ in $P$ as $\score_{\vec s}(P,W) = \sum_{A\in P} s_{|A\cap W|}$. Then, $\Thiele_{\vec s}$  chooses all $W\in\committee k$ with the maximum score. Special cases of Theile methods include  {\em proportional approval voting ($\pav$)}, where for every $i\le k$, $s_i= \sum_{j=1}^i\frac 1 j $   and {\em approval Chamberlin–Courant ($\CC$)}, where $\vec s = (0,1,\ldots,1)$. 

An {\em  approval-based committee scoring  (ABCS)} rule~\citep{Lackner2021:Consistent} is specified by a scoring function $\mathbf s:\calX_{m,k}\ra \mathbb R$, where  $\calX_{m,k} = \{(a,b): b\in \{0,\ldots,m\}, a\in \{\max(b+k-m,0),\ldots, \min(k,b)\}\}$. The rule is denoted by $\ABCS_{\mathbf s}$. For any $P\in (2^\ma)^*$ and any $W\in \committee k$, define $\score_{\mathbf s}(P,W) = \sum_{A\in P} \mathbf s(|A\cap W|,|A|)$. Then, 
$\ABCS_{\mathbf s}(P) =\arg\max_{W\in\committee k}\score_{\mathbf s}(P,W)$.

\myparagraph{Multi-winner rules for rank preferences.} 
A {\em committee scoring rule ($\CSR$)}~\citep{Elkind2017:Properties}, denoted by $\CSR_{\bw}$, is specified by a scoring function $\bw: [m]^k\ra \mathbb N$. For any $R\in \listset m$ and any $W\in\committee k$, define $\score_{\bw}(R,W) = \bw(i_1,\ldots,i_k)$, where $(i_1,\ldots,i_k)$ represents the non-decreasing order of the ranks of the alternatives in $W$ in  $R$. Then, for any  $P\in  (\listset m)^n$, define $\score_{\bw}(P,W) = \sum_{R\in P}\score_{\bw}(R,W)$ and $\CSR_\bw(P) = \arg\max_{W\in\committee k}\score_{\bw}(P,W)$. 

An {\em ordered weighted average (OWA)} rule~\citep{Skowron2016:Finding} is specified by a weight vector $\vec w = (w_1,\ldots, w_k)\in \mathbb R_{\ge 0}^k$. The standard (cardinal) OWA rule, denoted by $\OWA_{\vec w}: ({\mathbb R}^m)^n\ra \committee k$, takes $n$ agents' utilities of alternatives as input. For any $U = (\vec u_1,\ldots,\vec u_n)$, where for every $j\le n$, $\vec u_j\in \mathbb R^{m}$ such that $[\vec u_j]_i$ represents agent $j$'s utility for alternative $i$. For every $W\in\committee k$ and every $j\le n$, let $(u_j^1,\ldots,u_j^k)$ denote the ordering of voter $j$'s utilities for the $k$ alternatives in $W$ in non-increasing order. Then, define $\score_{\vec w}(\vec u_j, W) = \sum_{i=1}^k w_i u_j^i$ and $\score_{\vec w}(U, W) = \sum_{j\le n} \score_{\vec w}(\vec u_j, W)$. Finally, $\OWA_{\vec w}$ chooses the $k$-committee with the maximum total score, i.e., $\OWA_{\vec w}(U) = \arg\max_{W\in\committee k}\score_{\vec w}(U, W)$.  The ordinal version of OWA takes voters' rankings over $\ma$ as input and uses a given intrinsic utility vector $\vec u\in \mathbb R^m$ that converts all voters' ordinal preferences to utilities. The rule is denoted by $\OWA_{\vec w}^{\vec u}$.  For any $R\in\listset m$, the corresponding utility vector is denoted by $\vec u_R=(u_{i_1},\ldots,u_{i_m})$ where for every $t\le m$, $i_t$ is the rank of alternative $i$ in $R$. For any profile $P = (R_1,\ldots,R_n)$, define $U_P = (\vec u_{R_1},\ldots,\vec u_{R_n})$. Then, $\OWA_{\vec w}^{\vec u}(P) = \OWA_{\vec w}(U_{P})$.

\myparagraph{Proportionality axioms of ABC rules.} Given a profile $P$ of $n$ approval preferences, a subgroup of voters, denoted by $P'\subseteq P$, is called an {\em $\ell$-cohesive group} for some $\ell\in\mathbb N$, if $|P'|\geqslant \ell\cdot \frac{n}{k}$ and $|\bigcap_{A\in P'} A|\geqslant \ell$. Based on this definition,   axioms in JR-family are defined as follows:  we first define an axiom $X$, then view it  as a function that maps a profile to all $k$-committees satisfying $X$.  We say an ABC rule $r$ satisfies $X$ if for all profiles $P$, $r(P)\subseteq X(P)$.
\begin{itemize}
\item {\em Justified Representation (JR, \citep{Aziz2017:Justified})}:  $W\in \committee k$ satisfies JR  if for every $1$-cohesive group $P'$, there exists $A\in P'$, such that $|A\cap W|\geqslant 1$.  
\item {\em Extended Justified Representation (EJR, \citep{Aziz2017:Justified})}:  $W\in \committee k$ satisfies EJR  if for every $\ell$-cohesive group $P'$, there exists $A\in P'$, such that $|A\cap W|\geqslant \ell$. 
\item {\em Proportional Justified Representation (PJR, \citep{Sanchez-Fernandez2017:Proportional})}:  $W\in \committee k$ satisfies PJR  if for every $\ell$-cohesive group $P'$, $|W\cap (\bigcup_{A\in P'} A)|\geqslant \ell$. 
\end{itemize}
Another well-studied axiom for ABC rules is {\em core stability ($\core$, \citep{Aziz2017:Justified})}. $W$ is in the  $\core$ (or {\em is core-stable}) if for every subset of votes $P'\subseteq P$ and every set of alternatives $W'\subseteq \ma$ such that $\frac{|W'|}{k}\le \frac{|P'|}{n}$, there exists a voter $A\in P'$ with $|A\cap W'|\le |A\cap W|$. 

The definitions of other commonly-studied proportionality axioms, including {\em fully justified representation ($\fjr$)}, and the strengthened versions of $\ejr$ and $\pjr$ called $\ejrp$ and $\pjrp$, respectively~\citep{Brill2023:Robust}, as well as other voting rules for other decision spaces, can be found in Appendix~\ref{app:prelim}.

\section{Modeling: Linear Mappings, Rules,  Axioms, and Properties}
\label{sec:modeling}
Notice that both voting rules and proportionality axioms are defined as mappings from $\prefspace^*$ to $\decspace$. To build a unified theory of learning and analysis,  we introduce the following   terminology.

\begin{dfn}[{\bf Linear  mappings}{}]
Given finite preference space $\prefspace$ and decision space $\decspace$, a mapping $f:\prefspace^*\ra \decspace$ is said to be {\em linear}, if there exist a finite set of {\em separating hyperplanes} $\vH = \{\vec h_1,\ldots, \vec h_K\}\subseteq {\mathbb R}^{|\prefspace|}$ and a function $g:\{+,-,0\}^{K}\ra \decspace$, such that for any  $P\in\prefspace^*$, $f(P) = g(\signH(\hist(P)))$, where for every $k\le K$, 
$[\signH(\hist(P))]_k = \begin{cases}
+ & \text{if }\hist(P)\cdot \vec h_k>0\\
- & \text{if }\hist(P)\cdot \vec h_k<0\\
0 & \text{if }\hist(P)\cdot \vec h_k=0
\end{cases}$.

Let $\hyplinear$ denote linear mappings that can be represented by $K$ separating hyperplanes. When $\decspace = 2^{\committee k} $, we call $f$ a {\em linear multi-winner  mapping}; and when 
$\prefspace = 2^\ma$ and $\decspace = 2^{\committee k}$, we call $f$ a {\em linear ABC mapping}. 
\end{dfn} 
In other words, a linear mapping operates in two steps: first, it determines how $\hist(P)$ is positioned relative to each of the $K$ separating hyperplanes (whether it lies on the positive side, negative side, or on the hyperplane itself); then, based on these relative positions, it selects the appropriate decision through function $g$.



\subsection{Linear ABC Rules}
\label{sec:linear-rules}
In this subsection, we show that many commonly-studied ABC rules are linear. We start with defining a general class of ABC rules that extends ABCS rules and show that they are linear.
\begin{dfn}
\label{dfn:gabcs}
A {\em generalized approval-based committee scoring (GABCS)} rule $r$ is an ABC rule that is specified by a   scoring function $\mathbf s:2^\ma\times \committee k\ra {\mathbb R}$ such that for any profile $P\in (2^{\ma}) ^n$,  
$$r(P) \triangleq \arg\max\nolimits _{W\in\committee k} \sum\nolimits_{A\in P}\mathbf s(A,W)$$
\end{dfn}
\begin{thm}
\label{thm:linear-rule}
All GABCS  rules are linear.
\end{thm}
\begin{proof}
Let $\mathbf s$ denote the scoring function of the GABCS rule.  For any $i_1,i_2\le {m \choose k}$, we define $\vec h_{i_1,i_2}$ to be the score difference between $W_{i_1},W_{i_2}\in \committee k$. That is, for every $A\in 2^{\ma}$, the $A$-component of $\vec h_{i_1,i_2}$ is $ s(A,W_{i_1}) - s(A,W_{i_2})$. Then, the $g$ function chooses all $M\in \committee k$ such that for all other $W'\in \committee k$, the $(W,W')$ component of $\signH(P)$ is $0$ or $+$.
\end{proof}

Next, we propose generalizations of {\em sequential Thiele method}~\citep[Rule 5]{Lackner2023:Multi-Winner} and {\em reverse sequential Thiele method}~\citep[Rule 6]{Lackner2023:Multi-Winner}. The new rules sequentially apply multiple $\GABCS$ rules  to greedily add one alternative (respectively, remove one alternative) to the winning committee(s). 


\begin{dfn}
For every $i\le k$, let $\amap_i$ be a GABCS rule for $\committee i$ with scoring function $\mathbf s_i$  (which may use a tie-breaking mechanism to choose a single winning $i$-committee). The sequential combination of $\amap_1,\ldots,\amap_k$ chooses the winning committees in $k$ steps. Let $\calS_0=\{\emptyset\}$. For each step $i\le k$, let  $\calS_i  = \arg\max_{M\cup\{a\}: M\in\calS_{i-1}, a\in \ma\setminus M}\mathbf s_i(P,M\cup\{a\})$. The procedure outputs $\calS_k$.
\end{dfn}

The formal definition of reverse sequential rules (Definition~\ref{dfn:reverse-sequential}) and the proof of both sequential rules' linearity (Theorem~\ref{thm:seq-combination}) can be found in Appendix~\ref{app:modeling} . 


\begin{thm}
\label{thm:seq-combination}
Sequential rules and  reverse sequential  rules are linear.
\end{thm}




\subsection{Linear Proportionality Axioms}
\label{sec:linear-axiom}
All proportionality axioms mentioned in this paper are linear mappings (and are thus called {\em linear axioms}). We use $\core$ to illustrate this observation. For any $W\in\committee k$ and any $W'\subseteq \ma$, define
$$\calM_{W,W'}\triangleq \{A\subseteq \ma: |A\cap W'|>| A\cap W|\}$$
For any $G\in 2^{2^\ma}$,  let $\vec w_{G}\in \{0,1\}^{2^\ma}$ denote  the binary vector such that $[\vec w_{G}]_A=1$ if and only if $A\in G$; otherwise $[\vec w_{G}]_A=0$. Then,  for any profile $P$, $W\in \core(P)$ if and only if for all $G\in 2^{2^\ma}$, 
\begin{equation}
\label{eq:pa-inequality}\left(\vec w_{G} - \frac{\tau(W,G)}{k}\cdot \vec 1\right)\cdot \hist (P)<0,
\end{equation}
where  $\tau(W,G)\triangleq  \begin{cases}
|W'| & \text{if }G  = \calM_{W,W'}\text{ for some }W'\subseteq \ma \\ k+1 & \text{otherwise}\end{cases}$. 

In other words,  $\tau$ specifies a threshold for every potential deviating group $G$---if the total number of agents whose preferences are in $G$ reaches this threshold, then these agents constitute  a legitimate deviating group, ruling out $W$ from $\core$. If $G$ is never considered as a potential deviating group to $W$, then the threshold is set to $k+1$,  meaning that \eqref{eq:pa-inequality} always holds regardless of $P$. Notice that $\tau$ is well-defined because if $G= \calM_{W,W'}$, then such $W'$ is unique.  

Generalizing this observation, we  define the class of {\em group-threshold-based proportionality axioms} for general preference space $\prefspace$ as follows. 
\begin{dfn}
\label{dfn:GST-axiom}
An multi-winner mapping $f$ is a {\em group-satisfaction-threshold  proportionality axiom} ({\em GST axiom} for short), if there exists a group-satisfaction-threshold function $\tau:\committee k\times 2^{\prefspace}\ra \{0,\ldots,k+1\}$ such that for any $\prefspace$-profile $P$, $W\in f(P)$ if and only if inequality \eqref{eq:pa-inequality} holds for all $G\in 2^{\prefspace}$. The axiom is denoted by $\GST_\tau$. 
Let $\hypGST$ denote the set of all GST axioms.
\end{dfn}
The next theorem states that all proportionality axioms mentioned in this paper are GST axioms with $\prefspace = 2^\ma$, which is proved by explicitly defining the threshold function $\tau$ for each axiom. The proof can be found in Appendix~\ref{app:proof-thm:GST}.

 \begin{thm}
\label{thm:GST}
$\core,\jr,\pjr,\ejr,\pjrp,\ejrp$, and $\fjr$ are GST axioms.
\end{thm}
The linearity of \eqref{eq:pa-inequality} and the definition of GST axioms immediately imply the following theorem. 

\begin{thm}
\label{thm:linear-axiom}
All GST axioms are linear.
\end{thm}

\subsection{Operations on Linear Rules and Axioms}
\label{sec:linear-operations}
\begin{dfn}[{\bf Operations}{}] 
Let $\amap_1$ and $\amap_2$ denote any pair of $\prefspace^*\ra \decspace$ mappings. For any $P\in\prefspace^*$, define 
$(\amap_1\cap \amap_2)(P)\triangleq \amap_1(P)\cap \amap_2(P)$,
$$
 (\equalsq{\amap_1}{\amap_2})(P) \triangleq \begin{cases}\amap_1(P) & \text{if \ \ }\amap_1(P) = \amap_2(P)\\
\emptyset & \text{otherwise}
\end{cases}\text{, and }
(\subq{\amap_1}{\amap_2})(P) \triangleq \begin{cases} \amap_1(P) & \text{if \ \ }\amap_1(P) \subseteq \amap_2(P)\\
\emptyset & \text{otherwise}
\end{cases}$$
\end{dfn}
These operations allow us to analyze various important properties of rules and axioms  as summarized in Table~\ref{tab:properties}, where $P$ is a profile,  $r$, $r_1$, $r_2$ are rules, and  $\coX$, $\coX_1$, $\coX_2$ are axioms. 
\begin{table}[htp]
\centering
\begin{tabular}{|c|r@{: } l|}
\hline
\multirow{4}{*}{\bf \begin{tabular}{c}Properties   \\ about rules\end{tabular}}&
$|r(P)| = 1$ & $r$ is resolute at $P$\\
&  $|(\subq{r_1} r_2)(P)|>0$&  $r_1$ is a refinement of $r_2$ at $P$\\ 
& $|(r_1\cap r_2)(P)|>0$&  $r_1$ and $r_2$ winners overlap at $P$\\ 
& $|(\equalsq{r_1}r_2)(P)|>0$& $r_1$ and $r_2$ are the same at $P$ \\ 
\hline 
\hline 
\multirow{4}{*}{\bf \begin{tabular}{c} Properties\\ about axioms\end{tabular}}& $|\coX(P)|>0$&  $X$ is satisfied at $P$\\ 
& $|(\subq{\coX_1}\coX_2)(P)|>0$&  $X_1$ implies  $X_2$ at $P$\\ 
 & $|({\coX_1}\cap \coX_2)(P)|>0$&  $X_1$ and $X_2$ can both be  satisfied at $P$\\ 
 & $|(\equalsq{X_1}X_2)(P)|>0$& $X_1$ and $X_2$ are equivalent at $P$ \\
 \hline 
\hline 
\multirow{3}{*}{\bf \begin{tabular}{c} Properties about\\ axiomatic satisfaction\end{tabular}}&
 $|(\subq r \coX)(P)|>0$&  $r$ satisfies $X$ at $P$\\ 
& $|(r\cap\coX)(P)|>0$&  an $r$ co-winner satisfies $X$ at $P$\\ 
 & $|(\equalsq r \coX)(P)|>0$&  $r$ is characterized by $X$ at $P$\\
 \hline  
\end{tabular}
\caption{Properties of rules and axioms.
\label{tab:properties}}
\end{table}

\begin{thm}
\label{thm:preserve}
If  $\amap_1$ and $\amap_2$ are linear, then $\amap_1\cap \amap_2$,   $\equalsq{\amap_1}{\amap_2}$, and $\subq{\amap_1}{\amap_2}$ are linear.
\end{thm}
\begin{proof} Suppose $\amap_1$ (respectively, $\amap_2$) is specified by hyperplanes  $\vH_1$ and function $g_1$ (respectively, $\vH_2$ and function $g_2$). Then, each of  $\amap_1\cap \amap_2$,   $\equalsq{\amap_1}{\amap_2}$, and $\subq{\amap_1}{\amap_2}$ is linear by letting $\vH \triangleq \vH_1\cup \vH_2$ and defining the $g$ function that combines the outcomes of $g_1$ and $g_2$ w.r.t.~$\vH_1$ and $\vH_2$, respectively.
\end{proof}


\section{Learning: Sample Complexity of Linear Mappings}
\label{sec:learning}

We extend the learning-to-design framework introduced by~\citet{Procaccia09:Learnability} to our setting. Consider a decision maker seeking to design a voting rule through oracle consultation, where the oracle may be an expert or the stakeholders. Each query presents a profile $P$ to the oracle, which responds with the set of (co-)winners for $P$. These (profile, co-winners) pairs form the training data. Then, machine learning algorithms are applied to learn a voting rule that maps input profiles to their appropriate co-winners. This approach naturally extends to learning axioms. In practice, the real bottleneck is often the lack of  data rather than the shortage of runtime of the learning algorithm. To rigorously analyze the sample complexity---the number of data points required to learn either a voting rule or an axiom---below we first recall the relevant concepts from statistical learning theory using our terminology.

\myparagraph{The PAC learning framework}. There is a fixed but unknown distribution $\pi$ over $\prefspace^*$. We are given a set $\hypspace$ of functions in $\calD^{\prefspace^*}$.  Let $f\in\hypspace$ denote the unknown target of the learning process. A data point $(x,f(x))$ is obtained by first generating $x$ according to $\pi$, then applying $f$. The learning process takes multiple data points as input and outputs an $f'\in \hypspace$. Define the {\em true error rate} as 
$$\Pr\nolimits_{x\sim\pi}(f'(x)\ne f(x))$$
\begin{dfn}[{\bf Sample complexity}{}]
Given  $\hypspace\subseteq \decspace^{\prefspace^*}$, let $m_{\hypspace}: [0,1]^2\ra \mathbb N$ denote the {\em sample complexity} of $\hypspace$, such that for any $(\epsilon, \delta)\in (0,1)^2$,  $m_{\hypspace}(\epsilon,\delta)$ is the smallest number of data points needed to guarantee that with probability $1-\delta$, the true error rate of the learned function is at most $\epsilon$.
\end{dfn}
Prior work in machine learning has established a striking connection between sample complexity and the {\em Natarajan dimension}. 
\begin{dfn}[{\bf Shattering and Natarajan Dimension~\citep{Shalev-Shwartz2014:Understanding}}{}]
A set $\calP\subseteq \prefspace^*$ is {\em shattered} by a set of functions $\hypspace\subseteq \decspace^{\prefspace^*}$ if there exist   $f^0, f^1\in \decspace^{\prefspace^*}$ such that 
(1) for all $P\in \calP$, $f^0(P)\ne f^1(P)$, and (2) for every $B\subseteq \calP$, there exists $f\in\hypspace$ such that for all $P\in \calP$, $f(P) = \begin{cases}f^0(P) & \text{if }P\in B\\ f^1(P)& \text{otherwise}\end{cases}$.

$f^0$ and $f^1$ are called {\em witnesses}. The {\em Natarajan dimension} of $\hypspace$, denoted by $\Ndim(\hypspace)$, is the size of largest $\calP\subseteq \prefspace^*$ that is  shattered by $\hypspace$.
\end{dfn}

{\sc Theorem$^*$ }{\bf (The Multiclass Fundamental Theorem~\citep[Theorem 29.3]{Shalev-Shwartz2014:Understanding}).}{\em 
There exist absolute constants $C_1,C_2>0$ such that for any set of functions $\hypspace \subseteq \decspace^{\prefspace^*}$ and any $(\epsilon,\delta)\in (0,1)^2$
\begin{equation*}
\label{eq:sample-complexity} C_1\frac{\Ndim(\hypspace)+\log (1/\delta)}{\epsilon}\le m_{\hypspace}\le C_2\frac{\Ndim(\hypspace)\log \left(|\decspace|\Ndim(\hypspace)/\epsilon\right)+\log (1/\delta)}{\epsilon}
\end{equation*}
}
That is, the sample complexity is approximately $\Theta(\frac{\Ndim(\hypspace)+\log (1/\delta)}{\epsilon})$ up to a multiplicative factor of $\log \left(|\decspace|\Ndim(\hypspace)/\epsilon\right)$. Theorem$^*$ is the realizable case of the theorem, which assumes that the target function $f$ is in $\hypspace$. The full version of the theorem~\citep[Theorem 29.3]{Shalev-Shwartz2014:Understanding} also provides similar bounds for {\em agnostic learning} (when $f$ is not necessarily in $\hypspace$) and {\em uniform convergence rate} (the difference between the sample error rate and the true error rate of every function). 

In light of this connection, in the remainder of the paper we will use sample complexity and Natarajan dimension interchangeably. We start with the sample complexity for all linear mappings.


\begin{thm}[{\bf Sample complexity: linear mappings}{}] \ 
\label{thm:ndim-linear-mappings}
$$\frac 1 4 \left(K|\prefspace| + \sum\nolimits_{s=0}^{\min K, |\prefspace|} {K \choose s}2^s \right)\le\Ndim(\hyplinear)\le 4K|\prefspace|\log (12K) + 2\times  3^K\log |\decspace|$$
\end{thm}
Before presenting the proof, let us take a look at the asymptotic bounds. It follows that
$$\begin{cases}
 \Omega(K|\prefspace| + 3^K) & \text{when }  K\le |\prefspace| \\
\Omega(K|\prefspace| + 2^{|\prefspace|}(\frac{K}{|\prefspace|})^{|\prefspace|})& \text{ when } K> |\prefspace|
\end{cases}
\le\Ndim(\hyplinear)\le O(K|\prefspace|\log K + 3^K\log |\decspace|)$$
When $K\le |\prefspace|$, the bounds are asymptotically tight up to a $\log  \max (K , |\decspace|)$ factor, and in this case the sample complexity of $\hyplinear$  is  polynomial in $|\prefspace|$ and $\log |D|$ while being exponential in $K$. When $K>|\prefspace|$, the lower bound  is exponential in $|\prefspace|(\log K-\log |\prefspace| +1)$. 

The sample complexity bounds imply that, if the target function is a linear mapping, approximately $\Ndim(\hyplinear)$-many samples are necessary and sufficient to identify the function with high confident ($1-\delta$) and high accuracy ($1-\epsilon$). If the target function is possibly non-linear, i.e., in the agnostic learning setting, then approximately $\Ndim(\hyplinear)$-many samples are necessary and sufficient to identify the best linear mapping to approximate the target function. The precise connection between Natarajan dimension and sample complexity of agnostic learning can be found in the full version of The Multiclass Fundamental Theorem~\citep[Theorem 29.3]{Shalev-Shwartz2014:Understanding}). 
\begin{proof}
The upper bound is proved by upper-bounding the {\em growth function} of $\hyplinear$. For any $\hypspace\subseteq \decspace^{\prefspace^*}$ and any set $\calP= (x_1,\ldots, x_T)\in (\prefspace^*)^T$, let $\cal F(\calP)$ denote the set of all vectors in $\decspace^T$   obtained by applying all functions in $\cal F$, i.e.,
$$\cal F(\calP)\triangleq \{(f(x_1),\ldots, f(x_T)): f\in \cal F\}$$
The {\em growth function}, denoted by $\tau_{\cal F}(T)$,  is the maximum size of $\cal F (\calP)$ for all $|\calP| =T$, i.e., 
$$\tau_{\cal F}(T)\triangleq \max\nolimits_{\calP\in (\prefspace^*)^T} |\cal F(\calP)|$$
Clearly, for any  $\calP\in (\prefspace^*)^T$ that is shattered by $\cal F$, we have $2^T\le |\cal F(\calP)| \le \tau_{\cal F}(T)$. Therefore, the Natarajan dimension is bounded above by the maximum number $T$ such that $2^T\le  \tau_{\cal F}(T)$. 

To obtain an upper bound on the growth function of $\hyplinear$,  we first provide an upper bound on the number of sign vectors achieved by applying $K$ separating hyperplanes. The idea is similar to the proof of the upper bound on the sample complexity of ABCS rules by~\citet{Caragiannis2022:The-Complexity}, which is based on analyzing the {\em sign patterns} of polynomials~\citep{Alon1996:Tools,Warren1968:Lower}. Given $K^*$ real polynomial over $\iota\in\mathbb N$ variables, denoted by $\text{poly}_1,\ldots, \text{poly}_{K^*}$ , a sign pattern is a $K^*$-dimension vector in $\{+,-,0\}^{K^*}$
$$\sign(\text{poly}_1(\vec x),\ldots, \text{poly}_{K^*}(\vec x))$$
that is realized by some $\vec x\in \mathbb R^{\iota}$. \citet{Alon1996:Tools} and \citet{Warren1968:Lower} proved that the number of different sign patterns of $\kappa$ polynomials of degree $\zeta$ over $\iota$ variables is upper bounded by $\left(\dfrac{8e\zeta \kappa}{\iota}\right)^\iota$.

We   only use the $\zeta=1$ case of this upper bound. Fix a set of $T$ histograms over $\prefspace$, denoted by $\calP = (\vec x_1,\ldots,\vec x_T)\in(\mathbb  R^{|\prefspace|})^T$. For $K$ separating hyperplanes $(\vec h_1,\ldots,\vec h_K)\in\mathbb (R^{|\prefspace|})^K$, the {\em sign profile} $\sign(\vec x_1\cdot\vec h_1,\ldots, \vec x_1\cdot\vec h_K,\ldots, \vec x_T\cdot\vec h_1,\ldots, \vec x_T\cdot\vec h_K)$ is a sign pattern of  $KT$ polynomials of degree one over $K|\prefspace|$ variables $\vec h_1,\ldots,\vec h_K$. Therefore, following the upper bound by \citet{Alon1996:Tools} and \citet{Warren1968:Lower} with $\kappa = KT$, $\zeta = 1$, and $\iota = K|\prefspace|$, the number of different sign profiles is upper-bounded by $\left(\dfrac{8e KT}{K|\prefspace|}\right)^{K|\prefspace|} = \left(\dfrac{8e T}{|\prefspace|}\right)^{K|\prefspace|}$. 
Then, notice that the number of different $g$ functions is $(|\decspace|)^{3^K}$. Therefore, 
$$|\cal F(\calP)|\le \left(\dfrac{8e T}{|\prefspace|}\right)^{K|\prefspace|}\times (|\decspace|)^{3^K}$$
Assuming $|\prefspace|>16e$, we have $2^T\le \left(\dfrac{8e T}{|\prefspace|}\right)^{K|\prefspace|}\times (|\decspace|)^{3^K}\Leftrightarrow \frac{T}{|\prefspace|} \le K \log \frac{T}{|\prefspace|}+K \log\left( 8e  \right)+\frac{3^K\log |\decspace|}{|\prefspace|}
$. 
According to~\citep[Lemma A.2]{Shalev-Shwartz2014:Understanding}, for any numbers $\alpha>1$ and $\beta>0$, $x\ge 4\alpha\log(2\alpha)+2\beta \Rightarrow x \ge \alpha\log(x)+\beta$. Let $\alpha = K$, $\beta = K \log\left( 8e  \right)+\frac{3^K\log |\decspace|}{|\prefspace|}+1$, and $x=\frac{T}{|\prefspace|}$, we have
\begin{align*}
& \frac{T}{|\prefspace|}\le 4K\log (2K)+2(K \log\left( 8e  \right)+\frac{3^K\log |\decspace|}{|\prefspace|}+1)-1\\
\Rightarrow & T\le 4K|\prefspace|\log (12K) + 2\times {3^K\log |\decspace|} 
\end{align*}

{\bf The lower bound} follows after proving and  combining the following two inequalities.
\begin{center}
\begin{tabular}{| l | r |}
\hline
\begin{minipage}{.4\textwidth}
\begin{equation}
\label{ineq:lb1}
\frac 12 K|\prefspace| \le\Ndim(\hyplinear)
\end{equation}
\end{minipage}
&
\begin{minipage}{.5\textwidth}
\begin{equation}
\label{ineq:lb2}
 \sum\nolimits_{s=0}^{\min K, |\prefspace|} {K \choose s}2^s  \le\Ndim(\hyplinear)
\end{equation}
\end{minipage}\\
\hline
\end{tabular}
\end{center}
\eqref{ineq:lb1} is proved by specifying a set $\calP$ of $\frac 12 K|\prefspace|$ vectors (histograms) in $\mathbb R^{|\prefspace|}$ that is shattered by $\hyplinear$. All vectors in $\calP$ take value $1$ in the last dimension. Therefore, when defining $\calP$, we focus on the $|\prefspace|-1$ dimensional subspace $\{\vec x\in \mathbb R^{|\prefspace|}: x_{|\prefspace|} = 1\}$.  Because the VC dimension of half spaces of $(|\prefspace|-1)$-dimensional Euclidean space is $|\prefspace|-1$, there exist a set $\calP_1\subseteq \mathbb R^{|\prefspace|-1}$ with $|\calP_1|=|\prefspace|-1 $ that is shattered by half spaces. In fact, the construction guarantees that for every combinations of binary labels over $\calP_1$, there exists a half space that strictly separate the positive nodes and the negative nodes $\calP_1$. W.l.o.g.~suppose for all $\vec x\in \calP_1$, $0< [\vec x]_1<1$ (if not, we can apply linear transformations to $\calP_1$ to meet the condition).  For every $k\le \lceil \frac K 2 \rceil$, define $\calP_k \triangleq \{\vec x+(k,\ldots, 0):\forall \vec x\in \calP_1\}$. That is, $\calP_k$ is obtained from $\calP_1$ by shifting it along the $(1,0,\ldots,0)$ direction by distance $k$. Then, define  
$$\calP\triangleq \{(\vec x,1):\vec x\in \calP_1\cup\cdots\cup \calP_{\lceil \frac K 2 \rceil}\}$$
To see that $\calP$ is shattered by $\hyplinear$, it is convenient to consider half spaces in the hyperplane $x_{|\prefspace|} = 1$ because each of them unique corresponds to a hyperplane in $\mathbb R^{|\prefspace|}$. We use $\lceil \frac K 2 \rceil-1$ half spaces $x_1 =1 ,x_1 =2,\ldots, x_1 = \lceil \frac K 2 \rceil-1$ to separate $\calP_1,\ldots, \calP_{\lceil \frac K 2 \rceil}$,  respectively, and then use $\lceil \frac K 2 \rceil$ half spaces, one for each $\calP_k$, to shatter vectors in $\calP_k$. For any input, the $g$ function uses the first $\lceil \frac K 2 \rceil-1$ signs to detect which set $\calP_k$ the input is in, and then use the $k$-th half space in the latter half to decide which side the input is in.  See Figure~\ref{fig:illustration-proof} left subfigure for an illustration of $|\prefspace|=3$.

\begin{figure}[htp]
\centering
\begin{tabular}{cc}
\includegraphics[width=.7\textwidth]{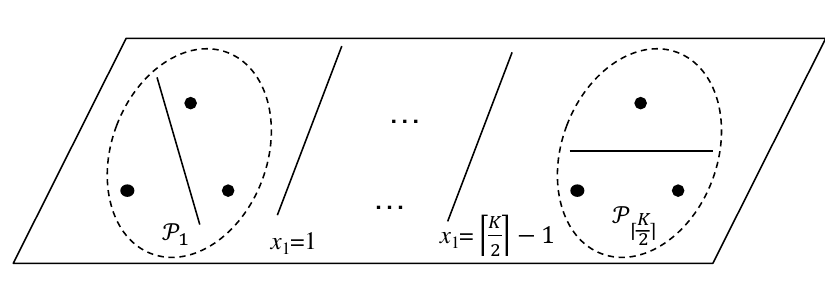} &
\includegraphics[width=.25\textwidth]{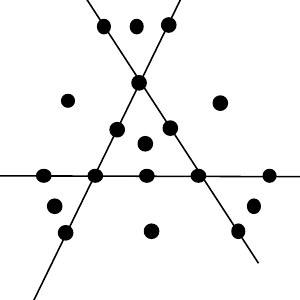} 
\end{tabular}
\caption{Illustrations of proofs. Left subfigure for \eqref{ineq:lb1}. Right subfigure for \eqref{ineq:lb2}.
\label{fig:illustration-proof}}
\end{figure}

To prove \eqref{ineq:lb2}, we consider the division of $\mathbb R^{|\prefspace|}$ into  subspaces using $K$ hyperplanes $\vec H$. For each subspace we choose a vector to form $\calP$. Then, fix the hyperplanes to be $\vec H$, each vector in $\calP$ has a different sign pattern, and therefore it is shattered by $\hyplinear$ by defining $g$ appropriately. See Figure~\ref{fig:illustration-proof} right subfigure for an illustration of $\calP$ for $|\prefspace | = 2$ and $K = 3$.

Next, we prove $|\calP| = \sum_{s=0}^{\min K, |\prefspace|} {K \choose s}2^s $. For any $d\in \mathbb N$ and any $k\ge 0$, let $A(d, k)$ denote the maximum number of subspaces (of all dimensions) that $k$ half planes can divide $\mathbb R^d$   into. \eqref{ineq:lb2} follows after the following claim, whose inductive proof can be found in Appendix~\ref{app:proof-claim:regions}.
\begin{claim}
\label{claim:regions}
$A(d, k)=\sum_{s=0}^{\min (k, d)} {k \choose s}2^s$.
\end{claim}
\end{proof}
For linear ABC mappings, $|\prefspace| = 2^m$, $|\decspace| = {m\choose k} $. Therefore, we have the following corollary.

\begin{coro}
The Natarajan dimension of ABC mappings is exponential in $m$ and $\min \{K,2^m\}$. 
\end{coro}

Unsurprisingly, due to the expressiveness of linear mappings, learning a  linear rule or a linear axiom  from this class requires (exponentially) many samples.  In the remainder of this section, we focus on learning linear mappings where $\decspace = 2^{\adecspace}$ for some ``atomic'' decision space 
$\adecspace$. All multi-winner rules and axioms studied in this paper belong to this class where $\adecspace = \committee k$. 

In each subsection below, we introduce a compact parameterized class and provide an upper bound on its sample complexity. While both classes can be used to learn linear mappings, the   {\em parameterized maximizers} (Definition~\ref{dfn:parameterized-maximzers}, Section~\ref{sec:sample-complexity-rules}) is specifically suitable for learning rules as many commonly studied classes of rules belong to this class. The more general {\em parameterized hyperplanes} (Definition~\ref{dfn:parameterized-lm}, Section~\ref{sec:sample-complexity-axioms}) can be used to learn  approximations of existing proportionality axioms. 

\subsection{Sample complexity of learning multi-winner rules}
\label{sec:sample-complexity-rules}
Before presenting the formal definitions, let us first take a closer look at Thiele  methods. Recall that each Thiele method is parameterized by $k+1$ numbers $\vec s$. Intuitively, this suggests that the class is simpler than $\hyplinear$. Also notice that the rule is a score maximizer in the sense that for each $W\in\committee k$, the profile $P$ is first mapped to a $(k+1)$-dimensional vector $\vec y$, where for every $i\le k+1$, the $i$-th dimension represents the occurrences of preference type $A$ in $P$ whose intersection with $W$ equals to $i$. Then, the score of $W$ in $P$ is $\vec y\cdot \vec s$. This motivates us to define {\em parameterized maximizers}, which resemble {\em general multiclass predictors}~\citep[Sec.~29.3.3]{Shalev-Shwartz2014:Understanding} in machine learning literature.


\begin{dfn}[{\bf Parameterized maximizers}{}]
\label{dfn:parameterized-maximzers} Given $\prefspace,\adecspace$,  a {\em class-sensitive feature mapping} $\psi:\prefspace^*\times \adecspace\ra \mathbb R^\eta$ for some $\eta\in\mathbb N$, and $\vec w\in \mathbb R^\eta$, let $f_{\vec w}: \prefspace^*\ra 2^\adecspace\setminus\{\emptyset\}$ denote the mapping such that for any $P\in \prefspace^*$, $f_{\vec w} (P)\triangleq \arg\max\nolimits_{d\in \adecspace} \psi(P,d)\cdot \vec w$. 
We call $f_{\vec w}$ a {\em parameterized maximizer} and let $\hypspace_{\psi}\triangleq \{f_{\vec w}: \vec w\in \mathbb R^\eta\}$ denote the set of all parameterized maximizers w.r.t.~$\psi$.
\end{dfn}
The next proposition shows that many social choice rules are parameterized maximizers. 
\begin{prop}[{\bf Voting rules as parameterized maximizers}{}]
\label{prop:gmp} 
The table lists commonly studied voting rules as parameterized maximizers, their notation, $\prefspace,\adecspace$, and $\eta$.
\begin{center}
\begin{tabular}{|c|c|c|c|c|}
\hline 
Class & Notation& $\prefspace$& $\adecspace$ & $\eta$\\
\hline

Thiele~\citep{Thiele1985:Om-Flerfoldsvalg} & $\hypThiele$& $2^\ma$ & $\committee k$ & $k-1$\\
ABCS~\citep{Lackner2021:Consistent} & $\hypABCS$ & $2^\ma$ & $\committee k$ & $|\calX_{m,k}|-1$\\
Generalized ABCS (Def.~\ref{dfn:gabcs}) & $\hypGABCS$& $2^\ma$ & $\committee k$ & $2^m\times {m\choose k}-1$\\
Committee scoring rule~\citep{Elkind2017:Properties}& $\hypCSR$& $\listset m$ & $\committee k$ & ${m\choose k} -1$\\
Ordinal OWA~\citep{Skowron2016:Finding}& $\hypOWAr^{\vec u}$& $\listset m$ & $\committee k$ & $k$\\
OWA~\citep{Skowron2016:Finding} & $\hypOWA$& $\mathbb R^m$ & $\committee k$ & $k$\\
Positional scoring rules (App.~\ref{app:prelim}) & $\hypPos$& $\listset m$ & $\ma$ & $m-1$\\
neutral SRSF~\citep{Conitzer09:Preference} (App.~\ref{app:prelim})& $\hypnSRSF$& $\listset m$ & $\listset m$ & $m!-1$\\
SRSF~\citep{Conitzer09:Preference} (App.~\ref{app:prelim})& $\hypSRSF$& $\listset m$ & $\listset m$ & $(m!)^2-1$\\
\hline
\end{tabular}
\end{center}
\end{prop}
\begin{proof}
For each class of voting rules in the table, we will explicitly construct the class-sensitive feature mapping. All mappings $\psi$ in our constructions are additive, in the sense that for any pair of profiles $P_1,P_2$, we have $\psi(P_1\cup P_2) = \psi(P_1)+\psi(P_2)$. Therefore, it suffices to define $\psi$ on a single $R\in \prefspace$. 

For the class of Thiele methods, for any $A\subseteq \ma$ and $W\in \committee k$, we define $\psi(A,W)$ to be the $k$-dimensional binary vector that takes $1$ on the $|A\cap W|$-th coordinate and takes $0$ otherwise---specifically, when $|A\cap W|=0$, $\psi(A,W) = \vec 0$. Any  $\Thiele_{\vec s}$ is in this class by letting $\vec w = (s_1,\ldots,s_k)$. The proof for other classes of rules in the statement of the proposition is done similarly and  can be found in Appendix~\ref{app:proof-prop:gmp}.
\end{proof}

Next, we prove a sample complexity upper bound for  parameterized maximizers that does not depend on $|\decspace|$, and then show that the upper bounds are quite accurate for the parameterized maximizers mentioned in Proposition~\ref{prop:gmp} by proving asymptotically matching (sometimes almost tight) lower bounds.

\begin{thm} [{\bf Sample complexity: parameterized maximizers}{}]
\label{thm:ndim-parameterized-maximizer}
$\Ndim(\hypspace_{\psi}) \le \eta $.
\end{thm}
 
\begin{sketch}
The proof is built upon the proof of \citep[Theorem 29.7]{Shalev-Shwartz2014:Understanding}, which upper-bounds the Natarajan dimension of a similar class of classifiers by revealing an insightful connection between linear multiclass classifiers and linear binary classifiers. The main difference is that  \citep[Theorem 29.7]{Shalev-Shwartz2014:Understanding} implicitly assumes that there is a unique $d\in \adecspace$ that maximizes $\psi(P,d)\cdot \vec w$, while our setting tackles the general case where multiple decisions in $\adecspace$ can achieve the maximum $\psi(P,d)\cdot \vec w$. 

Let us briefly recall the idea behind the proof  of~\citep[Theorem 29.7]{Shalev-Shwartz2014:Understanding}.  Let $\calP= (P_1,\ldots, P_T)\in  \prefspace^T$ denote any set  shattered by $\hypspace_{\psi}$ and let $f^0$ and $f^1$ be the witnesses. For every $t\le T$,  define 
\begin{equation}
\vec x_t\triangleq \psi(P_t,f^1(P_t)) -  \psi(P_t,f^0(P_t))
\label{eq:reduction}
\end{equation}

However, this conversion of $P_t$ to $\vec x_t$ implicitly assumes that $|f^1(P)| = |f^0(P)|=1$, otherwise $\psi(P_t,f^1(P_t))$ is not well-defined. 
Then, \citet{Shalev-Shwartz2014:Understanding} mentioned that for any $\vec w\in \mathbb R^\eta$ and any $P$ with $f_{\vec w}(P) = f^0(P)$ (respectively, $= f^1(P)$), $[\psi(P_t,f^1(P_t)) -  \psi(P_t,f^0(P_t))]\cdot \vec w<0$ (respectively, $>0$). This argument also implicitly assumes that $f^0(P)$ or $f^1(P)$ is the unique decision $d$ that maximizes $\psi(P_t,d)\cdot \vec w$. Under these assumptions, it then follows that $(\vec x_1,\ldots,\vec x_T)$ can be shattered by half spaces in $\mathbb R^\eta$, whose VC dimension is $\eta$. Therefore, $T\le \eta$.

This proof does not immediately work for set-valued classifiers even if we extend the definition of feature mapping to $\psi(P_t,D)$ for $D\subseteq \adecspace$, such that $\psi(P_t,D) = \sum_{d\in D}\psi(P_t,d)$. The complication arises when one of $f^0(P)$ and $f^1(P)$ is a subset of the other. When $f_{\vec w}(P_j) = f^0(P_t)\subset f^1(P_t)$ or $f_{\vec w}(P_j) = f^1(P_t)\subset f^0(P_t)$, we have $(\psi(P_t,f^1(P_t)) -  \psi(P_t,f^0(P_t)))\cdot\vec w =0$, which means that it is unclear whether the $0$ case of linear binary classifier should be classified as $0$ or $1$.  To address this complication, we define $\vec x_j^*$ that is similar to \eqref{eq:reduction}:
\begin{equation}
\vec x_t^*\triangleq \begin{cases} \psi(P_t,f^1(P_t)) -  \psi(P_t,f^0(P_t)) & \text{if } f^1(P_t)\not\subseteq f^0(P_t)\\
\psi(P_t,f^0(P_t)) -  \psi(P_t,f^1(P_t))  & \text{otherwise, i.e., } f^1(P_t)\subseteq f^0(P_t)
\end{cases}
\label{eq:reductionnew}
\end{equation}
The rest of the proof shows that $\{\vec x_1^*,\ldots,\vec x_T^*\}$ is shattered by linear binary classifiers in $\mathbb R^\eta$  and can be found in Appendix~\ref{app:proof-thm:ndim-parameterized-maximizer}, which   concludes the proof of Theorem~\ref{thm:ndim-parameterized-maximizer}.
\end{sketch}

\begin{thm}[{\bf Sample complexity lower bounds}{}]
\label{thm:ndim-lower} 
The following table summarizes lower bounds on the Natarajan dimension of some classes of rules.
\renewcommand{\arraystretch}{1.2}
\begin{center}
\begin{tabular}{|c|c|c|c|c|c|c|c|}
\hline 
$\hypThiele$& $\hypGABCS$& $\hypCSR$& $\hypOWAr$ \& $\hypOWA$ & $\hypPos$& $\hypnSRSF$& $\hypSRSF$\\
\hline 
 $  k-1$& $ (2^m-2)\left ({m\choose k }-1\right)$& ${m\choose k}-2$ & $k-2$& $ k-2$& $m!/2 -1$ & $ (m!-1)(m! -2)/2$ \\
\hline
\end{tabular} 
\end{center}
\renewcommand{\arraystretch}{1}
\end{thm}
\begin{sketch} For each class $\hypspace$ in the theorem statement, we construct a set $\calP$ shattered by $\hypspace$ such that $|\calP|$ is  the corresponding lower bound. The high-level idea behind the construction for each $P\in \calP$ is that under $P$, only two decisions are tied for the top choice, and then the tie-breaking between them is determined by the difference between two parameters.

\myparagraph{$\hypThiele$.} Let $W^0= \{1,3,\ldots, k+1\}$ and $W^1= \{2,3,\ldots, k+1\}$. Let $\calP =  \{P_1,\ldots, P_{k-1}\}$ such that for every $i\le k-1$,
$$P_i \triangleq 3k\times \{W^0,W^1\}\cup \{\{1,3,\ldots, i+2\}, \{2\}\}$$
To see that $\calP$ is shattered by $\hypThiele$, let $f^0$ (respectively, $f^1$) denote the function that always output $\{W^0\}$ (respectively, $\{W^1\}$) on all profiles in $\calP$. Then, given any $B\subseteq \calP$, define $\Thiele_{\vec s}$ as follows, where $\vec s = (s_0,\ldots, s_k)$: let $s_0=0$ and $s_1 = 2$, and for every $i\le k-1$, define $s_{i+1} = \begin{cases}
s_i +1&\text{if }P_i\in B\\
s_i+ 3&\text{otherwise}\end{cases}$. Then, for every $i\le k-1$, because $s_k-s_0< 3k$, for every $W\in \committee k$, we have $\score_{\vec s}(P_i,W^1) - \score_{\vec s}(P_i,W)\ge 3k(s_k-B_{k-1}) - (s_k-s_0) >0$.  According to the definition of $\vec s$, 
$$\score_{\vec s}(P_i,W^0) - \score_{\vec s}(P_i,W^1) = s_{i+1} - s_i - (s_1-s_0) = \begin{cases}
-1&\text{if }P_i\in B\\
1&\text{otherwise}\end{cases}$$ 
Therefore, $\Thiele_{\vec s}(P_i) = \begin{cases}
W^0 = f^0(P_i)&\text{if }P_i\in B\\
W^1 = f^1(P_i)&\text{otherwise}\end{cases}$. Hence  $\calP$ is shattered by $\hypThiele$, which means that $\Ndim(\hypThiele) \ge |\calP| = k-1$. The rest of the proof can be found in Appendix~\ref{app:proof-thm:ndim-lower}.
\end{sketch}

\subsection{Sample complexity of learning  axioms}
\label{sec:sample-complexity-axioms}
We start with proving upper and lower bounds for GST axioms.
\begin{thm}[{\bf Sample complexity: GST axioms}] \ 
\label{thm:ndim-pa}
$\sqrt{\frac{2}{\pi}}\cdot \frac{ 2^{|\prefspace|}}{|\prefspace|}\le \Ndim(\hypGST)\le 2^{|\prefspace|}{m\choose k}\log (k+2)$.
\end{thm}
\begin{proof}
Notice that $2^{\Ndim(\hypGST)}\le |\hypGST| \le  (k+2)^{{m\choose k }2^{|\prefspace|}}$, which means that $\Ndim(\hypGST)\le 2^{|\prefspace|}{m\choose k }\log (k+2)$.   For the lower bound,  we construct a set $\calP$ with $|\calP|\ge  {|\prefspace| \choose \lfloor |\prefspace|/2\rfloor} $ that is shattered by $\hypGST$. Let $\calS^*\subseteq 2^\prefspace$ denote the set of all subsets of $\prefspace$ with $\lfloor |\prefspace|/2\rfloor$ elements. That is, $\calS^* = \{G \subseteq \prefspace: |G| = \lfloor |\prefspace|/2\rfloor\}$. Then, define $\calP = \{P_{G}: G\in\calS^*\}$, where  $P_{G} = G$. For every $P_{G}\in \calP$, let $f^0(P_G) = \committee k \setminus \{1,\ldots,k\}$ and let $f^1(P_G) = \committee k$. For any $B\subseteq \calP$, define $\GST_{\tau}$ as follows, where $\tau:\committee k \times 2^\prefspace\ra \{0,\ldots,k+1\}$ is a group-satisfaction-threshold function (Definition~\ref{dfn:GST-axiom}).  For every $W\in \committee k$ and every $G'\in  2^\prefspace$, define $\tau(W, G') = \begin{cases}k & \text{if }W= \{1,\ldots, k\}\text{ and }G'\in B\\
k+1&\text{otherwise} \end{cases}$.

Then, for every $P_{G}\in \calP$, every $W\in \committee k$, and every $G'\in 2^\prefspace$, we have 
\begin{align*}
\left(\vec w_{G'} - \frac{\tau(W,G')}{k}\cdot \vec 1\right)\cdot \hist (P_{G})  &=  \frac{|G'\cap G|}{|G|}n -  \begin{cases}  n&   \text{if }W= \{1,\ldots, k\}\text{ and }G'\in B\\
\frac{k+1}{k} n &\text{otherwise} \end{cases}\\ & \begin{cases}  =0&   \text{if }W= \{1,\ldots, k\}\text{ and }G' = G\in B\\
<0 &\text{otherwise} \end{cases} 
\end{align*}
This means that $\GST_{\tau}(P_G) = \begin{cases}  \committee k\setminus \{\{1,\ldots,k\}\} = f^0(P_G)&   \text{if } P_G\in B\\
\committee k  = f^1(P_G)  &\text{otherwise} \end{cases} $, which proves that $\calP$ is shattered by $\hypGST$. Therefore, $\Ndim(\hypGST)\ge |\calP| = {|\prefspace| \choose \lfloor |\prefspace|/2\rfloor}$. The lower bound follows after the following claim, whose proof can be found in Appendix~\ref{app:proof-claim:lb-combinatorics}.
\begin{claim}
\label{claim:lb-combinatorics} For any $\eta\in\mathbb N$, 
${\eta \choose \lfloor \eta/2\rfloor} \ge \sqrt{\frac{2}{\pi}}\cdot \frac{2^\eta}{\sqrt \eta}$.
\end{claim}

\end{proof}

Notice that all $\GST$ axioms for ABC mapping discussed in this paper are ``symmetric'' in the sense that the group threshold function $\tau$ is invariant to permutations over $\ma$. This is a natural group of axioms that can be viewed as treating the candidates equally (c.f.~neutrality in single-winner voting literature). Formally, we define symmetric $\GST$ axioms as follows.
\begin{dfn}[{\bf Symmetric GST axioms}{}]
An ABC mapping $\GST_\tau\in\hypGST$ is {\em symmetric}, if for every permutation $\sigma$ over $\ma$,  every $W\in \committee k$ , and every $G\in 2^ \prefspace$, $\tau(W,G) = \tau(\sigma(W),\sigma(G))$. Let $\hypsGST$ denote the set of all symmetric GST axioms.
\end{dfn}
One may wonder if $\hypsGST$ requires less data to learn than  $\hypGST$. The following theorem shows that while the upper and lower bounds are reduced by a factor of  $m\choose k$ and $m!$ respectively, both bounds are still doubly exponential in $m$. The proof can be found in Appendix~\ref{app:proof-thm:ndim-spa}.

\begin{thm}[{\bf Sample complexity: symmetric GST axioms}{}]
\label{thm:ndim-spa} \ 
\begin{center}
$\sqrt{\frac{2}{\pi}}\cdot \frac{ 2^{2^m}}{2^m\cdot m!}\le \Ndim(\hypsGST)\le 2^{2^m} \log (k+2)$
\end{center}
\end{thm}

Recall from  Theorem~\ref{thm:ndim-parameterized-maximizer} that a compact representation of a linear mapping can significantly reduce its sample complexity. We do not see a way to model 
GST axioms as parameterized maximizers  to apply  Theorem~\ref{thm:ndim-parameterized-maximizer}. Still, we can adopt the idea  to compactly represent the relative position between the histogram of the input profile and the hyperplanes. This leads to the following definition.

\begin{dfn}[{\bf Parameterized hyperplanes}{}]
\label{dfn:parameterized-lm} Given $\prefspace,\decspace$, $K\in\mathbb N$, a {\em hyperplane-sensitive feature mapping} $\psi:\prefspace^*\times [K]\ra \mathbb R^\eta$ for some $\eta\in\mathbb N$, and a set $\cal G$ of functions from $\{+,-,0\}^K$ to $\decspace$, for any $\prefspace$-profile $P$, any $\vec w\in \mathbb R^\eta$, and any $g\in \cal G$, we define 
$$f_{\vec w,g} (P)\triangleq g( \sign(\psi(P,1)\cdot \vec w),\ldots, \sign(\psi(P,K)\cdot \vec w))$$
We call $f_{\vec w,g}$ a {\em parameterized hyperplane} and let $\hypspace_{\psi,\cal G} \triangleq \{f_{\vec w, g}: \vec w\in \mathbb R^\eta, g\in\cal G\}$.
\end{dfn}
That is, for any $f_{\vec w, g}\in \hypspace_{\psi,\cal G}$ and any profile $P$, the decision is made in two steps. First, for each $k\le K$, we use $\psi$ to compute the ``feature vector'' of $P$ for hyperplane $k$, which is an $\eta$-dimensional vector. Then, $g$ makes a decision based on the signs of the dot products of all feature vectors with $\vec w$. Next, we give an upper bound on the sample complexity of parameterized hyperplanes.


\begin{thm}[{\bf Sample complexity: parameterized hyperplanes}{}]\ 
\label{thm:ndim-linear-feature} 
$$\Ndim(\hypspace_{\psi,\cal G})\le 2 \eta\log\left( 8e K  \right)+2\log |\cal G|= O(\eta \log K +\log |\cal G|)$$
\end{thm}
\begin{proof} The proof is similar to the proof of the upper bound of Theorem~\ref{thm:ndim-linear-mappings}, which is based on upper-bounding the growth function. The main difference is that the analysis of the first term will be based on $\eta$-dimensional vectors.  Formally, given  $\calP = (P_1,\ldots,P_T)\in  (\prefspace^*)^T$ and $\vec w\in \mathbb R^\eta$,  the sign profile $\sign(\psi(P_1,1)\cdot\vec w,\ldots, \psi(P_1,K)\cdot\vec w,\ldots, \psi(P_T,1)\cdot\vec w,\ldots, \psi(P_T,K)\cdot\vec w)$ is a sign pattern of  $KT$ polynomials of degree one over $\eta$ variables $\vec w$. Therefore, following the upper bound by \citet{Alon1996:Tools} and \citet{Warren1968:Lower} by letting $\kappa = KT$, $\zeta = 1$, and $\iota = \eta$, the number of different sign profiles is upper bounded by $\left(\dfrac{8e KT}{\eta}\right)^{\eta}$. Therefore, 
 $|\cal F(\calP)|\le \left(\dfrac{8e KT}{\eta}\right)^{\eta}\times |\cal G|,$ 
which is an upper bound on the growth function. Consequently, 
\begin{equation}
\label{ineq:plm}
2^T\le \left(\dfrac{8e KT}{\eta}\right)^{\eta}\times |\cal G|\Leftrightarrow \frac{T}{\eta} \le  \log \frac{T}{\eta}+ \log\left( 8e K  \right)+\frac{\log |\cal G|}{\eta}
\end{equation}
Notice that $\log\left( 8e K  \right)+\frac{\log |\cal G|}{\eta}>1$. If $\frac{T}{\eta}> 2 (\log\left( 8e K  \right)+\frac{\log |\cal G|}{\eta})$, then $\frac{T}{\eta}>2$, which means that $\log \frac{T}{\eta}< \frac{T}{\eta}/2$. This implies that $\frac{T}{\eta} - \log \frac{T}{\eta}>  \frac{T}{2\eta}\ge \log\left( 8e K  \right)+\frac{\log |\cal G|}{\eta}$. In other words, we would have
$\frac{T}{\eta} > \log \frac{T}{\eta}+ \log\left( 8e K  \right)+\frac{\log |\cal G|}{\eta} 
$, 
which is a contradiction to \eqref{ineq:plm}. Therefore, $\frac{T}{\eta}\le  2 (\log\left( 8e K  \right)+\frac{\log |\cal G|}{\eta})$, which means that 
$T\le 2 \eta\log\left( 8e K  \right)+2\log |\cal G|$. 
\end{proof}
Notice that a parameterized maximizer of $\eta$ parameters can be represented as a parameterized hyperplane with $|\adecspace|\choose 2$ hyperplanes using $\eta$ parameters and $|\calG|=1$---each representing the score difference of a pair of elements in $\adecspace$. Therefore, Theorem~\ref{thm:ndim-linear-feature} implies an upper bound of $2 \eta\log\left( 8e {|\adecspace|\choose 2}  \right)+2\log |\cal G|$ on the Natarajan dimension of the parameterized maximizer, which is not as good as the $\eta$ bound guaranteed by Theorem~\ref{thm:ndim-parameterized-maximizer}. 

Next, we recall the definition of approximate $\core$~\citep{Peters2020:Proportionality} and then show how to apply Theorem~\ref{thm:ndim-linear-feature} to prove that it has polynomial sample complexity.

\begin{dfn}[{\bf Approximate $\core$~\citep{Peters2020:Proportionality}}{}]
For any $\alpha,\beta\ge 0$, given a profile $P$, a $k$-committee $W$ is an $(\alpha,\beta)$-$\core$ if for every $G\in 2^{\prefspace}$,  $\left(\vec w_{G} - \frac{\tau_{(\alpha,\beta)}(W,G)}{k}\cdot \vec 1\right)\cdot \hist (P)<0$, where
\begin{align}
&\tau_{(\alpha,\beta)}(W,G)\triangleq  \begin{cases}
\beta\cdot |W'| & \text{if }G  = \{A\subseteq \ma: |A\cap W'| >\alpha\cdot |A\cap W|\} \text{ for some }W'\subseteq \ma \\ k+1 & \text{otherwise}\end{cases}\label{eq:a-core}
\end{align}

Let $\core_{(\alpha,\beta)}$ denote the mapping that outputs all $k$-committees in the $(\alpha,\beta)$-$\core$. Let $\hypspace_{a\core}$ denote the set of  $\core_{(\alpha,\beta)}$ for all $\alpha,\beta\ge 0$. 
\end{dfn}

\begin{thm}[{\bf Sample complexity of approximate $\core$}{}]\ 
\label{thm:ndim-acore}
$$\Ndim(\hypspace_{\acore})\le 4m + (8k+4)\log m + 4\log(8e) = O(m+ k\log m)$$
\end{thm}
\begin{proof}
While $\alpha$ has infinitely many choices, effectively it only leads to finitely many types of constraints on the potential deviating group in \eqref{eq:a-core}, as $|A\cap W'|$ is an integer. Therefore, we will first discretize $\alpha$, define a hyperplane-sensitive feature mapping, and then apply Theorem~\ref{thm:ndim-linear-feature} to prove the desired upper bound on $\Ndim(\hypspace_{\acore})$.

More precisely, for every $\alpha>0$, we define a function $\lambda_{\alpha}: \{0,\ldots, k\}\ra \{1,\ldots, m\}$ such that for every $\ell \in  \{0,\ldots, k\}$, $\lambda_\alpha(\ell)$ is the largest integer that is smaller or equal to $\alpha\ell$.  Let $\Lambda \triangleq \{\lambda_\alpha: \alpha >0 \}$ denote the set of all such functions. For every $\lambda\in \Lambda$, every $W\in \committee k$, and every $W'\subseteq \ma$, we define a hyperplane indexed by $(\lambda, W, W')$. Therefore, there are $K = |\Lambda|{m\choose k}2^m$ hyperplanes in total. Define $\psi$ to be an additive mapping such that for every $A\subseteq \ma$ and every $(\lambda, W, W')\in \Lambda\times \committee k\times 2^\ma$, 
$$\psi(A, (\lambda, W,W')) \triangleq \left(\begin{cases}1 &\text{if }|A\cap W'|>\lambda(|A\cap W|)\\
0& \text{otherwise}\end{cases}, -\frac{|W'|}{k}\right)$$
Let $\vec w = (1,\beta)$. It follows that for any profile $P$, 
$$\psi(P, (\lambda, W,W'))\cdot \vec w =  |\{A\in P: |A\cap W'|>\lambda(|A\cap W|)\}|  -\beta \frac{|W'|}{k} |P|$$
Define $\calG =\{g\}$, where $W\in g(P)$ if and only if for all $W'\subseteq \ma$, $\psi(P, (\lambda, W,W'))\cdot \vec w<0$. It follows that $\hypspace_{\acore}\subseteq \hypspace_{\psi,\cal G}$. Therefore, $\Ndim(\hypspace_{\acore})\le \Ndim(\hypspace_{\psi,\cal G})$, and according to Theorem~\ref{thm:ndim-linear-feature}, the latter is upper bounded by $2 \eta\log\left( 8e K  \right)+2\log |\cal G|$. Notice that $\eta =2$, $K = |\Lambda|{m\choose k}2^m\le m^{k+1}{m\choose k}2^m$. Therefore, $\Ndim(\hypspace_{\acore})\le 4\log\left( 8e m^{k+1}{m\choose k}2^m\right) \le 4m + (8k+4)\log m + 4\log(8e) = O(m\log m)$, which completes the proof.
\end{proof}

\begin{dfn}[{\bf Approximate GST axioms}{}]
\label{dfn:approx-GST}
For any $\GST_\tau\in \hypGST$, we define the class of {\em approximate $\GST_\tau$}, denoted by $\hypspace_{\approx \GST_\tau}$, as follows: each $f_{\beta}\in \hypspace_{\approx \GST_\tau}$  is parameterized by a number $\beta\in \mathbb R_{\ge 0}$, such that for any  $P\in \prefspace^*$, $W\in f_\beta(P)$ if and only if for all $G\in 2^{\prefspace}$ with $\tau(W,G)\ne k+1$, 
\begin{equation}
\label{eq:aGST}
\left(\vec w_{G} - \beta\cdot \frac{\tau(W,G)}{k}\cdot \vec 1\right)\cdot \hist (P)<0
\end{equation} 
\end{dfn}
When $\beta =1$, $f_\beta$ degenerates to its original GST axiom. $\beta<1$ makes the axiom harder to satisfy and $\beta>1$ makes the axiom easier to satisfy. 
In the definition we require $\tau(W,G)\ne k+1$ because $k+1$ is introduced to model  groups that are never considered to be a valid deviating group regardless of the total number of agents whose preferences are in $G$. 
Definition~\ref{dfn:approx-GST} generalizes the  approximate version of EJR by~\citet{Halpern2023:Representation} to other GST axioms.  

Given any $\GST_\tau$ and any $W$, we say that $G\in 2^\prefspace$ is a {\em potential deviating group} if $\tau(W,G)\le k$. Let $\PD_\tau(W)$ denote the set of all potential deviating groups w.r.t.~$W$.
  
\begin{thm}[{\bf Sample complexity of approximate GST axioms}{}]
\label{thm:ndim-aGST}\ 
\begin{align*}
\forall \GST_\tau\in \hypGST, \Ndim(\hypspace_{\approx \GST_\tau})\le & 4 \log(\sum\nolimits_{W\in\committee k}|\PD_\tau(W)|) + 4\log (8e)\\
\forall \GST_\tau\in \hypsGST, \Ndim(\hypspace_{\approx  \GST_\tau})\le &4 \log( |\PD_\tau(\{1,\ldots,k\})|) + 4\log (8e)
\end{align*}
\end{thm}
 The proof is similar to the proof of Theorem~\ref{thm:ndim-acore} and can be found in Appendix~\ref{app:proof-thm:ndim-aGST}.   Theorem~\ref{thm:ndim-aGST} says that the sample complexity is upper-bounded by the log of the (total) number of potential deviating groups. Notice that all proportionality axioms mentioned in Theorem~\ref{thm:GST} are symmetric GST and their numbers of potential deviating groups are no more than $2^m\times 2^k$ (by $\pjr$). 
 Therefore, we have the following corollary.

\begin{coro}
\label{coro:ndim-axioms}
For any $X\in \{\core,\jr,\pjr,\ejr,\pjrp,\ejrp,\fjr\}$, $\Ndim(\hypspace_{\approx X})\le 4(m+k+\log (8e))$. 
\end{coro}

\section{Analysis: Likelihood of Properties of Linear Mappings}
\label{sec:analysis}

In this section, we consider the likelihood of various properties of linear mappings. For the purpose of illustration, we focus on ABC mappings and i.i.d.~models. We first propose a new class of models for approval preferences. 

\begin{dfn}[{\bf Independent approval distribution}{}]
\label{dfn:iad}
For any $\vec p\in [0,1]^m$, let $\pi_{\vec p}$ denote the distribution over $2^\ma$ such that for every $A\subseteq \ma$, $\Pr_{\pi_{\vec p}}(A) = \prod_{i\in A} p_i\prod_{i\notin A} (1-p_i)$.
\end{dfn}
When $\vec p = (p,\ldots, p)$ for some $p\in [0,1]$, $\pi_{\vec p}$ becomes the {\em $p$-Impartial Culture} model~\citep{Bredereck2019:An-Experimental,Elkind2023:Justifying}, denoted by $\IC_p$. 
For any $\vec p\in [0,1]^m$, let $\top_k(\vec p)$ denote the set of all $k$-committees whose corresponding elements in $\vec p$ are larger than or equal to all other elements. For example, $\top_2(0.5, 0.5,0.7) = \{\{1,3\}, \{2,3\}\}$. In the remainder of this section, we assume $\vec p\in (0,1)^m$ for the purpose of illustration.

\subsection{Likelihood of properties about rules}
\label{sec:likelihood-rules}
\begin{thm}
\label{thm:Thiele}
For any fixed $m$, $k$, $\vec p\in (0,1)^m$, and any   $\Thiele_{\vec s}$ with integer scoring vector $\vec s$, 
$$\Pr\nolimits_{P\sim (\pi_{\vec p})^n} (|\Thiele_{\vec s}(P)| = 1)  =  \begin{cases}
1-\exp(-\Omega(n)) & \text{if }|\top_k(\pi_{\vec p})|=1\\
1- \Theta(\frac {1}{\sqrt n})&\text{otherwise}
\end{cases}$$
\end{thm}
\begin{sketch} Recall that Thiele  methods are linear and testing whether a linear rule is resolute is linear as well (Table~\ref{tab:properties}). Therefore,   the event ``$\Thiele_{\vec s}$ is resolute at a profile $P$'' can be represented by the union of finitely many systems of linear inequalities whose variables are the occurrences of different types of preferences in $P$. In other words, the event can be represented by a union of finitely many polyhedra in $|\prefspace|$-dimensional space, such that the event holds if and only if $\hist(P)$ is in this union. Then, we prove the theorem by taking the {\em polyhedral approach}~\citep{Xia2021:How-Likely}, which provides a   dichotomy on the likelihood of the event. 

\myparagraph{Modeling.} We model and characterize the complement event, i.e., ``$\Thiele_{\vec s}$ is irresolute at a profile $P$''. Given a profile $P$, let $\hist(P) = (x_{A}:A\in\prefspace)$. For any pair of $k$-committees $W_1,W_2$, we define the following system of linear inequalities, denoted by $\text{LP}^{W_1,W_2}$, to represent the event ``$W_1$ and $W_2$ have the highest Thiele scores''.
$$\text{LP}^{W_1,W_2} = \left\{\begin{minipage}{0.8\textwidth}
\begin{align*}
& \sum\nolimits_{A\in \prefspace} (s_{|A\cap W_1|}-s_{|A\cap W_2|})x_A = 0 & \text{$W_1$ and  $W_2$ have the same score}\\
\forall W\in \committee k,  &\sum\nolimits_{A\in \prefspace} (s_{|A\cap W|}-s_{|A\cap W_1|})x_A \le  0 & \text{The score is the highest}
\end{align*}
\end{minipage}\right.$$

Let $\ppoly{W_1,W_2}\subseteq \mathbb R^{|\prefspace|}$  denote the set of all points in $R^{|\prefspace|}$ that satisfy these inequalities and let $\calU = \bigcup_{W_1,W_2\in \committee k} \ppoly{W_1,W_2}$. Clearly, $\Thiele_{\vec s}$ is irresolute at $P$ if and only if $\hist(P)\in \calU$. 

\myparagraph{Analyzing likelihood.} We first recall the i.i.d.~case of~\citep[Theorem 1]{Xia2021:How-Likely} in our notation. Given any polyhedron $\poly = \{\vec x\in \mathbb R^{|\prefspace|}: \ba\cdot \invert{\vec x}\le \invert{\vec b} \}$, where $\ba$ is an integer matrix,  let $\polynint$ denote the non-negative integer points $\vec x\in \poly$ such that $\vec x \cdot 1 = n$ and let $\polyz$ denote the {\em recess cone} of $\poly$, i.e., $\polyz = \{\vec x\in \mathbb R^{|\prefspace|}: \ba\cdot \invert{\vec x}\le \invert{\vec 0} \}$. Then,

\begin{equation}
\label{eq:polyhedral}
\Pr\nolimits_{P\sim (\pi_{\vec p})^n}\left(\hist(P)\in \poly\right)=\left\{\begin{array}{ll}0 &\text{if } \polynint=\emptyset\\
\exp(-\Theta(n)) &\text{if } \polynint \ne \emptyset \text{ and }\pi_{\vec p}\notin \polyz\\
\Theta\left((\sqrt n)^{\dim(\polyz)-|\prefspace|}\right) &\text{otherwise}
\end{array},\right.
\end{equation}
where $\dim(\polyz)$  is the dimension of $\polyz$. We then apply \eqref{eq:polyhedral} to  every $\ppoly{W_1,W_2}$ and combine the results. 
The remaining proof verifies that the $0$ case does not hold, interprets $\pi_{\vec p}\notin \polyz$ as $\{W_1,W_2\}\not\subseteq \top_k(\vec\pi)$, and proves  $\dim(\ppolyz{W_1,W_2})= |\prefspace|-1$. The full proof can be found in Appendix~\ref{app:proof-thm:Thiele}.
\end{sketch}

The proof of Theorem~\ref{thm:Thiele}, especially the way to apply the polyhedral approach, is similar to the proofs in~\citep{Xia2021:How-Likely}. This illustrates the  usefulness of the polyhedral approach for linear axioms, rules, and properties---a dichotomy of likelihood exists and can be obtained by  verifying similar conditions as in \eqref{eq:polyhedral} and characterizing the dimension of a recess cone. These can still be challenging but they are much easier than bounding the probabilities. The following theorem studies the $\subseteq_{\scriptstyle?}$,  $\cap$ , and $=_{\scriptstyle?}$, operations between two different rules, which correspond to one rule implies the other rule, the the two rules can be simultaneously satisfied, and the two rules being equal (Table~\ref{tab:properties}). The theorem states the probability either converges to $1$ with exponential rate, or is $\Theta(1)$ away from   $0$ or $1$ under any independent approval distribution.

\begin{thm}
\label{thm:Thiele-refinement}
For any fixed $m$, $k$, $\vec p\in (0,1)^m$, and any pair $\Thiele_{\vec s_1}$ and $\Thiele_{\vec s_2}$ with integer scoring vector $\vec s_1$ and $s_2$ that are not linear transformations of each other, for any sufficiently large $n$,
$$\Pr\nolimits_{P\sim (\pi_{\vec p})^n} (|(\subq{\Thiele_{\vec s_1}} \Thiele_{\vec s_2})(P)  | >0)  =  \begin{cases}
1-\exp(-\Omega(n)) & \text{if }|\top_k(\pi_{\vec p})|=1\\
 \Theta(1)\wedge (1- \Theta(1))&\text{otherwise}
\end{cases}$$
$$\Pr\nolimits_{P\sim (\pi_{\vec p})^n} (|(\equalsq{\Thiele_{\vec s_1}}{\Thiele_{\vec s_2})(P)} | >0)  =  \begin{cases}
1-\exp(-\Omega(n)) & \text{if }|\top_k(\pi_{\vec p})|=1\\
 \Theta(1)\wedge (1- \Theta(1))&\text{otherwise}
\end{cases}$$
$$\Pr\nolimits_{P\sim (\pi_{\vec p})^n} (| (\Thiele_{\vec s_1}\cap {\Thiele_{\vec s_2})(P)} | >0)  =  \begin{cases}
1-\exp(-\Omega(n)) & \text{if }|\top_k(\pi_{\vec p})|=1\\
 \Theta(1)\wedge (1- \Theta(1))&\text{otherwise}
\end{cases}$$
\end{thm}
\begin{proof}We first prove the $\subq{\Thiele_{\vec s_1}}{\Thiele_{\vec s_2}}$ part of the theorem. 
The $1-\exp(-\Omega(n))$ case   follows after the proof of the $1-\exp(-\Omega(n))$ case  of Theorem~\ref{thm:Thiele}, which shows that the  only $k$-committee in $\top_k(\vec p)$ is the unique winner under any given Thiele method. Then we apply the union bound to $\Thiele_{\vec s_1}$ and $\Thiele_{\vec s_2}$ and combine the results.

The $\Theta(1)\wedge (1- \Theta(1))$ case follows after combining the following two inequalities. For any $W_1,W_2\in \top_k(\vec p)$ with $W_1\ne W_2$, 
\begin{align}
\text{\bf\boldmath  Same w.p.~$\Theta(1)$: }& \Pr\nolimits_{P\sim(\pi_{\vec p})^n} ( \Thiele_{\vec s_1}(P)= \Thiele_{\vec s_2}(P)   = \{W_1\})  = \Theta(1) \label{inequ:theta-1}
\end{align}
There exist $W_1,W_2\in \top_k(\vec p)$ such that 
\begin{align}
\text{\bf\boldmath  Different w.p.~$\Theta(1)$: }& \Pr\nolimits_{P\sim(\pi_{\vec p})^n} ( \Thiele_{\vec s_1}(P)= \{W_1\} \text{ and } \Thiele_{\vec s_2}(P)  = \{W_2\} )  = \Theta(1) \label{inequ:1-theta-1}
\end{align}
Both will be proved by applying the  polyhedral approach. The rest of the proof can be found in Appendix~\ref{app:proof-thm:Thiele-refinement}.
\end{proof}

\subsection{Likelihood of properties about axioms}
\label{sec:likelihood-axioms}

In this section, we present results on likelihood   of properties about axioms in Table~\ref{tab:properties}. Our first theorem studies non-emptiness of $\core$.

\begin{thm}
\label{thm:core-iad}
For any fixed $m$, any fixed $k<m$, and any fixed $\vec p\in (0,1)^m$,
$$\Pr\nolimits_{P\sim(\pi_{\vec p})^n}\left(\top_k(\vec p)\subseteq \core(P)\right) = 1-\exp(-\Omega(n))$$
\end{thm}
\begin{sketch}
Again, we apply the polyhedral approach to prove the theorem. 

\myparagraph{Modeling.} In light of the linearity of $\core$, for any $k$-committee $W$ and any $W'\subseteq \ma$, we define LP$^{W,W'}$ to model the event that $W$ is not in $\core$ because a qualified group wants to deviate to $W'$. More precisely, LP$^{W,W'}$ consists of the following single inequality that corresponds to \eqref{eq:pa-inequality}. 
$$\text{LP}^{W,W'} = \left\{ \frac{|W'|}{k}\sum\nolimits_{A\in\prefspace} x_A- \sum\nolimits_{A\in \calM_{W,W'}} x_A  \le 0,  \right.$$
where we recall that $\prefspace = 2^\ma$ for $\core$ and $\calM_{W,W'} = \{A\in \prefspace: |A\cap W'| > |A\cap W|\}$. 
Let $\ppoly{W,W'}$ denote all vectors that satisfy $\text{LP}^{W,W'}$ and  let $\calU^W = \bigcup_{W'\in\prefspace} \ppoly{W,W'}$. Clearly, $W\notin\core(P)$ if and only if $\hist(P)\in \calU^W$.

\myparagraph{Analyzing likelihood.} We now apply \eqref{eq:polyhedral} (the i.i.d.~case of~\citep[Theorem 1]{Xia2021:How-Likely}) to analyze the probability for $\hist(P)$ to be in $\ppoly{W,W'}$ under distribution $\pi_{\vec p}$. We only need to consider $W'\not \subseteq W$ with $|W'|\le k$. For each such $W'$ and every $n\in\mathbb N$, let $P^*=n\times \{W'\setminus W\}$. Then, we have $\hist(P^*)\in \ppoly{W,W'}$, which means that $\ppolynint{W,W'}\ne\emptyset$. Therefore, the $0$ case of \eqref{eq:polyhedral} does not hold. 

 Next, we prove that the exponential case of \eqref{eq:polyhedral} holds for all $W'\not \subseteq W$ with $|W'|\le k$. That is, $\pi_{\vec p}\notin \ppolyz{W,W'}=\ppoly{W,W'}$. In other words,  we need to prove   $\left(\vec w_{\calM_{W,W'}}-\frac {|W'|}{k}\cdot \vec 1\right)\cdot \pi_{\vec p}<0,$ 
which is equivalent to 
\begin{equation}
\label{eq:core-p}
\Pr\nolimits_{A\sim\pi_{\vec p}}(A\in \calM_{W,W'})< \frac {|W'|}{k}
\end{equation}
Notice that every $A\in \calM_{W,W'}$ can be enumerated by considering the combination of the following three sets $A_1$, $A_2$, and $A_3$: 
$$\underbrace{A\cap (W\setminus W')}_{A_1},\underbrace{ A\cap (W'\setminus W)}_{A_2}, \underbrace{A\cap [(W\cap W')\cup (\neg W\cap \neg W')]}_{A_3}$$
Clearly, $|A_1|<|A_2|$ and $A_3$ can be any subset of $[(W\cap W')\cup (\neg W\cap \neg W')]$. For every set $A'\subseteq \ma$, let $B_{A',\vec p}$ denote the Poisson binomial random variable that is the sum of $|A'|$ independent Bernoulli trials  with success probabilities $\{p_i: i\in A'\}$. Then, 
\begin{align*}
&\Pr\nolimits_{A\sim\pi_p}(A\in \calM_{W,W'}) =   \Pr(|A_1|<|A_2|)=  \Pr(B_{W\setminus W',\vec p} < B_{W'\setminus W,\vec p})
 \end{align*}
Because $\frac{|W'\setminus W|}{|W\setminus W'|}\le \frac{|W'|}{k}$, to prove \eqref{eq:core-p}, it suffices to prove $\Pr(B_{W\setminus W',\vec p} < B_{W'\setminus W,\vec p}) <\frac{|W'\setminus W|}{|W\setminus W'|}$. 
Let $p$ denote the minimum probability in $\vec p$ indexed by $W\setminus W'$, that is, $p = \min_{i\in W\setminus W'} p_i$. This means that for all $i\in W\setminus W'$ we have $p_i\ge p$, and for every $i'\in W'\setminus W$ we have $p_{i'}\le p$ (because $W\in\top_k(\vec p)$). For any $n'\in\mathbb N$, let  $B_{n',p}$ denote the binomial random variable $(n',p)$, i.e., the sum of $n'$ independent binary  random variables, each of which takes $1$ with probability $p$. Let $k_1=|W\setminus W'| $ and $k_2 = |W'\setminus W|$ to simplify the notation.  We first prove
\begin{equation}
\label{ineq:S-p}
\Pr(B_{k_1,p} < B_{k_2,p})\cdot \frac {k_1}{k_2} <1 
\end{equation}
We then prove $\Pr(B_{W\setminus W',\vec p} < B_{W'\setminus W,\vec p})<\Pr(B_{|W\setminus W'|,p} < B_{|W'\setminus W|,p})$, which would  conclude the proof of Theorem~\ref{thm:core-iad}. The full proof can be found in Appendix~\ref{app:proof-thm:core-iad}. 
\end{sketch}

The constant in Theorem~\ref{thm:core-iad} depends on $m,k$, and $\vec p$. Next, we prove a similar theorem for $\ejrp$.

\begin{thm}
\label{thm:ejrp-iad}
For any fixed $m$, any fixed $k<m$, and any fixed $\vec p\in (0,1)^m$,
$$\Pr\nolimits_{P\sim(\pi_{\vec p})^n}\left(\top_k(\vec p)\subseteq \ejrp(P)\right) = 1-\exp(-\Omega(n))$$
\end{thm}
\begin{proof}
The proof is similar to that of Theorem~\ref{thm:core-iad} with a different modeling and application of~\eqref{eq:polyhedral}. 

\myparagraph{Modeling.} In light of the linearity of $\ejrp$ as a GST axiom (proof of Theorem~\ref{thm:GST}), for any $k$-committee $W$, any $a\in \ma\setminus W$, and any $\ell \le k$, we define LP$_W^{a,\ell}$ to model the event that $W$ does not satisfy $\ejrp$ because a qualified group of size at least $\frac \ell k n$ approving $a$. 

$$\text{LP}_W^{a,\ell} = \left\{ \frac{\ell}{k}\sum\nolimits_{A\in\prefspace} x_A- \sum\nolimits_{A\in \calM_W^{a,\ell}} x_A  \le 0,  \right.$$
where we recall that $$\calM_W^{a,\ell}\triangleq \{M\in 2^\ma: a\in M\text{ and }|M\cap W| \le \ell -1\}$$ 
Let $\pspoly{W}{a,\ell}$ denote all vectors that satisfy LP$_W^{a,\ell}$ and  let $\calU^{W} = \bigcup_{a\in \ma\setminus W,\ell\le k} \pspoly{W}{a,\ell}$. Clearly, $W\notin\ejrp(P)$ if and only if $\hist(P)\in \calU^W$.

\myparagraph{Analyzing likelihood.} We  apply \eqref{eq:polyhedral}  to analyze the probability for $\hist(P)$ to be in $\pspoly{W}{a,\ell}$ under distribution $\pi_{\vec p}$.  For each such $W'$ and every $n\in\mathbb N$, let $P=n\times \{a\}$. Then, $\hist(P)\in \pspoly{W}{a,\ell}$, which means that $\pspolynint{W}{a,\ell}\ne\emptyset$. Therefore, the $0$ case of \eqref{eq:polyhedral} does not hold. 

 Next, we prove that the exponential case of \eqref{eq:polyhedral} holds for all $a\notin W$ and $\ell\le k$. That is, $\pi_{\vec p}\notin \pspolyz{W}{a,\ell}=\pspoly{W}{a,\ell}$. In other words,  we need to prove   $\left(\vec w_{\calM_W^{a,\ell}}-\frac {\ell}{k}\cdot \vec 1\right)\cdot \pi_{\vec p}<0,$ 
which is equivalent to 
\begin{equation}
\label{eq:ejrp-p}
\Pr\nolimits_{A\sim\pi_{\vec p}}(A\in \calM_W^{a,\ell})< \frac {\ell}{k}
\end{equation}
Notice that every $A\in  \calM_W^{a,\ell}$ can be enumerated by considering the combination of following two sets $A_1$ and $A_2$ (recall that $a\in A$): 
$$\underbrace{A\cap W}_{A_1},\underbrace{ A\cap (\ma\setminus( W\cup\{a\})) }_{A_2}$$
Clearly, $|A_1|<\ell$ and $A_2$ can be any subset of $\ma\setminus( W\cup\{a\}) $.   Let  $p_a$ denote the $a$ component of $\vec p$. Then, 
\begin{align*}
&\Pr\nolimits_{A\sim\pi_p}(A\in \calM_W^{a,\ell}) =   \Pr(B_{W,\vec p} < \ell)\times p_a
 \end{align*}
 Let $p$ denote the minimum probability in $\vec p$ indexed by $W$, that is, $p = \min_{i\in W} p_i$. This means that for all $i\in W $ we have $p_i\ge p\ge p_a$   (because $W\in\top_k(\vec p)$). Because $B_{W,\vec p}$ first-order stochastically dominates $B_{k,p}$, we have $\Pr(B_{W,\vec p} < \ell)\times p_a < \Pr(B_{k,p} < \ell)\times p$. 
 Therefore, it suffices to prove  
\begin{equation*} 
\Pr(B_{k,p} < \ell)\times p <\frac{\ell}{k} 
\end{equation*}
This holds for all $\ell\le k$ because a $k$-committee that satisfies $\ejrp$ always exists for any profile. Let $P^*$ denote the (fractional) profile that corresponds to $\IC_p$. Then, any $k$-committee $W$ in $\ejrp(P^*)$ verifies this  inequality for all $\ell\le k$. This concludes the proof. 
\end{proof}

Applying Theorem~\ref{thm:core-iad} and~\ref{thm:ejrp-iad} to  $\IC_p$, where $\vec p = (p,\ldots, p)$ for some $p\in (0,1)$,  we have the following corollary.
\begin{coro}
\label{coro:core-ic}
For any fixed $m$, any fixed $k<m$, and any fixed $p\in (0,1)$,  
$$\Pr\nolimits_{P\sim\IC_p}\left(\core(P) = {\committee k}\right) = 1-\exp(-\Theta(n))$$
$$\Pr\nolimits_{P\sim\IC_p}\left(\ejrp(P) = {\committee k}\right) = 1-\exp(-\Theta(n))$$
\end{coro}

\myparagraph{Discussions.} Theorem~\ref{thm:core-iad} and Corollary~\ref{coro:core-ic} are good news, because even though whether $\core$ is always non-empty still remains open question~\citep{Lackner2023:Multi-Winner}, Theorem~\ref{thm:core-iad} shows that under any independent approval model ($\pi_{\vec p}$), it is very likely that $\core$ is non-empty. Corollary~\ref{coro:core-ic} further shows that all $k$-committee are in the CORE and satisfy $\ejrp$. In other words, any ABC rule would satisfy $\core$ with high probability under $\IC$, which means that it satisfies weaker axioms such as $\ejr$, $\pjr$,   $\jr$, and $\pjrp$ as well. 

Next, we examine the probability for $\jr$ to imply $\core$ and the probability for $\jr$ to be the same as $\core$.  Recall that it is known that $\core$ always implies $\jr$ and sometimes $\jr$ does not imply $\core$. 
\begin{prop}
\label{prop:jr-core}
For every fixed $m\ge 4$ and $k\ge 3$, there exists $\vec p\in (0,1)^m$ such that for every sufficiently large $n$, 
 $\Pr\nolimits_{P\sim(\pi_{\vec p})^n}\left(|(\subq{\jr}{\core})(P)|=0\right) = 1-\exp(-\Omega(n))$ and 
 $$\Pr\nolimits_{P\sim(\pi_{\vec p})^n}\left(|(\equalsq{\jr}\core)(P)|=0\right) = 1-\exp(-\Omega(n))$$
\end{prop}
\begin{sketch}
Let $\epsilon = \frac{1}{2k^2}$ and let $\vec p = (\underbrace{\epsilon,\ldots,\epsilon}_{k-1}, \underbrace{1-\epsilon,\ldots,1-\epsilon}_{m-k+1})$. Let $W^* = \{1,\ldots, k\}$. The proposition follows after proving that with high probability $W^*\in \jr(P)$ and $W^*\notin \core(P)$.  The proof is again based on applying the polyhedral approach and can be found in Appendix~\ref{app:proof-prop:jr-core}.
\end{sketch}

\subsection{Likelihood of properties about axiomatic satisfaction}
\label{sec:likelihood-axiomatic-satisfaction}
In this subsection we present examples of likelihood analysis of axiomatic satisfaction (Table~\ref{tab:properties}). While it is known that Thiele methods (or more generally, any existing multi-winner rule) fails $\core$, the following theorem says that they satisfy $\core$ with high probability under $\pi_{\vec p}$.

\begin{thm}
\label{thm:Thiele-core}
For any fixed $m$, any fixed $k<m$,  any fixed $\vec p\in (0,1)^m$, any  $\Thiele_{\vec s}$,
$$\Pr\nolimits_{P\sim(\pi_{\vec p})^n}\left(|(\subq{\Thiele_{\vec s}}\core)(P)|>0\right) = 1-\exp(-\Omega(n))$$
\end{thm}
To prove the theorem, we first prove that with high probability $\Thiele_{\vec s}(P) \subseteq \top_k(\pi_{\vec p})$. The theorem then follows after Theorem~\ref{thm:core-iad}. The full proof can be found in Appendix~\ref{app:proof-thm:Thiele-core}.

Notice that when $\Thiele_{\vec s}(P)\subseteq \core(P)$, we must have $\Thiele_{\vec s}(P)\cap \core(P)\ne \emptyset$. Therefore, we have the following corollary of Theorem~\ref{thm:Thiele-core}.
\begin{coro}
For any fixed $m$, any fixed $k<m$,  any fixed $\vec p\in (0,1)^m$, any  $\Thiele_{\vec s}$,
$$\Pr\nolimits_{P\sim(\pi_{\vec p})^n}\left(| ({\Thiele_{\vec s}}\cap\core)(P)|>0\right) = 1-\exp(-\Omega(n))$$
\end{coro}
The next theorem provides a dichotomy on characterizing Thiele methods by GST axioms. 
\begin{thm} 
\label{thm:Thiele-GST}
For any fixed $m$, any fixed $k<m$,  any fixed $\vec p\in (0,1)^m$, any  $\Thiele_{\vec s}$, 
\begin{itemize}
\item if $|\top_k(\vec p)|=1$, then there exists a GST axiom $\GST_{\tau}$ such that 
$$\Pr\nolimits_{P\sim(\pi_{\vec p})^n}\left(|(\equalsq{\Thiele_{\vec s}}{\GST_{\tau}})(P)|>0\right) = 1-\exp(-\Omega(n))$$
\item otherwise (i.e., $|\top_k(\vec p)|>1$), for every $\GST_{\tau}$,
$$\Pr\nolimits_{P\sim(\pi_{\vec p})^n}\left(|(\equalsq{\Thiele_{\vec s}}{\GST_{\tau}})(P)|=0\right) = \Theta(1) $$
\end{itemize}
\end{thm}
Notice that the intersection of any pair of GST axioms is also a GST axiom. Therefore, the $\Theta(1)$ case of Theorem~\ref{thm:Thiele-GST} also applies  combinations of GST axioms as well.
\begin{proof}
When $|\top_k(\vec p)|=1$, let $\top_k(\vec p) = \{W^*\}$. Let $\GST_\tau$ be the GST axioms that always chooses $W^*$ by setting $\tau(W^*,G) = k+1$ and $\tau(W,G) = 0$ for all $G\in2^{2^\ma}$ and all $W\in\committee k\setminus\{W^*\}$. This case of the theorem then follows after the proof of the exponential case of Theorem~\ref{thm:Thiele}.

When $|\top_k(\vec p)|>1$, let $W_1,W_2\in \top_k(\vec p)$ be such that $|W_1\cap W_2| = k-1$. W.l.o.g.~suppose $W_1 = \{1,3,\ldots, k+2\}$ and $W_2 = \{2,3,\ldots, k+2\}$. We prove that at least one of the following two events happens with probability $\Theta(1)$ when $P$ is generated from $(\pi_{\vec p})^n$.
\begin{itemize}
\item {\bf Event 1:} $W_1\in \Thiele_{\vec s}(P)$ and $W_1\notin \GST_\tau(P)$.
\item {\bf Event 2:} $ \score_{\vec s}(W_1,P)> \score_{\vec s}(W_2,P)$ and $W_2\in \GST_\tau(P)$.
\end{itemize}
Event 2 implies that $W_2\in \GST_\tau(P)$ but $W_2\notin \Thiele_{\vec s}(P)$ (because $W_1$'s score is higher, which does not necessarily mean that $W_1$ is a Thiele winner).
Clearly, if at least one of Event 1 and Event 2 holds, then $\Thiele_{\vec s}$ is not characterized by $\GST_\tau$ at $P$.  The rest of the proof takes the polyhedral approach to prove that at least Event 1 and/or Event 2 hold with probability $\Theta(1)$ and the proof can be found in Appendix~\ref{app:proof-thm:Thiele-GST}.
\end{proof}

\section{Future Work}

Many promising research directions emerge from our work. An immediate challenge in modeling is determining the linearity of other widely-studied ABC rules, axioms, and properties. Based on decision boundary analysis in the proof of Proposition~\ref{prop:phrag} (Appendix~\ref{app:phragmen}), we conjecture that $\phrag$-like rules, such as the method of equal shares~\citep{Peters2020:Proportionality}, are non-linear. Our sample complexity results can be applied when we try to learn the best approximation of non-linear rules within the (sub)class of linear rules, by applying the agnostic part of The Multiclass Fundamental Theorem~\citep[Theorem 29.3]{Shalev-Shwartz2014:Understanding}). The polyhedral approach exemplified in Section~\ref{sec:analysis} can be applied to upper- (respectively, lower-) bound the likelihood of properties of non-linear rules w.r.t.~linear or non-linear axioms by analyzing a superset (respectively, subset) of the properties in histogram space.  On the algorithmic front, key open questions include developing efficient algorithms for testing properties outlined in Table~\ref{tab:properties} and learning axioms or rules from samples. For parameterized maximizers with $\eta$ parameters, learning can be efficiently done by solving a system of linear equations with $\eta$ variable. For general linear mappings, we believe that existing algorithms for perceptron tree learning could prove valuable in practical applications.
For likelihood analysis, an important next step is completing the probabilistic characterization of rules, axioms, and properties presented in Table~\ref{tab:properties}. This analysis could extend to other properties such as manipulability and could be conducted under more general semi-random models~\citep{Xia2020:The-Smoothed,Xia2021:How-Likely}. The polyhedral approach used in Section~\ref{sec:analysis} shows promise for these investigations.
A broader, open-ended challenge lies in extending our linear theory to other social choice domains, particularly participatory budgeting~\citep{Aziz2021:Participatory}.  

\section*{Acknowledgements}
We thank Ratip Emin Berker, Mingyu Guo, Emanuel Tewolde, and all anonymous reviewers for helpful comments and suggestions.

\bibliographystyle{plainnat}
\bibliography{../../references}

\newpage
\appendix

\section{Additional Preliminaries}
\label{app:prelim}


\myparagraph{FJR.} Given any set $T$ of alternatives and any natural number $\beta$, a group of voters $P'$ is called {\em weakly $(\beta,T)$-cohesive}, if for every $R\in P'$ we have $|R\cap T|\ge \beta$, and $|P'|\ge \frac{|T|n}{k}$. A $k$-committee $W$ satisfies  {\em fully justified representation ($\fjr$)} at a profile $P$, if for every $\beta\in\mathbb N$, every $T\subseteq \ma$, and every $(\beta,T)$-cohesive group $P'\subseteq P$, there exists a voter $R\in P'$ such that $|R\cap W|\ge \beta$, that is, $R$ is reasonably satisfied with $W$.

\myparagraph{$\ejrp$ and $\pjrp$.}  EJR and PJR are recently strengthened to $\ejrp$ and $\pjrp$ by modifying the cohesiveness requirement~\citep{Brill2023:Robust}. $W\in \committee k$ satisfies $\ejrp$ if there is no $a\in\ma\setminus W$, a subgroup of voters $P'$ with $|P'|\ge \frac{\ell n}{k}$, such that for every $A\in P'$, $a\in A$ and $|A\cap W|< \ell$. $W\in \committee k$ satisfies $\pjrp$ if there is no $a\in\ma\setminus W$, a subgroup of voters $P'$ with $|P'|\ge \frac{\ell n}{k}$, such that for every $A\in P'$, $a\in A$, and $|\bigcup_{A\in P'}A\cap W|< \ell$. 

\myparagraph{\bf Positional scoring rules.}  An {\em (integer) positional scoring rule}  is specified by an integer scoring vector $\vec s=(s_1,\ldots,s_m)\in{\mathbb Z}^m$ with $s_1\ge s_2\ge \cdots\ge s_m$ and $s_1>s_m$. The rule is denoted by $\Pos_{\vec s}$. For any alternative $a$ and any linear order $R\in\ml(\ma)$, we let $\score_{\vec s}(R,a)=s_i$, where $i$ is the rank of $a$ in $R$. Given a profile $P$, let $\score_{\vec s}(P,a) = \sum_{R\in P} \score_{\vec s}(R,a)$. Then, $\Pos_{\vec s} = \arg\max_{a\in \ma} \score_{\vec s}(P,a)$.  For example, {\em plurality} uses the scoring vector $(1,0,\ldots,0)$, {\em Borda} uses the scoring vector $(m-1,m-2,\ldots,0)$, and {\em veto} uses the scoring vector $(1,\ldots,1,0)$. 

\begin{dfn}[{\bf SRSF and neutral SRSF~\citep{Conitzer09:Preference}}{}]
When $\prefspace = \decspace = \listset m$, a {\em simple rank scoring function} $f$ is defined by a scoring function $\mathbf s:\listset m\times \listset m\ra \mathbb R$, such that for any profile $P$,  $f(P) = \arg\max_{R^*\in \listset m} \sum_{R\in P} \mathbf s(R,R^*)$. We say that $\mathbf s$ is {\em neutral}, if for every permutation $\sigma$ over $\ma$ and any $R,R'\in \listset m$, $\mathbf s(R,R') = \mathbf s(\sigma(R),\sigma(R'))$. Let $\hypspace_\text{SRSF}$ and $\hypspace_\text{nSRSF}$ denote the set of all SRSFs and the set of all SRSFs based on neutral scoring functions, respectively.
\end{dfn}
It has been proved that $\hypspace_\text{nSRSF}$ coincide with  the set of all neutral SRSFs~\citep[Lemma~2]{Conitzer09:Preference}.

\section{Materials for Section~\ref{sec:modeling}}
\label{app:modeling}
\begin{dfn}[{\bf Reverse sequential rule}{}]
\label{dfn:reverse-sequential}
For every $i\le m- k$, let $\amap_i$ be a GABCS rule for $\committee {m-i}$ with scoring function $\mathbf s_i$  (which may use a tie-breaking mechanism). The reverse sequential combination of $\amap_1,\ldots,\amap_{m-k}$ chooses the winning committees in $m-k$ steps. Let $\calS_0=\{\ma\}$. For each $i\le k$, let 
$$\calS_i  = \arg\max\nolimits_{M\setminus\{a\}: M\in\calS_{i-1}, a\in M}\mathbf s_i(P,M\setminus\{a\})$$
Then the procedure outputs $\calS_k$.
\end{dfn}

%


\subsection{$\phrag$'s sequential rule is non-linear}
\label{app:phragmen}

The following proposition shows that   $\phrag$'s sequential rule, denoted by seq-$\phrag$~\citep[Rule 9]{Lackner2023:Multi-Winner}, is linear for $k=1$ but non-linear for all $k\ge 1$.
\begin{prop}
\label{prop:phrag} For every $m\ge 3$ and $k\le m-1$, 
seq-$\phrag$ is linear for $k=1$ but non-linear for every $k\ge 2$.
\end{prop}
\begin{proof}
We first show that seq-$\phrag$ satisfies homogeneity. Therefore, its domain can be extended to ${\mathbb Q}_{\ge 0}^{m!}$. Then, we show that the decision boundary between two committees in a certain region cannot be represented by finitely many linear classifiers.

\myparagraph{Extension to ${\mathbb Q}_{\ge 0}^{m!}$.} seq-$\phrag$ satisfies homogeneity because when executing the rule using its continuous version on $kP$, the order of selection of committee members is the same as in $P$, except that the speed  $k$ times faster. Since seq-$\phrag$ satisfies anonymity,  for any $\vec x\in {\mathbb Q}_{\ge 0}^{m!}$, we can define seq-$\phrag(\vec x) = $seq-$\phrag(t\vec x)$ for any $t$ such that $t\vec x\in {\mathbb Z}_{\ge 0}^{m!}$.

\myparagraph{Non-linear decision boundary.} We first prove the $k=2$ case. Consider the following polyhedron $\poly^*$ where $x_{\{1\}}+x_{\{1,2\}}+x_{\{3\}}=1$, $x_{\{1\}}\ge 2/3$, and all other $x$'s are $0$. For all $\vec x\in\poly^*$, clearly $1$ must be a cowinner and whether the other cowinner is $2$ or $3$ depends on the following comparison, based on the discrete procedure of seq-$\phrag$:
$$\frac{1+ \frac{x_{\{1,2\}}}{x_{\{1\}}+x_{\{1,2\}}}}{x_{\{1,2\}}}\text{ v.s. }\frac{1}{x_{\{3\}}}$$
When   LHS$<$RHS, $2$ is the other cowinner; when   LHS$>$RHS, $3$ is the other cowinner; and  when LHS$=$RHS, the other cowinner depends on the tie-breaking rule. Notice that  LHS equals to  $\frac{1}{x_{\{1,2\}}}+\frac{1}{1-x_{3}}$. 
Therefore, a 
sufficient conditions for the seq-$\phrag$ being $\{1,2\}$ and $\{1,3\}$ are as follows (using the discrete rule definition).

\begin{minipage}{0.35\textwidth}
\begin{align*}
 \{1,2\}: & x_{\{1,2\}} > \frac{x_{\{3\}}(1-x_{\{3\}})}{1-2x_{\{3\}}}\\
\{1,3\}: & x_{\{1,2\}} < \frac{x_{\{3\}}(1-x_{\{3\}})}{1-2x_{\{3\}}}
\end{align*}
\end{minipage}
\begin{minipage}{0.6\textwidth}
\includegraphics[width=.5\textwidth]{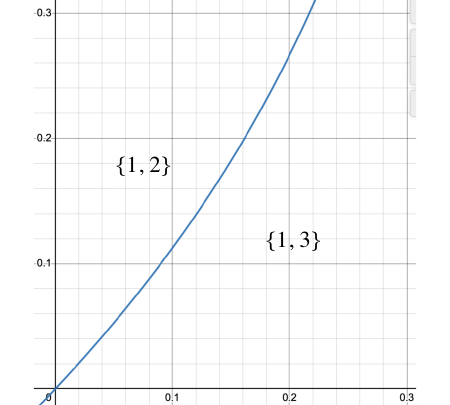}
\end{minipage}

In other words, the decision boundary between $\{1,2\}$ and $\{1,3\}$ is the non-linear curve $ x_{\{1,2\}} = \frac{x_{\{3\}}(1-x_{\{3\}})}{1-2x_{\{3\}}}$ illustrated in the figure on the right. It is not hard to verify that for all $x_{\{3\}}<\frac13$, $\left(\frac{x_{\{3\}}(1-x_{\{3\}})}{1-2x_{\{3\}}}\right)'' = \frac{2}{(1-2x_{\{3\}})^3}>0$, which means that the decision boundary is convex. Intuitively, this cannot be represented by combining finitely many separating hyperplanes, which means that seq-$\phrag$ is non-linear.

Formally, for the sake of contradiction, suppose seq-$\phrag$ can be represented by $K$ separating hyperplanes. Then, these planes partition the whole space into no more than $3^K$ convex regions. We chose $3^K+1$ points in $\poly^*$, denoted by $\{\vec x_1,\ldots,\vec x_{3^K+1}\}$, that are close to the decision boundary, such that 
\begin{itemize}
\item for every $\vec x_{i}$, seq-$\phrag(\vec x_i) = \{1,3\}$; and 
\item for every $\vec x_{i}, \vec x_{j}$, there exists a point  $\vec x$ on the line segment between them such that seq-$\phrag(\vec x_i) = \{1,2\}$.
\end{itemize}
The existence of such points are guaranteed by the convexity of the decision boundary. Then, due to the pigeon hole principle, there exist $\vec x_{i}$ and $\vec x_{j}$ that are in the same convex region, which means that the seq-$\phrag$ winner  of any point on the line segment between them should be $\{1,3\}$ as well, which is a contradiction. This proves that seq-$\phrag$ is non-linear for $k=2$.

The proof for general $k\ge 2$ proceeds by focusing on the polyhedron where only $x_{\{1,m-k+3,\ldots,m\}}$, $x_{\{1,2,m-k+3,\ldots,m\}}$ and $x_{\{3,m-k+3,\ldots,m\}}$ are non-zero. Or in other words, the $k-2$  alternatives $\{m-k+3,\ldots,m\}$ are approved by everyone, which means that they are cowinners by seq-$\phrag$  and does not affect the selection of the remaining (two) cowinners as every voter pays an equal price to chose them.
\end{proof}

\subsection{Proof of Theorem~\ref{thm:seq-combination}}
\appThmnoname{thm:seq-combination}{
Sequential rules and  reverse sequential  rules are linear.
}
\begin{proof}The proof is similar to the proof that STV is a generalized scoring rule~\citep{Xia08:Generalized}. To see that a sequential rule is linear, for every $i\le k$, we enumerate all possible combinations of round $i-1$ winners, and for each combination, use hyperplanes to test whether a given $i$-committee is a round $i$ co-winner. The $g$ function mimics the execution of the sequential rule. The proof for reverse sequential rules is similar. 
\end{proof}

\subsection{Proof of Theorem~\ref{thm:GST}}
\label{app:proof-thm:GST}
\appThmnoname {thm:GST}
{
$\core,\jr,\pjr,\ejr,\pjrp,\ejrp$, and $\fjr$ are GST axioms.
}
\begin{proof}
The theorem is proved by explicitly constructing the threshold function $\tau$ for each axiom. The case of $\core$ has been shown in the beginning of Section~\ref{sec:linear-axiom}. 

\myparagraph{\boldmath\bf  $\jr$:} 
Define
$$\calM_W^a\triangleq \{M\in 2^\ma\setminus \emptyset: a\in M\text{ and }M\cap W = \emptyset\}$$
Clearly, $W$  satisfies $\jr$ at $P$ if and only if for all $a\not\in W$, 
$$\vec w_{\calM_W^a} \cdot \hist(P)< \frac k n \text{, which is equivalent to }\left(\vec w_{\calM_W^a}-\frac 1 k\cdot \vec 1 \right)\cdot \hist(P)< 0$$   

Then, define the threshold function $\tau_\jr$ for $\jr$ as  
$$\tau_\jr(W,G)\triangleq  \begin{cases}
1 & \text{if }G  = \calM_W^a\text{ for some }a\in \ma\setminus W \\ k+1 & \text{otherwise}\end{cases}$$

\myparagraph{\boldmath\bf  $\ejr$:} Recall  that  $W\in\committee k$ satisfies $\ejr$ at a profile $P$ if and only if there do not exist a subset of votes $P'\subseteq P$ and $\ell \le m$, such that
\begin{enumerate}
\item [(i)] $|\bigcap_{V\in P'}|\ge \ell $, 
\item [(ii)]  $\forall V\in P'$, $|V\cap W|<\ell$,
\item [(iii)]  $P'$ contains at least $ \frac {\ell n}{ k}$ votes.
\end{enumerate}
The high-level idea behind the definition of $\tau$ is similar to that of $\jr$. Each   group is enumerated by specifying the intersection in (i), denoted by $A\subseteq  \ma$. Then, for every $W\in\committee k$ and every non-empty $A\subseteq  \ma$, let $\calM_W^A\subseteq 2^\ma$ denote the set of committees that contain $A$ and do not overlap with $W$ by more than $|A| -1$ alternatives, that is
$$\calM_W^A\triangleq \{M\in 2^\ma: A\subseteq M\text{ and }|M\cap W| \le |A| -1\}$$  
Then, define the threshold function $\tau_\ejr$ for $\ejr$ as  
$$\tau_\ejr(W,G)\triangleq  \begin{cases}
|A| & \text{if }G  = \calM_W^A\text{ for some }A\subseteq  \ma \\ k+1 & \text{otherwise}\end{cases}$$

\myparagraph{\boldmath\bf  $\ejrp$:} According to Definition~10 of~\citep{Brill2023:Robust}, the  difference between $\ejrp$ and $\ejr$ is that the violation of the former requires the group to approve a common alternative that is not in $W$, while the later requires the group to approve a common set of at least $\ell$ alternatives. Therefore,  for every $W\in\committee k$, every  $a\in\ma\setminus W$, and every $\ell\le k$, define $\calM_W^{a,\ell}\subseteq 2^\ma$ to be the set of committees that contain $a$ and do not overlap with $W$ by more than $\ell -1$ alternatives. That is,
$$\calM_W^{a,\ell}\triangleq \{M\in 2^\ma: a\in M\text{ and }|M\cap W| \le \ell -1\}$$  
Then, define the threshold function $\tau_\ejrp$ for $\ejrp$ as  
$$\tau_\ejrp(W,G)\triangleq  \begin{cases}
\ell & \text{if }G  = \calM_W^{a,\ell}\text{ for some }a \in  \ma \text{ and } \ell\le k \\ k+1 & \text{otherwise}\end{cases}$$

\myparagraph{\boldmath\bf  $\pjr$:} Recall that  $W\in\committee k$ satisfies $\pjr$ at a profile $P$ if and only if there do not exist a  subset of votes $P'\subseteq P$ and  $W'\subseteq W$ with  $|W'| = \ell-1$, such that
\begin{enumerate}
\item [(i)] $|\bigcap_{V\in P'}|\ge \ell $, 
\item [(ii)] $\forall V\in P'$, $V\cap W\subseteq W'$,
\item [(iii)] $P'$ contains at least $ \frac{\ell n}{k}$ votes.
\end{enumerate}
To enumerate the  group, we will enumerate  a set $A$ of $\ell$ commonly approved alternatives by the $\ell$-cohesive group for (i), and the set $W'\subset W$ with $|W'| = \ell-1 = |A|-1$ for  (ii), such that the intersection of $W$ and every type of preferences in the $\ell$-cohesive group is contained in $W'$. More precisely, for every $W\in\committee k$, every non-empty $A\subseteq  \ma$, and every $W'\subseteq W$ with $|W'| = |A|-1$,  define $\calM_W^{W',A}\subseteq 2^\ma$ to be the set of committees that contain $A$ and the overlaps with $W$ are contained in $W'$, that is
$$\calM_W^{W',A}\triangleq \{M\in 2^\ma: A\subseteq M\text{ and }M\cap W \subseteq W'\}$$  
Then, define the threshold function $\tau_\pjr$ for $\pjr$ as  
$$\tau_\pjr(W,G)\triangleq  \begin{cases}
|W'|+1 & \text{if }G  = \calM_W^{W',A}\text{ for some }A\in 2^\ma\setminus\emptyset \text{ and } W'\subseteq W\text{ with }|W'|=|A|-1 \\ k+1 & \text{otherwise}\end{cases}$$

\myparagraph{\boldmath\bf  $\pjrp$:} According to Observation~3 of~\citep{Brill2023:Robust}, the only difference between $\pjrp$ and $\pjr$ is that the violation of the former requires a sufficiently large group  who approve a common alternative that is not in $W$, while the later requires such a group to approve a common set of at least $\ell$ alternatives. Therefore, the threshold function of $\pjrp$ is similar to that of  $\pjr$  except that we only need to enumerate $A$ with $|A| = 1$, while $W'$ can be any subset of $W$. Formally, define the threshold function $\tau_\pjrp$ for $\pjrp$ as  
$$\tau_\pjrp(W,G)\triangleq  \begin{cases}
|A| & \text{if }G  = \calM_W^{W',\{a\}}\text{ for some }a\in \ma\setminus W \text{ and } W'\subseteq W \\ k +1 & \text{otherwise}\end{cases}$$

\myparagraph{\boldmath\bf  $\fjr$:} Recall that  $W\in\committee k$ satisfies $\fjr$ at a profile $P$ if and only if there do not exist a subset of votes $P'\subseteq P$, $\beta\in\mathbb N$, and $T\subseteq\ma$, such that
\begin{enumerate}
\item  [(i)] $\forall V\in P'$, $|V\cap T|\ge \beta$,
\item  [(ii)] $\forall V\in P'$, $|V\cap W|<\beta$,
\item  [(iii)] $P'$ contains at least $ \frac{|T| n}{k}$ votes.
\end{enumerate}
Define
$$\calM_W^{\beta,T}\triangleq \{M\in 2^\ma: |M\cap T|\ge \beta\text{ and }|M\cap W| <\beta\}$$  
Then, define the threshold function $\tau_\fjr$ for $\fjr$ as  
$$\tau_\fjr(W,G)\triangleq  \begin{cases}
|T| & \text{if }G  = \calM_{W}^{\beta,T}\text{ for some }\beta\in\mathbb N \text{ and } T\subseteq\ma \\ k+1 & \text{otherwise}\end{cases}$$
\end{proof}

\subsection{Remaining proof for Proposition~\ref{prop:gmp}}
\label{app:proof-prop:gmp}
\begin{proof}
We continue the proof sketch in the main text by defining $\psi$ for other rules.

\begin{itemize} 
\item {\bf \ABCS{} ($\hypABCS$).}  Define $\psi(A,W)$ to be the $(|\calX_{m,k}|-1)$-dimensional binary vector such that for any $(a,b) \in\calX_{m,k}\setminus\{(0,0)\}$, the $(a,b)$-th coordinate of $\psi(A,W)$ takes $1$ if and only if $|A\cap W|=a$ and $|A| =b$; otherwise it takes $0$. 
\item {\bf \GABCS ($\hypGABCS$).} Define $\psi(A,W)$ to be the $(2^m\times {m\choose k}-1)$-dimensional binary vector such that for any $(A,W) \in2^\ma\times \committee k \setminus\{(\emptyset,\{1,\ldots, k\})\}$, the $(A,W)$-th coordinate of $\psi(A,W)$ takes $1$   and other coordinates take $0$. 
\item{\bf Committee scoring rule ($\hypCSR$).} Note that any committee scoring rule can be equivalently parameterized by the score of the set of ranks of alternatives in $W$  in $R$, and there are ${m \choose k}$ of such sets. Define $\psi(R,W)$ to be the $( {m\choose k}-1)$-dimensional binary vector such that for any $T\subseteq \{1,\ldots, m\}$ with $T\ne \{1,\ldots,k\}$, the $T$-th coordinate of $\psi(R,W)$ takes $1$ if and only if $T = \{\rank(R,a):a\in W\}$; otherwise it takes $0$. 
\item {\bf Neutral SRSF ($\hypnSRSF$).} Let $R_1  = [1\succ \cdots\succ m]$. For any $R \in \listset m$, let $\sigma_{R}$ denote the permutation over $\listset m$ that maps $R_1$ to $R$. For any $R,R'\in \listset m$, define $\psi(R,R')$ to be the $(m!-1)$-dimensional binary vector that are indexed by $\listset m\setminus \{R_1\}$, such that the $\sigma_{R}^{-1}(R')$-th coordinate is $1$ if and only if $\sigma_{R}^{-1}(R')\ne R_1$; otherwise it takes $0$. 
\item  {\bf  SRSF ($\hypSRSF$).} The definition is similar to that of neutral SRSF. Define $\psi(R,R')$ to be the $((m!)^2-1)$-dimensional binary vector that takes $1$ on the $(R,R')$-th coordinate except when $(R,R') = (R_1,R_1)$; otherwise it takes $0$. 
\end{itemize}
A similar normalization for OWA and ordinal OWA does not work, because the weight vectors are already implicitly normalized, where the alternatives that do not appear in the $k$-committee under consideration can be viewed as having $0$ weight.
\begin{itemize}  
\item {\bf OWA ($\hypOWA$).} For any $\vec u\in \mathbb R^m$ and any $W\in\committee k$, define $\psi(\vec u,W)$ to be the $k$-dimensional vector whose $i$-th coordinate is the $i$-th largest value  in multiset $\{u_{i'}:i'\in W\}$. 
\item{\bf Ordinal OWA ($\hypOWAr^{\vec u}$).} For any $R\in \listset m$ and any $W\in\committee k$, define $\psi(R,W)$ to be the $k$-dimensional vector whose $i$-th coordinate is the $i$-th largest value  in multiset $\{u_{i'}:i'\in W\}$.\end{itemize}
\end{proof}

\section{Materials for Section~\ref{sec:learning}}

\subsection{Proof of Claim~\ref{claim:regions}}
\label{app:proof-claim:regions}
\appClaimnoname {claim:regions}
{
$A(d, k)=\sum_{s=0}^{\min k, d} {k \choose s}2^s$.
}
\begin{proof}  
Consider adding one hyperplanes to the existing $k$ hyperplanes, the number of regions is increased by two times the number of regions in the new hyperplane (which is an $(d-1)$-dimensional Euclidean space) divided by the existing $k'$ hyperplanes. Therefore, we have $A(d,k+1) \le A(d,k)+ A(d-1,k)$. In fact, the equation holds when the $k+1$ hyperplanes are in general positions. Therefore, we have 
$$A(d,k+1) = A(d,k)+ 2\times A(d-1,k)$$
We now prove the claim by induction. Clearly the claim holds for $k=0$ because when no hyperplane is used, the only subspace is $\mathbb R^{d}$.  Then, we have
\begin{align*}
A(d,k+1) = & \sum\nolimits_{s=0}^{\min k, d} {k \choose s}2^s + 2\times \sum\nolimits_{s=0}^{\min k, d-1} {k \choose s}2^s
\end{align*}
When $k+1\le d$, the right hand side is $3\times \sum\nolimits_{s=0}^{k} {k \choose s}2^s = 3^{k+1} = \sum\nolimits_{s=0}^{k+1} {k+1 \choose s}2^s$. When $k+1> d$, we have 
\begin{align*}
A(d,k+1) = & \sum\nolimits_{s=0}^{d} {k \choose s}2^s + 2\times \sum\nolimits_{s=0}^{d-1} {k \choose s}2^s
= {k\choose 0} + \sum\nolimits_{s=1}^{d} \left({k \choose s}+{k \choose s-1}\right)2^s\\
=& \sum\nolimits_{s=0}^{d} {k+1\choose s} 2^s
\end{align*}
This proves the claim.
\end{proof}

\subsection{Remaining Proof of Theorem~\ref{thm:ndim-parameterized-maximizer}}
\label{app:proof-thm:ndim-parameterized-maximizer}
\begin{proof}
Next, we prove that $(\vec x_1^*,\ldots,\vec x_T^*)$ is shattered by linear binary classifiers in $\mathbb R^\eta$. Let $\vec w\in \mathbb R^\eta$ be such that $f_{\vec w}$ equals to $f^0$ or $f^1$ on each of $(P_1,\ldots, P_T)$. For every $t\le T$, let $\gamma_t = \max_{d\in\adecspace}\psi(P,d)$. There are four cases.
\begin{itemize}
\item $f^1(P_t)\not\subseteq f^0(P_t)$ and $f_{\vec w}(P_t) = f^1(P_t)$. Then, by definition  $\psi(P_t,f^1(P_t))\cdot \vec w  = \gamma_t \ge \psi(P_t,f^0(P_t))\cdot \vec w$, which is equivalent to $\vec x_t^*\cdot\vec w\ge 0$.
\item $f^1(P_t)\not\subseteq f^0(P_t)$ and  $f_{\vec w}(P_t) = f^0(P_t)$.   Therefore, there exists $d\notin f^0(P_t)$, $d\in f^1(P_t)$, which means that $\psi(P_t,d)\cdot \vec w<\gamma_t$. This means that  
$$\psi(P_t,f^1(P_t))\cdot \vec w  < \gamma_t = \psi(P_t,f^0(P_t))\cdot \vec w,$$
which is equivalent to $\vec x_t^*\cdot\vec w< 0$. 
\item $f^1(P_t)\subseteq f^0(P_t)$ and $f_{\vec w}(P_t) = f^1(P_t)$. Recall that $f^1(P_t)\ne f^0(P_t)$. Therefore, there exists  $d\in f^0(P_t)$ and  $d\notin f^1(P_t)$, which means that $\psi(P_t,d)\cdot \vec w<\gamma_t$. We have 
\begin{align*}
&\psi(P_t,f^0(P_t))\cdot \vec w  < \gamma_t = \psi(P_t,f^1(P_t))\cdot \vec w \Leftrightarrow (\psi(P_t,f^0(P_t)) - \psi(P_t,f^1(P_t)))\cdot \vec w<0 \\
\Leftrightarrow &\vec x_t^*\cdot\vec w< 0
\end{align*}
\item $f^1(P_t)\subseteq f^0(P_t)$ and $f_{\vec w}(P_t) = f^0(P_t)$. By definition we have
\begin{align*}
&\psi(P_t,f^0(P_t))\cdot \vec w  = \gamma_t \ge  \psi(P_t,f^1(P_t))\cdot \vec w \Leftrightarrow (\psi(P_t,f^0(P_t)) - \psi(P_t,f^1(P_t)))\cdot \vec w\ge 0 \\
\Leftrightarrow &\vec x_t^*\cdot\vec w\ge 0
\end{align*}
\end{itemize}
In other words, comparing the outcomes of $f_{\vec w}$ on $(P_1,\ldots,P_T)$ and the signs of $\vec w$ on $(\vec x_1^*, \ldots,\vec x_T^*)$, we notice that they agree on all $t\le T$ such that $f^1(P_t)\not\subseteq f^0(P_t)$ and are opposite to each other on all other $t$'s (i.e., those such that $f^1(P_t)\subseteq f^0(P_t)$). Because $(P_1,\ldots,P_T)$ is shattered by $\hypspace_{\psi}$,  every combination of $T$ signs can be realized by an $f_{\vec w} \in \hypspace_{\psi}$. This means that every combination of $T$ signs over $(\vec x_1^*,\ldots,\vec x_T^*)$ can be realized by a $\vec w\in\mathbb R^\eta$. That is, $(\vec x_1^*,\ldots,\vec x_T^*)$ is shattered by linear binary classifiers in $\mathbb R^\eta$. This proves that $T\le \eta$ and completes the proof of the theorem.
\end{proof}

\subsection{Remaining Proof of Theorem~\ref{thm:ndim-lower}}
\label{app:proof-thm:ndim-lower}

\begin{proof}  

\myparagraph{$\hypGABCS$.} Let $W^0 = \{1,\ldots, k\}$. Let $\calP= \{P_{A,W}:  W\in \committee k\setminus \{W^0\}, A\in 2^\ma\setminus \{W,W^0\}\}$, where $P_{A,W} = \{W,W^0,A\}$.
To see that $\calP$ is shattered by $\hypGABCS$, for every  $P_{A,W}\in \calP$, let $f^0 (P_{A,W}) = \{W^0\}$ and  $f^1 (P_{A,W}) = \{W\}$. Then, for every $B\subseteq \calP$, define  $\GABCS_{\vec w}$ as follows, where elements in $\vec w$ are indexed by $(A,W)\in 2^\ma \times \committee k$. 
\begin{itemize}
\item For every $A\in 2^\ma\setminus\{W^0\}$, let $w_{A,W^0} = 2$. Let $w_{W^0,W^0} = 6$.
\item For every $W\in\committee k$ and every $A\in \committee k\setminus \{W,W^0\}$, let $w_{A,W} = \begin{cases}
1&\text{if }P_{A,W}\in B\\
3&\text{otherwise}\end{cases}$. Let $w_{W,W} = 6$ and $w_{W^0,W} = 2$. 
\end{itemize}
  Then, for every $W\in \committee k\setminus \{W^0\}$ and every $A\in 2^\ma\setminus \{W,W^0\}$,   we first prove that $f_{\vec w}(P_{A,W})\subseteq \{W,W^0\}$. Notice that for every $W'\notin \{W,W^0\}$, 
\begin{align*}
\score(P_{A,W}, W) - \score(P_{A,W}, W') & =  w_{W,W} - w_{W,W'} - (w_{W^0,W} - w_{W^0,W'}) + w_{A,W} - w_{A,W'}\\
& \ge    6-3 - (2-2)  + 1-3 >0
\end{align*}
Also, 
\begin{align*}
\score(P_{A,W}, W) - \score(P_{A,W}, W^0) & =  w_{W,W} - w_{W,W^0} - (w_{W^0,W} - w_{W^0,W^0}) + w_{A,W} - w_{A,W^0}\\
& \ge   6-2 - (6-2)  + w_{A,W}-2 = \begin{cases}
-1&\text{if }P_{A,W}\in B\\
1&\text{otherwise}\end{cases}
\end{align*}
This means that $f_{\vec w} (P_{A,W}) = \begin{cases}
\{W^0\} = f^0(P_{A,W})&\text{if }P_{A,W}\in B\\
\{W \}= f^1(P_{A,W})&\text{otherwise}\end{cases}$. 
Therefore, $\calP$ is shattered by $\hypGABCS$, which means that $\Ndim(\hypGABCS) \ge |\calP| = (2^m-2)({m\choose k }-1)$.

\myparagraph{$\hypCSR$.} Let $\calS_{m,k}\subseteq 2^{[m]}$ denote the set of all $k$ subsets in $[m]$, which represent the combinations of $k$ {\em positions}. We define a directed graph whose vertices are $\calS_{m,k}$ and there is an edge $S_1\ra S_2$ if and only if there exists $S\subseteq [m]$ and $i$ such that $S_1 = S\cup \{i \}$  and $S_2 = S\cup \{i+1\}$.  Let $\calT$ denote an arbitrary but fixed spanning tree of this graph. For example the graph for $m=5$, $k=3$ and a spanning tree are illustrated in Figure~\ref{fig:CSR}. 
\begin{figure}[htp]
\centering
\includegraphics[width = .3\textwidth]{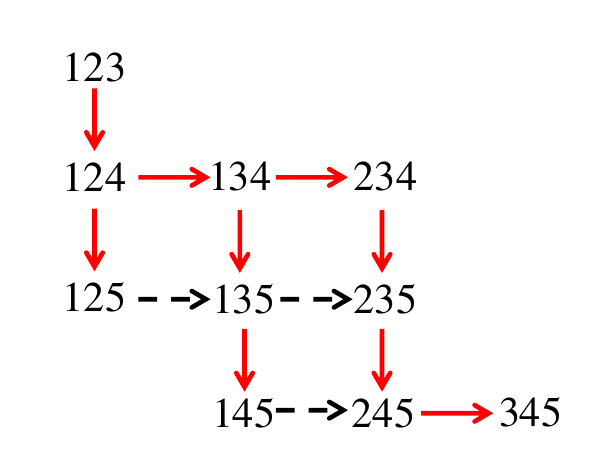}
\caption{The graph over $\calS_{m,k}$. $\calT$ contains solid edges. $\calT' = \calT - (123,124)$. \label{fig:CSR}}
\end{figure}

Notice that $\calT$ must contain the edge  $\{1,\ldots,k\}\ra \{1,\ldots,k-1,k+1\}$) because this is the only outgoing edge from $\{1,\ldots,k\}$. Let $\calT^-$  denote the tree obtained from $\calT$ by removing this edge. It follows that $\{1,\ldots,k-1,k+1\}$ is the root of $\calT$.  Then, for every edge $S_1\ra S_2\in \calT^-$ we define a profile $P_{S_1\ra S_2}$ as follows.  Let $S_1 = S\cup \{i\}$  and $S_2 = S\cup \{i+1\}$. Let $R_1$ (respectively, $R_2$) denote an arbitrary linear order where alternatives $\{3,4,\ldots, k+1\}$ are ranked in top-$|S|$ positions and alternative $1$ (respectively, $2$) is ranked at the $|S|$-th position. Let $R$ denote an arbitrary linear order where alternative $1$ (respectively, $2$) is ranked at the $i$-th position (respectively, $(i+1)$-th position) and the ranks of alternatives $\{3,4,\ldots, k+1\}$ are $S$. Then, define
$$P_{S_1\ra S_2} \triangleq 2\times \{R_1,R_2\}\cup \{R\}$$
For example, consider $m=5$, $k=3$, and the $235\ra 245$ edge in $\calT'$ in Figure~\ref{fig:CSR}. We have $i=3$ and  $S = \{2,5\}$. $W^0 = \{1,3,4\}$, $W^1 = \{2,3,4\}$, $R_1 = 3\succ 4\succ 1\succ 2\succ 5$, $R_2 = 3\succ 4\succ 2\succ 1\succ 5$. In $R$, alternatives $\{3,4\}$ should be ranked at the top two positions, alternative $1$ should be ranked at the 3rd  position, and alternative $2$ should be ranked at the 4th position. Therefore, one choice of $R$ is $[3\succ 4\succ 1\succ 2\succ 5]$.

Let $\calP = \{P_{S_1\ra S_2}: S_1\ra S_2\in \calT^-\}$.  To see that $\calP$ is shattered by $\hypCSR$, let $W^0 = \{1,3,4,\ldots, k+1\}$ and $W^1 = \{2,3, 4,\ldots, k+1\}$. For every  $P_{S_1\ra S_2}\in \calP$, let $f^0 (P_{S_1\ra S_2}) = \{W^0\}$ and  $f^1 (P_{S_1\ra S_2}) = \{W^1\}$. Then, recall that any $\CSR$'s can be equivalently specified by a function $\mathbf s: \calS_{m,k}\ra\mathbb R$. For every $B\subseteq \calP$,  define $\CSR_{\mathbf s}$ as follows. 
\begin{itemize}
\item $\mathbf s(\{1,\ldots,k\})=8{m\choose k}$ and $\mathbf s(\{1,\ldots,k-1,k+1\})=0$. 
\item For every $S_1\ra S_2 \in \calT^-$, let $ s(S_1) = \begin{cases}
\mathbf s(S_2)+ 1&\text{if }P_{S_1\ra S_2}\in B\\
\mathbf s(S_2)-1&\text{otherwise}\end{cases}$. 
\end{itemize}
Because $\calT^-$ is a spanning tree over $\calS_{m,k}\setminus\{\{1,\ldots,k\}\}$, $\mathbf s$ is well-defined. Also, for every $S'\in\calS_{m,k}\setminus\{1,\ldots,k\}$, we have $|\mathbf s(S')|\le {m\choose k}$. For every $S_1\ra S_2\in \calT^-$ and any $W\in\committee k\setminus\{W^0,W^1\}$, we have 
\begin{align*}
&\score_{\mathbf s}(P_{S_1\ra S_2},W^0)-\score_{\mathbf s}(P_{S_1\ra S_2},W)\ge  2\mathbf s(\{1,\ldots,k\}) - 7\times {m\choose k} - \mathbf s(\{1,\ldots,k\}) > 0
\end{align*}
Also, 
\begin{align*}
\score_{\mathbf s}(P_{S_1\ra S_2},W^0) = &2(\mathbf s(\{1,\ldots,k\}) + \mathbf s(\{1,\ldots,k-1,k+1\}))+\mathbf s(S_1)\\
\score_{\mathbf s}(P_{S_1\ra S_2},W^1) = &2(\mathbf s(\{1,\ldots,k\}) + \mathbf s(\{1,\ldots,k-1,k+1\}))+\mathbf s(S_2),
\end{align*}
which means that 
\begin{align*}
\score_{\mathbf s}(P_{S_1\ra S_2},W^0) -\score_{\mathbf s}(P_{S_1\ra S_2},W^1) = \mathbf s(S_1) - \mathbf s(S_2) = \begin{cases}
1&\text{if }P_{S_1\ra S_2}\in B\\
-1&\text{otherwise}\end{cases}
\end{align*} 
Therefore, $\CSR (P_{S_1\ra S_2}) = \begin{cases}
\{W^0\} = f^0(P_{S_1\ra S_2})&\text{if }P_{S_1\ra S_2}\in B\\
\{W^1 \}= f^1(P_{S_1\ra S_2})&\text{otherwise}\end{cases}$. 
Therefore, $\calP$ is shattered by $\hypCSR$, which means that $\Ndim(\hypCSR) \ge |\calP| = {m\choose k }-2$.

\myparagraph{$\hypOWAr^{\vec u}$ and $\hypOWA$.} Since $\hypOWAr^{\vec u}\subseteq \hypOWA$, it suffices to prove the lower bound for $\hypOWAr^{\vec u}$. Let $\vec u = (u_1,\ldots, u_k)$ denote the utility vector for $\hypOWAr$ and let $W^0= \{1,3,\ldots, k+1\}$, $W^1= \{2,3,\ldots, k+1\}$. Let $\calP = \{P_1,\ldots, P_{k-1}\}$ such that for all $i\le k-1$,
\begin{align*}
P_i \triangleq &3\times \{3\succ \cdots\succ k+1\succ 1\succ\others,3\succ \cdots\succ k+1\succ 2\succ\others\}\\
&\cup \{3\succ \cdots \succ i+1\succ 1\succ 2\succ\others\}
\end{align*}
To see that $\calP$ is shattered by $\hypOWAr$, let $f^0$ (respectively, $f^1$) denote the function that always output $\{W^0\}$ (respectively, $\{W^1\}$) on all profiles in $\calP$. Then, for every $B\subseteq \calP$, we define  $\OWA_{\vec w}^{\vec u}$ as follows, where $\vec w = (s_1,\ldots, s_k)$: let $s_1=6k$ and for every $i\le k-1$, define 
$$w_{i+1} = \begin{cases}
\frac{u_iw_i-1}{u_{i+1}}&\text{if }P_i\in B\\
\frac{u_iw_i+1}{u_{i+1}}&\text{otherwise}\end{cases}$$
Next, we prove that for every $P_i$, under $\OWA_{\vec w}^{\vec u}$, the score  of every alternative in $\{3,\ldots,k+1\}$ is strictly higher than the scores of $1,2$, which are strictly higher than the score of any  other alternative. According to the definition of $\vec w$, for every $i\le k-1$, we have $u_1w_1-k\le u_iw_i\le u_1w_1+k$. Notice  
$$\score_{\vec w}^{\vec u}(P_i,1) = 3u_kw_k + u_iw_i, \score_{\vec w}^{\vec u}(P_i,2) = 3u_kw_k + u_{i+1}w_{i+1}$$
For every $\ell_1 \in \{3,\ldots,k+1\}$ and every $\ell_1 \in \{ k+2,\ldots, m\}$, 
\begin{align*}
&\score_{\vec w}^{\vec u}(P_i,\ell_1)\ge 6\times (u_1w_1-k) \ge 4(u_1w_1+k) &\text{because }u_1\ge 1\text{ and }w_1>5k\\
\ge & \score_{\vec w}^{\vec u}(P_i,1) > u_1w_1+k \ge \score_{\vec w}^{\vec u}(P_i,\ell_2)
\end{align*}
This proves that $\{3,\ldots, k+1\} \subseteq \OWA_{\vec w}^{\vec u}(P_i)\subseteq \{1,\ldots, k+1\}$. Then, notice that 
$$\score_{\vec w}^{\vec u}(P_i,1) - \score_{\vec w}^{\vec u}(P_i,2) = u_iw_i-u_{i+1}w_{i+1} =\begin{cases}
1 & \text{if }P_i\in B\\
-1 & \text{otherwise}
\end{cases}$$
Therefore, $\OWA_{\vec w}^{\vec u}(P_i) = \begin{cases}
\{W^0\} & \text{if }P_i\in B\\
\{W^1\} & \text{otherwise}
\end{cases}$, which proves that $\calP$ is shattered by $\hypOWAr^{\vec u}$, and therefore $\Ndim(\hypOWA)\ge \Ndim(\hypOWAr^{\vec u})\ge |\calP| = k-1$.

\myparagraph{$\hypPos$.}  Let $\calP =  \{P_1,\ldots, P_{k-2}\}$ such that for all $i\le k-2$,
\begin{align*}
P_i = & 2m\times \{1\succ 2\succ\others, 2\succ 1\succ \others\}\cup \{3\succ \cdots\succ j+1\succ 1\succ 2\succ \others, \others\succ 2\succ 1 \} 
\end{align*}
To see that $\calP$ is shattered by $\hypPos$, let $f^0$ (respectively, $f^1$) denote the function that always output $\{1\}$ (respectively, $\{2\}$) on all profiles in $\calP$. Then, for every $B\subseteq \calP$, define a positional scoring rule $\Pos_{\vec s}$ with scoring vector $\vec s = (s_1,\ldots,s_m)$ as follows. Let $s_m=0, s_{m-1} = 2$, and for every $i\le m-1$, define $s_i = \begin{cases}s_{i+1}+ 3 &\text{if }P_i\in \calP\\
s_{i+1}+ 1 &\text{otherwise}\end{cases}$.  Then, for every $\ell \in \{3,\ldots, m\}$, we have 
$$\score_{\vec s}(P_i, 1) - \score_{\vec s}(P_i, \ell ) \ge 8m - 2\times 3m>0\text{, and}$$
$$\score_{\vec s}(P_i, 1) - \score_{\vec s}(P_i, 2) = (s_i-s_{i+1}) - (s_2 - s_1) = \begin{cases}1 &\text{if }P_i\in \calP\\
-1 &\text{otherwise}\end{cases},$$
which means that $\Pos_{\vec s}(P_i) = \begin{cases}\{1\} &\text{if }P_i\in B\\
\{2\} &\text{otherwise}\end{cases}$. Therefore, $\calP$ is shattered by $\hypPos$, which means that $\Ndim(\hypPos) \ge |\calP| = k-2$. 

\myparagraph{$\hypnSRSF$.} Fix $R^* = [1\succ 2\succ \others]$ and $R' = [2\succ 1\succ \others]$, where alternatives in ``$\others$'' are ordered alphabetically. For any  rankings $R$, let $\sigma_{R}$ denote the permutation over $\listset m$ that maps $R^*$ to $R$, that is, $\sigma_R(R^*) = R$. Then, let $\listset m = L_1\cup\cdots\cup L_T$ denote the  partition of $\listset m$ such that  for every $t\le T$, $|L_t|$ is $1$ or $2$; $L_t = \{R\}$ if and only if $\sigma_{R} = \sigma_R^{-1}$; and $L_t = \{R_1,R_2\}$ if and only if $\sigma_{R_1} = \sigma_{R_2}^{-1}$. For example, when $m=3$, the partition is 
$$\listset 3 = \{123\}\cup \{213\}\cup \{132\}\cup \{321\}\cup \{231,312\}$$
It follows that $|T|> \frac{m!}{2}$ and $\{R^*\}$ and $\{R'\}$ are elements in this partition. W.l.o.g.~let $L_{T} = \{R^*\}$ and  $L_{T-1} = \{R'\}$. For every $t\le T$, let $R_t\in L_t$ denote an arbitrary but fixed ranking in $L_t$. Then, define $\calP = \{P_t:1\le t\le T-2\}$, where 
$$P_t = 3\times \{R^*,R_t\}\cup\{R'\}$$
To see that $\calP$ is shattered by $\hypnSRSF$, for every $t\le T-2$, define   $f^0(P_t) = \{R^*\}$ and $f^1(P_t) = \{R_t\}$. Then, for every $B\subseteq \calP$, define $\SRSF_{\mathbf s}$ as follows, where $\mathbf s:\listset m\times \listset m\ra \mathbb R$ is a neutral scoring function, which means that it suffices to specify ${\mathbf s}(R^*,R)$ for all $R\in\listset m$ and then extend it to the full domain of $\mathbf s$ via neutrality. 
Let  ${\mathbf s}(R^*,R^*) = 8$, ${\mathbf s}(R^*,R') = 2$, and for every $t\le T-2$ and every $R\in L_t$, let ${\mathbf s}(R^*,R)=\begin{cases} 1 &\text{if } P_t\in B\\
3&\text{otherwise}\end{cases}$.

For every $t\le T-2$ and every $R\in \listset m \setminus ( \{R^t,R^*\})$, we have 
\begin{align*}
&\score_{\mathbf s}(P_t, R^*) - \score_{\mathbf s}(P_t, R)\\
= & 3\times ({\mathbf s}(R^*,R^*)+ {\mathbf s}(R_t,R^*)) + {\mathbf s}(R',R^*) - 3\times ({\mathbf s}(R^*,R)+ {\mathbf s}(R_t,R)) - {\mathbf s}(R',R)\\
\ge &3\times (8+1)+1 -3\times(3+3) - 8>0
\end{align*}
Also, 
\begin{align}
&\score_{\mathbf s}(P_t, R^*) - \score_{\mathbf s}(P_t, R_t)\notag\\
= & 3\times ({\mathbf s}(R^*,R^*)+ {\mathbf s}(R_t,R^*)) + {\mathbf s}(R',R^*) - 3\times ({\mathbf s}(R^*,R_t)+ {\mathbf s}(R_t,R_t)) - {\mathbf s}(R',R_t)\notag\\
= & 3\times({\mathbf s}(R^*,R^*) - {\mathbf s}(R_t,R_t)) + 3\times ({\mathbf s}(R_t,R^*) - {\mathbf s}(R^*,R_t)) + {\mathbf s}(R',R^*) -  {\mathbf s}(R',R_t)\notag\\
= &  {\mathbf s}(R',R^*) -  {\mathbf s}(R',R_t) \label{eq:neutrality}\\
 = &  \begin{cases} 1 &\text{if } P_t\in B\\
-1&\text{otherwise}\end{cases}\notag
\end{align}
\eqref{eq:neutrality} holds because due to neutrality, ${\mathbf s}(R^*,R^*) - {\mathbf s}(R_t,R_t) = 0$, and ${\mathbf s}(R_t,R^*) - {\mathbf s}(R^*,R_t) = {\mathbf s}(R^*,\sigma_{R_t}^{-1}(R_t)) - {\mathbf s}(R^*,R_t)$. Recall that according to the definition of $R_t$ (and $L_t$), we have $\sigma_{R_t}^{-1}(R^*) = \sigma_{R_t}(R^*) = R_t$, which means that ${\mathbf s}(R^*,\sigma_{R_t}^{-1}(R^*)) - {\mathbf s}(R^*,R_t) = 0$. Therefore, $\calP$ is shattered by $\hypnSRSF$, which means that $\Ndim(\hypnSRSF) \ge |\calP| \ge m!/2 -1$.

\myparagraph{$\hypSRSF$.} Let $\listset m = \{R_1,\ldots,R_{m!}\}$. Let $\calP = \{P_{t_1,t_2}: t_1>t_2\}$, where $P_{t_1,t_2} = \{R_{t_1},R_{t_2}\}$. To see that $\calP$ is shattered by $\hypSRSF$, define two functions $f^0,f^1$ such that for every $P_{t_1,t_2}\in \calP$, $f^0(P_{t_1,t_2}) = \{R_{t_1}\}$  and $f^1(P_{t_1,t_2}) = \{R_{t_2}\}$.  For every $B\subseteq \calP$, define $\SRSF_{\mathbf s}$ as follows, where $\mathbf s:\listset m\times \listset m\ra \mathbb R$.
\begin{itemize}
\item For every $R\in\listset m$, $\mathbf s (R,R) = 6$.
\item For every $t_1>t_2$, $\mathbf s (R_{t_1},R_{t_2}) =  \begin{cases} 3 &\text{if } P_t\in B\\
1&\text{otherwise}\end{cases}$ and let $\mathbf s (R_{t_2},R_{t_1}) = 2$. 
\end{itemize}
For every $R\in \listset m \setminus \{R_{t_1}, R_{t_2}\}$, we have 
\begin{align*}
\score_{\mathbf s}(P_{t_1,t_2},R_{t_1}) -\score_{\mathbf s}(P_{t_1,t_2},R) \ge 6+1 - (3+3) >0
\end{align*}
Also, 
\begin{align*}
\score_{\mathbf s}(P_{t_1,t_2},R_{t_1}) -\score_{\mathbf s}(P_{t_1,t_2},R_{t_2})  = \mathbf s (R_{t_2}, R_{t_1}) -  \mathbf s (R_{t_1}, R_{t_2}) = \begin{cases} 1 &\text{if } P_t\in B\\
-1&\text{otherwise}\end{cases}
\end{align*}
Therefore, $\calP$ is shattered by $\hypSRSF$, which means that $\Ndim(\hypSRSF) \ge |\calP| \ge (m!-1)(m! -2)/2$. 
\end{proof}

\subsection{Proof of Claim~\ref{claim:lb-combinatorics}}
\label{app:proof-claim:lb-combinatorics}
\appClaimnoname{claim:lb-combinatorics} {For any $\eta\in\mathbb N$, 
${\eta \choose \lfloor \eta/2\rfloor} \ge \sqrt{\frac{2}{\pi}}\cdot \frac{2^\eta}{\sqrt \eta}$.
}

\begin{proof}  It is easy to verify that the inequality holds when $\eta=1$ holds. In the rest of the proof we assume $\eta\ge 2$. We prove the claim by applying Robbins' refinement of Stirling's approximation~\citep{Robbins1955:Remark}. When $\eta$ is an even number, we have 
\begin{align*}
{\eta \choose \lfloor \eta/2\rfloor}  = \frac{\eta!}{\frac \eta 2 !} \ge &\frac{\sqrt{2\pi \eta}\left(\frac{\eta}{e}\right)^\eta e^{\frac{1}{12\eta+1}}}{\left(\sqrt{\pi \eta}\left(\frac{ \eta/2 }{e}\right)^{ \eta/2 }e^{\frac{1}{12 \eta/2 }}\right)^2}
=  2^\eta \sqrt {\frac{2}{ \pi\eta}}e^{{\frac{1}{12\eta+1} -  \frac{1}{3 \eta}}} >  \sqrt{\frac{2}{\pi}}\frac{2^\eta}{\sqrt \eta}
\end{align*}
When $\eta$ is an odd number, we have
\begin{align*}
{\eta \choose \lfloor \eta/2\rfloor}  = & \frac{\eta!}{\frac {\eta +1}{2} (\frac {\eta -1}{2} !)^2} \ge  \frac{\sqrt{2\pi \eta}\left(\frac{\eta}{e}\right)^\eta e^{\frac{1}{12\eta+1}}}{\frac{\eta+1}{2}\left(\sqrt{\pi (\eta-1)}\left(\frac{ (\eta-1)/2 }{e}\right)^{(\eta-1)/2 }e^{\frac{1}{12 (\eta-1)/2 }}\right)^2}\\
= &  2^\eta \sqrt {\frac{2}{ \pi\eta}}\times e^{{\frac{1}{12\eta+1} -  \frac{1}{3 (\eta-1)}}} \times \frac{\eta^{2}}{(\eta-1)(\eta+1)}\times \left (\frac{\eta}{\eta-1}\right)^{\eta-1}\times \frac {1}{e}>  \sqrt{\frac{2}{\pi}}\frac{2^\eta}{\sqrt \eta}
\end{align*}
The last inequality holds because $\eta^2>(\eta-1)(\eta +1)$ and when $\eta\ge 2$, $\left (\frac{\eta}{\eta-1}\right)^{\eta-1}=\left (1+\frac{1}{\eta-1}\right)^{\eta-1}>e$.
\end{proof}

\subsection{Proof of Theorem~\ref{thm:ndim-spa}}
\label{app:proof-thm:ndim-spa}
\appThm{\bf Sample complexity: symmetric GST axioms}
{thm:ndim-spa}{
$$\sqrt{\frac{2}{\pi}}\cdot \frac{ 2^{2^m}}{2^m\cdot m!}\le \Ndim(\hypsGST)\le 2^{2^m} \log (k+2)$$
}
\begin{proof} The upper bound follows after noticing that $|\hypsGST|\le (k+2)^{2^{2^m}}$, which means that $\Ndim(\hypsGST) \le \log (|\hypsGST|) = 2^{2^m}\log (k+2)$. To prove the lower bound, like the proof of Theorem~\ref{thm:ndim-pa}, let $\calS^*\subseteq 2^\prefspace$ denote the set of all subsets of $\prefspace$ with $\lfloor |\prefspace|/2\rfloor$ elements.  We  partition $\calS^*$ according to equivalence relationship $\equiv$, such that $G\equiv G'$ if and only if there exists a permutation $\sigma$ over $\ma$ such that $\sigma(G) = G'$. Let $\calS^* = L_1\cup\ldots\cup L_{T}$ denote the partition.  For every $j\le T$, let $G_j\in L_j$ denote an arbitrary but fixed element. Then let $\calP= \{P_1,\ldots, P_{T}\}$ where $P_j = G_j$. For every $P_j\in \calP$, let $f^0(P_j) = \emptyset$ and let $f^1(P_j) = \committee k$. For any $B\subseteq \calP$, define $\GST_{\tau}$ as follows, where $\tau:\committee k \times 2^\prefspace\ra \{0,\ldots,k+1\}$ is a symmetric group-satisfaction-threshold function.  For every $G\in 2^\prefspace$, define $\tau(\{1,\ldots,k\}, G) = \begin{cases}k & \text{if }G\in L_j\text{ s.t. }P_j\in B\\
k+1&\text{otherwise} \end{cases}$. Then, for every permutation $\sigma$ over $\ma$ and every $G\in 2^ \prefspace$, define $\tau(\sigma(\{1,\ldots,k\}), \sigma(G)) = \tau(\{1,\ldots,k\}, G)$. Because of the equivalence relationship $\equiv$, if $G\in L_i$ for some $i$, then for all permutations $\sigma$ we have $\sigma(G)\in L_i$. Therefore, for every $W\in \committee k$ and every $G\in 2^\prefspace$, we have
$$\tau(W, G) = \begin{cases}k & \text{if }G\in L_j\text{ s.t. }P_j\in B\\
k+1&\text{otherwise} \end{cases}$$

Then, for every $P_j\in \calP$, every $W\in\committee k$, and every $G\in 2^ \prefspace$, let $n_j = |P_j|$, we have 
\begin{align*}
\left(\vec w_{G} - \frac{\tau(W,G)}{k}\cdot \vec 1\right)\cdot \hist (P_j)  &=  \frac{|G_j\cap G|}{|G_j|}n_j -  \begin{cases}  n_j&   \text{if } G\in L_j\text{ s.t. }L_j\in B\\
\frac{k+1}{k} n_j &\text{otherwise} \end{cases}\\ & \begin{cases}  =0&   \text{if } G  = G_j\text{ and }L_j\in B\\
<0 &\text{otherwise} \end{cases} 
\end{align*}
This means that $\GST_{\tau}(P_j) = \begin{cases}  \emptyset = f^0(P_j)&   \text{if } P_j\in B\\
\committee k  = f^1(P_j)  &\text{otherwise} \end{cases} $, which proves that $\calP$ is shattered by $\hypsGST$. Therefore, $\Ndim(\hypsGST) \ge |\calP| = T\ge |\calS^*|/m! = {|\prefspace|\choose \lfloor \prefspace/2\rfloor} / m!$. The lower bound follows after Claim~\ref{claim:lb-combinatorics}.
\end{proof}

\subsection{Proof of Theorem~\ref{thm:ndim-aGST}}
\label{app:proof-thm:ndim-aGST}
\appThm{\bf Sample complexity of approximate GST axioms}{thm:ndim-aGST}
{
\begin{align*}
\forall \GST_\tau\in \hypGST, \Ndim(\hypspace_{\approx \GST_\tau})\le & 4 \log(\sum\nolimits_{W\in\committee k}|\PD_\tau(W)|) + 4\log (8e)\\
\forall \GST_\tau\in \hypsGST, \Ndim(\hypspace_{\approx  \GST_\tau})\le &4 \log( |\PD_\tau(\{1,\ldots,k\})|) + 4\log (8e)
\end{align*}
}

\begin{proof}The proof is similar to the proof of Theorem~\ref{thm:ndim-acore}. Let $\GST_\tau\in \hypGST$. For every  $W\in \committee k$  and every $G\in \PD_\tau(W)$, we define a hyperplane indexed by $(W,G)$.  Define $\psi$ to be an additive mapping such that for every $A\in \prefspace$, 
$$\psi(A, (W,G)) \triangleq \left(\begin{cases}1 &\text{if }A\in G\\
0& \text{otherwise}\end{cases}, -\frac{\tau(W,G)}{k}\right)$$
Let $\vec w = (1,\beta)$. It follows that for any profile $P$, 
$$\psi(A, (W,G))\cdot \vec w =  |\{A\in P: A\in G\}|  -\beta \frac{\tau(W,G)}{k} |P|$$
Define $\calG =\{g\}$, where $W\in g(P)$ if and only if for all $G\in \PD_\tau(W)$, $\psi(P, (W,G))\cdot \vec w<0$. It follows that $\hypspace_{\approx \GST_\tau}\subseteq \hypspace_{\psi,\cal G}$. Therefore, $\Ndim(\hypspace_{\approx \GST_\tau})\le \Ndim(\hypspace_{\psi,\cal G})$, and according to Theorem~\ref{thm:ndim-linear-feature}, the latter is upper bounded by $2 \eta\log\left( 8e K  \right)+2\log |\cal G|$. Notice that $\eta =2$, $K = \sum_{W\in\committee k}|\PD_\tau(W)|$, and $|\calG|=1$. Therefore, $\Ndim(\hypspace_{\approx \GST_\tau})\le 4 \log(\sum_{W\in\committee k}|\PD_\tau(W)|) + 4\log (8e) $.

The proof for $\GST_\tau\in \hypsGST$ is similar, except that we only need $K=|\PD_\tau(W)|$ for any $W\in \committee k$ (which are the same). Therefore, $\Ndim(\hypspace_{\approx \GST_\tau})\le 4 \log( |\PD_\tau(\{1,\ldots,k\})|) + 4\log (8e) $.
\end{proof}

\section{Materials for Section~\ref{sec:analysis}}
 
\subsection{Proof of Theorem~\ref{thm:Thiele}}
\label{app:proof-thm:Thiele}
\appThmnoname{thm:Thiele}
{
For any fixed $m$, $k$, $\vec p\in (0,1)^m$, and any   $\Thiele_{\vec s}$ with integer scoring vector $\vec s$, 
$$\Pr\nolimits_{P\sim (\pi_{\vec p})^n} (|\Thiele_{\vec s}(P)| = 1)  =  \begin{cases}
1-\exp(-\Omega(n)) & \text{if }|\top_k(\pi_{\vec p})|=1\\
1- \Theta(\frac {1}{\sqrt n})&\text{otherwise}
\end{cases}$$
}
\begin{proof}
{\bf Intuition.} Recall that Thiele's methods are linear and testing whether a linear rule is resolute is linear as well (Table~\ref{tab:properties}). Therefore, given a Thiele  method $\Thiele_{\vec s}$, the event ``$\Thiele_{\vec s}$ is resolute at a profile $P$'' can be represented by the union of finitely many systems of linear inequalities whose variables are the occurrences of different types of votes in $P$. In other words, the event can be represented by a union of finitely many polyhedra in $|\prefspace|$-dimensional space, such that the event holds if and only if $\hist(P)$ is in this union. Then, we prove the theorem by taking the {\em polyhedral approach}~\citep{Xia2021:How-Likely}, which provides a tight dichotomy on the likelihood of the event. 

\myparagraph{Modeling.} We model and characterize the complement event, i.e., ``$\Thiele_{\vec s}$ is irresolute at a profile $P$''. Given a profile $P$, let $\hist(P) = (x_{A}:A\in\prefspace)$. For any pair of $k$-committees $W_1,W_2$, we the following system of linear inequalities, denoted by $\text{LP}^{W_1,W_2}$, to represent the event ``$W_1$ and $W_2$ have the highest Thiele scores''.
$$\text{LP}^{W_1,W_2} = \left\{\begin{minipage}{0.8\textwidth}
\begin{align*}
& \sum\nolimits_{A\in \prefspace} (s_{|A\cap W_1|}-s_{|A\cap W_2|})x_A = 0 & \text{$W_1$ and  $W_2$ have the same score}\\
\forall W\in \committee k,  &\sum\nolimits_{A\in \prefspace} (s_{|A\cap W|}-s_{|A\cap W_1|})x_A \le  0 & \text{The score is the highest}
\end{align*}
\end{minipage}\right.$$

Let $\ppoly{W_1,W_2}\subseteq \mathbb R^{|\prefspace|}$  denote the set of all points in $R^{|\prefspace|}$ that satisfy these inequalities and let $\calU = \bigcup_{W_1,W_2\in \committee k} \ppoly{W_1,W_2}$. Clearly, $\Thiele_{\vec s}$ is irresolute at $P$ if and only if $\hist(P)\in \calU$. 

\myparagraph{Characterizing likelihood.} We first recall the i.i.d.~case of~\citep[Theorem 1]{Xia2021:How-Likely} in our notation. Given any polyhedron $\poly = \{\vec x\in \mathbb R^{|\prefspace|}: \ba\cdot \invert{\vec x}\le \invert{\vec b} \}$, where $\ba$ is an integer matrix,  let $\polynint$ denote the non-negative integer points $\vec x\in \poly$ such that $\vec x \cdot 1 = n$ and let $\polyz$ denote the {\em recess cone} of $\poly$, i.e., $\polyz = \{\vec x\in \mathbb R^{|\prefspace|}: \ba\cdot \invert{\vec x}\le \invert{\vec 0} \}$. Then,

\begin{equation*}
\Pr\nolimits_{P\sim (\pi_{\vec p})^n}\left(\hist(P)\in \poly\right)=\left\{\begin{array}{ll}0 &\text{if } \polynint=\emptyset\\
\exp(-\Theta(n)) &\text{if } \polynint \ne \emptyset \text{ and }\pi_{\vec p}\notin \polyz\\
\Theta\left((\sqrt n)^{\dim(\polyz)-|\prefspace|}\right) &\text{otherwise}
\end{array},\right.
\end{equation*}
where $\dim(\polyz)$  is the dimension of $\polyz$. We will apply this theorem to every $\ppoly{W_1,W_2}$ and combine the result to obtain the characterization in the statement of the theorem. First, notice that the $0$ case of \eqref{eq:polyhedral} does not hold for $\poly = \ppoly{W_1,W_2}$, because let $P^*$ denote $n$ copies of $\emptyset$, then $\Thiele_{\vec s}(P)=\committee k$, which means that $\hist(P^*)\in \ppolynint{W_1,W_2}$.

The condition for the $\exp(-\Theta(n))$ case of \eqref{eq:polyhedral} holds for $\poly = \ppoly{W_1,W_2}$ if and only if $\pi_{\vec p}\notin \ppolyz{W_1,W_2}$, which is the same as $\ppoly{W_1,W_2}$ because $\vec b=\vec 0$  in $\ppoly{W_1,W_2}$. This is equivalent to requiring that at least one of $W_1$ and $W_2$ is not a  co-winner of $\Thiele_{\vec s}$ at the fractional profile $\pi_{\vec p}$, where the multiplicity of each vote $A$ is $\Pr_{\pi_{\vec p}}(A)$. The following claim shows that $\Thiele_{\vec s}(\pi_{\vec p}) = \top_k(\vec p)$.
\begin{claim}
\label{claim:thiele-co-winners}
For any $\vec p\in (0,1)^m$ and any  $\Thiele_{\vec s}$, we have $\Thiele_{\vec s}(\pi_{\vec p}) = \top_k(\vec p)$.
\end{claim}
\begin{proof}
We first prove that for any $W_1\in \top_k(\vec p)$ and $W_2\in \committee k\setminus \top_k(\vec p)$, the Thiele score of $W_1$ is strictly higher than that of $W_2$ in $\pi_{\vec p}$. W.l.o.g.~suppose  $W_1=\{1,\ldots,k\}$ and $p_1\ge p_2\ge\cdots\ge p_k$. Let $W_2 = \{i_1,\ldots, i_k\}$ with $p_{i_1}\ge p_{i_2}\ge\cdots \ge p_{i_k}$. It follows that for every $t\le k$ we have $p_t\ge p_{i_t}$, and at least one of the inequality is strict.  
For any $t\in [m]$, let $X_t$ denote the independent Bernoulli trial such that $\Pr(X_t=1) = p_t$. Let $Y_1 = \sum_{t=1}^k X_t$ and $Y_2 = \sum_{t=1}^k X_{i_t}$. We prove the following claim.
\begin{claim}
\label{claim:dominance-Y}
For every $0\le t\le k$, $\Pr(Y_1\ge t)>\Pr(Y_2\ge t)$.
\end{claim}
\begin{proof}
For any $k'\le k$, let $Y_1^{k'} = \sum_{t=1}^{k'} X_t$ and   $Y_2^{k'} = \sum_{t=1}^{k'} X_{i_t}$. Let $k^*$ denote the smallest number such that $p_{k^*}> p_{t_{k^*}}$. We use induction to prove that for any $k^*\le \ell\le k$ and any $1\le t\le \ell$, 
\begin{equation}
\label{ineq:prob-dominance}
\Pr(Y_1^{\ell}\ge t)>\Pr(Y_2^{\ell}\ge t)
\end{equation}
We first prove the base case $\ell = k^*$. Let $Y^* = \sum_{t=1}^{k^*-1} X_t$. It follows that $Y_1^{k^*} = Y^*+ X_{k^*}$ and $Y_2^{k^*} = Y^*+ X_{i_{k^*}}$. Therefore, for every $1\le t\le k^*$,
\begin{align*}
 &\Pr(Y_1^{k^*}\ge t) - \Pr(Y_2^{k^*}\ge t)\\
 = &\Pr(X_{k^*}=1)\Pr(Y^*\ge t-1) +  \Pr(X_{k^*}=0)\Pr(Y^*\ge t)\\
&\hspace{10mm}-(\Pr(X_{i_{k^*}}=1)\Pr(Y^*\ge t-1) +  \Pr(X_{i_{k^*}}=0)\Pr(Y^*\ge t))\\
= &(\Pr(X_{k^*}=1) - \Pr(X_{i_{k^*}}=1))(\Pr(Y^*\ge t-1) - \Pr(Y^*\ge t)) >0
\end{align*}
The last inequality holds because $\vec p\in (0,1)^m$.  This proves \eqref{ineq:prob-dominance} for $\ell=k^*$. Suppose \eqref{ineq:prob-dominance} holds for $\ell = k'$. When $\ell = k'+1$, for every $1\le t\le k'+1$, we have
\begin{align}
 \Pr(Y_1^{k'+1}\ge t)    = &\Pr(X_{k'+1}=1)\Pr(Y_1^{k'}\ge t-1) +  \Pr(X_{k'+1}=0)\Pr(Y_1^{k'}\ge t)\notag\\
 \ge & \Pr(X_{i_{k'+1}}=1)\Pr(Y_1^{k'}\ge t-1)+\Pr(X_{i_{k'+1}}=0)\Pr(Y_1^{k'}\ge t)\label{eq:switch}\\
 >&\Pr(X_{i_{k'+1}}=1)\Pr(Y_2^{k'}\ge t-1)+\Pr(X_{i_{k'+1}}=0)\Pr(Y_2^{k'}\ge t)\label{eq:switch2}\\
 =& \Pr(Y_2^{k'+1}\ge t)\notag
 \end{align}
 \eqref{eq:switch} holds because $\Pr(X_{k'+1}=1) \ge \Pr(X_{i_{k'+1}}=1) $ and $\Pr(Y_1^{k'}\ge t-1) > \Pr(Y_1^{k'}\ge t)$.  \eqref{eq:switch2} holds because $\Pr(X_{i_{k'+1}}=1)>0$, $\Pr(X_{i_{k'+1}}=0)>0$, and 
 \begin{itemize}
\item when $t=1$, $\Pr(Y_1^{k'}\ge t-1) = \Pr(Y_2^{k'}\ge t-1)=1$ and $\Pr(Y_1^{k'}\ge t) > \Pr(Y_2^{k'}\ge t)$ by the induction hypothesis;
\item when $t=k'+1$, $\Pr(Y_1^{k'}\ge t-1) > \Pr(Y_2^{k'}\ge t-1)$ by the induction hypothesis and $\Pr(Y_1^{k'}\ge t) = \Pr(Y_2^{k'}\ge t)=0$;
\item otherwise $\Pr(Y_1^{k'}\ge t-1) > \Pr(Y_2^{k'}\ge t-1)$ and $\Pr(Y_1^{k'}\ge t) > \Pr(Y_2^{k'}\ge t)$ by the induction hypothesis.
 \end{itemize}
\end{proof}
Notice that we can calculate $\score_{\vec s}(\pi_{\vec p},W^1)$ as 
$$\score_{\vec s}(\pi_{\vec p},W_1) = \sum\nolimits_{i=0}^k s_i\Pr(Y_1 = i) = \sum\nolimits_{i=0}^k (s_{i}-s_{i-1})\Pr(Y_1\ge i),$$
where we let $s_{-1}=0$. Similarly, 
$$\score_{\vec s}(\pi_{\vec p},W_2) =   \sum\nolimits_{i=0}^k (s_{i}-s_{i-1})\Pr(Y_2\ge i)$$
Therefore,
\begin{align*}
\score_{\vec s}(\pi_{\vec p},W_1) - \score_{\vec s}(\pi_{\vec p},W_2) = \sum\nolimits_{i=0}^k (s_{i}-s_{i-1})(\Pr(Y_1\ge i)-\Pr(Y_2\ge i)) >0
\end{align*}
The last inequality holds because of Claim~\ref{claim:dominance-Y} and $s_k>s_0$, which means that  $s_{i}-s_{i-1}>0$ for some $i\le k$. 

For any $W_2\in \top_k(\pi_{\vec p})$, using the same terminology above we have that for every $t\le k$, have $p_t = p_{i_t}$. This means that $Y_1 = Y_2$, which means that $\score_{\vec s}(\pi_{\vec p},W_1) - \score_{\vec s}(\pi_{\vec p},W_2)=0$ following a similar calculation. This completes the proof of Claim~\ref{claim:thiele-co-winners}.
\end{proof}
Following Claim~\ref{claim:thiele-co-winners}, the $\exp(-\Theta(n))$ case of \eqref{eq:polyhedral} holds for $\poly = \ppoly{W_1,W_2}$ if and only if $\{W_1,W_2\}\subseteq \top_k(\vec p)$. Therefore, \eqref{eq:polyhedral} can be simplified to
\begin{equation}
\label{eq:polyhedral-simplified}
\Pr\nolimits_{P\sim (\pi_{\vec p})^n}\left(\hist(P)\in \ppoly{W_1,W_2}\right)=\left\{\begin{array}{ll} 
\exp(-\Theta(n)) &\text{if } \{W_1,W_2\}\not\subseteq \top_k(\pi_{\vec p})\\
\Theta\left((\sqrt n)^{\dim(\ppoly{W_1,W_2})-|\prefspace|}\right) &\text{otherwise}
\end{array}\right.
\end{equation}
Next, we use \eqref{eq:polyhedral-simplified} to characterize the probability for $\Thiele_{\vec s}$ to be (ir)resolute.

\myparagraph{The $1-\exp(-\Theta(n))$ case of the theorem.} When $|\top_k(\vec p)|=1$, for every pair of $k$-committees $W_1,W_2$, according to \eqref{eq:polyhedral-simplified}, the probability for both of them to be co-winners is exponentially small. Notice that there are constantly many such pairs (as we fix $m$ and $k$). Therefore, by the union bound, the probability for $\Thiele_{\vec s}$ to be resolute is $1-\exp(-\Theta(n))$.

\myparagraph{The $1-  \Theta(\frac{1}{\sqrt n})$ case of the theorem.} In this case $|\top_k(\vec p)|\ge 2$. Again, according to \eqref{eq:polyhedral-simplified}, for every pair $\{W_1,W_2\}\not\subseteq \top_k(\vec p)$, the probability for $W_1$ and $W_2$ to be co-winners is exponentially small. When $\{W_1,W_2\}\subseteq \top_k(\vec p)$, notice that $\dim(\ppoly{W_1,W_2})\le  |\prefspace|-1$ because it contains at least one equation in $\text{LP}^{W_1,W_2}$. Next, we show that there exist $\{W_1^*,W_2^*\}\subseteq \top_k(\vec p)$ such that $\dim(\ppoly{W_1^*,W_2^*})\ge   |\prefspace|-1$. W.l.o.g.~suppose $W_1^* = \{1,\ldots, k\}$. According to the definition of $\top_k(\vec p)$, there exists $W_2^*\in \top_k(\vec p)$ that differs $W_1$ only on the alternative with the smallest probability among all alternatives in $W_1$.  W.l.o.g.~let $W_2^*=\{1,\ldots, k-1,k+1\}$. Let $\ell\le k$ denote the largest number such that $s_{\ell}>s_{\ell-1}$. We will define a profile $P^*$ and prove that $\Thiele_{\vec s}(P^*) = \{W_1^*,W_2^*\}$. 
\begin{itemize}
\item When $\ell = k$, define $P^* = \{W_1^*,W_2^*\}$. Clearly $\Thiele_{\vec s}(P^*) = \{W_1^*,W_2^*\}$.
\item When $\ell \le k-1$, let $P^+$ denote the set of all $\ell$-subsets of $\{1,\ldots, k-1\}$ and let $P^-$ denote the set of all $(\ell-1)$-subsets of $\{1,\ldots, k-1\}$. Define $P_1 = \{A\cup \{k\}: A\in P^-\}$ and  $P_2 = \{A\cup \{k+1\}: A\in P^-\}$. Finally, define 
$$P^* = 2|P^-|\times P^+ + P_1+ P_2$$
The Thiele scores of $W_1^*$ and $W_2^*$ are equal and are $(2|P^+|+1)|P^-|s_{\ell}+|P^-|s_{\ell-1}$. For any $W\in\committee k$ such that $\{1,\ldots, k-1\}\subseteq W$, its Thiele score is no more than $2|P^+||P^-|s_{\ell}+2|P^-|s_{\ell-1}$. For any $W\in\committee k$ such that $\{k,k+1\}\subseteq W$, its Thiele score is no more than $2|P^+||P^-|s_{\ell}+2|P^-|s_{\ell-1}$. Therefore, $\Thiele_{\vec s}(P^*) = \{W_1^*,W_2^*\}$.
\end{itemize}
It follows from the construction above that $\hist(P^*)\in \ppoly{W_1^*,W_2^*}$, and all inequalities in LP$^{W_1^*,W_2^*}$ are strict except the first equation. This means that $\dim(\ppoly{W_1^*,W_2^*})\ge |\prefspace|-1$. Therefore, according to \eqref{eq:polyhedral-simplified}, $\Pr\nolimits_{P\sim (\pi_{\vec p})^n}\left(\hist(P)\in \ppoly{W_1^*,W_2^*}\right) = \Theta(\frac{1}{\sqrt n})$.  The $1-  \Theta(\frac{1}{\sqrt n})$ case of the theorem follows after combining (constantly many) applications of \eqref{eq:polyhedral-simplified} to all $(W_1,W_2)$ pairs.  
\end{proof}

\subsection{Remaining proof of Theorem~\ref{thm:Thiele-refinement}}
\label{app:proof-thm:Thiele-refinement}
\begin{proof}
\myparagraph{Modeling for \eqref{inequ:theta-1}.}  For any $\vec s_1$, $\vec s_2$, and any  $W^*\in \committee k$, we define the following system of linear inequalities, denoted by $\text{LP}_{\vec s_1,\vec s_2}^{W^*}$, to represent the event ``$W^*$ is the unique winner under $\Thiele_{\vec s_1}$ and $\Thiele_{\vec s_1}$''. Recall that for Thiele methods $\prefspace = 2^\ma$.
$$\text{LP}_{\vec s_1,\vec s_2}^{W^*} = \left\{\begin{minipage}{0.8\textwidth}
\begin{align*}
 \forall W\in \committee k\setminus\{W^*\},  &\sum\nolimits_{A\in \prefspace} ([\vec s_1]_{|A\cap W|}-[\vec s_1]_{|A\cap W^*|})x_A \le  -1 \\
  \forall W\in \committee k\setminus\{W^*\},  &\sum\nolimits_{A\in \prefspace} ([\vec s_2]_{|A\cap W|}-[\vec s_2]_{|A\cap W^*|})x_A \le  -1 
\end{align*}
\end{minipage}\right.$$

Let $\pspoly{W^*}{\vec s_1,\vec s_2}\subseteq \mathbb R^{|\prefspace|}$  denote the set of all points in $R^{|\prefspace|}$ that satisfy these inequalities. Clearly, $\Thiele_{\vec s_1}$ and $\Thiele_{\vec s_2}$ both choose $W^*$ as the unique winner  at $P$ if and only if $\hist(P)\in \pspoly{W^*}{\vec s_1,\vec s_2}$. 

\myparagraph{Analyzing likelihood in \eqref{inequ:theta-1}.} We will apply \eqref{eq:polyhedral} to analyze the likelihood. We will show that the $0$ case does not hold for every sufficiently large $n$ following the proof of the polynomial case. To verify that the exponential case does not hold, notice that $\hist(P)\in \pspolyz{W^*}{\vec s_1,\vec s_2}$ if and only if $W^*$ is a co-winner of $P$ under both $\Thiele_{\vec s_1}$ and $\Thiele_{\vec s_2}$. It follows after Claim~\ref{claim:thiele-co-winners} that $\pi_{\vec p} \in \pspolyz{W^*}{\vec s_1,\vec s_2}$, which proves that the exponential case does not hold. To prove that the polynomial case holds and $\dim(\pspolyz{W^*}{\vec s_1,\vec s_2}) = |\prefspace| = 2^m$, notice that the $\vec b$ vector of $\pspoly{W^*}{\vec s_1,\vec s_2}$ is non-positive. Therefore, it suffices to prove the following claim.
\begin{claim}
\label{claim:exist-uniuque-W-star}
Fix $W^*\in\committee m$. For every sufficiently large $n$, there exists a profile $P^*\in ( 2^\ma)^n$ such that for every $\Thiele_{\vec s}$, $\Thiele_{\vec s}(P^*) = \{W^*\}$.
\end{claim}
\begin{proof} We first prove the claim for $n=2^m-1$ and then use the construction to prove the claim for general $n$. 
Define  $P^*\triangleq 2^{W^*}$. That is, $P^*$ consists of all subsets of $W^*$. For any $0\le \ell \le k-1$ and any $W\in\committee k$, define $P^*_{W,\ell} \triangleq \{A\in P^*: |A\cap W|>\ell\}$. We prove 
\begin{equation}
\label{eq:W-star-dominates}\forall W\in\committee k \setminus \{W^*\}, \forall 0\le \ell \le k-1, 
|P^*_{W^*,\ell}| > |P^*_{W,\ell}|
\end{equation}
This is because for every $A\in P^*_{W,\ell}$, we have $A\in P^*_{W^*,\ell}$. Also, let $A' \in P^*_{W^*,\ell}$ be an arbitrary set with $|A'|=\ell$ and $A'$ contains at least one alternative in $W^*\setminus W$. Then, $A'\notin P^*_{W,\ell}$. This means that $P^*_{W,\ell}\subsetneq P^*_{W^*,\ell}$, which proves~\eqref{eq:W-star-dominates}.
W.l.o.g.~assume that $s_0=0$. Then, we have
\begin{align*}
\score_{\vec s}(P^*,W^*) - \score_{\vec s}(P^*,W) = \sum\nolimits_{\ell =0}^{k-1}(s_{\ell+1}-s_{\ell})(|P^*_{W^*,\ell}| - |P^*_{W,\ell}|)\ge s_k-s_0>0
\end{align*}
This proves Claim~\ref{claim:exist-uniuque-W-star} for $n=2^m-1$. For any $n> 2^m-1$, add $n-2^m+1$ copies of $\emptyset$. This proves the claim for general $n$.
\end{proof}
It follows from Claim~\ref{claim:exist-uniuque-W-star} that the $0$ case of~\eqref{eq:polyhedral} does not hold for all $n\ge 2^m-1$. Moreover, $\hist(P^*)$ (where $P^*$ is the profile guaranteed by Claim~\ref{claim:exist-uniuque-W-star}) is an interior point of $ \pspolyz{W^*}{\vec s_1,\vec s_2} $, which means that $\dim(\pspolyz{W^*}{\vec s_1,\vec s_2}) = 2^m$. This proves~\eqref{inequ:theta-1}. 

\myparagraph{Modeling for \eqref{inequ:1-theta-1}.}  For any $\vec s_1$, $\vec s_2$, and any  $W_1,W_2\in \committee k$, we define the following system of linear inequalities, denoted by $\text{LP}_{\vec s_1,\vec s_2}^{W_1,W_2}$, to represent the event ``$W_1$ is the unique winner under $\Thiele_{\vec s_1}$ and $W_2$ is the unique winner under $\Thiele_{\vec s_2}$''.
$$\text{LP}_{\vec s_1,\vec s_2}^{W_1,W_2} = \left\{\begin{minipage}{0.8\textwidth}
\begin{align*}
 \forall W\in \committee k\setminus\{W_1\},  &\sum\nolimits_{A\in \prefspace} ([\vec s_1]_{|A\cap W|}-[\vec s_1]_{|A\cap W_1|})x_A \le  -1 \\
  \forall W\in \committee k\setminus\{W_2\},  &\sum\nolimits_{A\in \prefspace} ([\vec s_2]_{|A\cap W|}-[\vec s_2]_{|A\cap W_2|})x_A \le  -1 
\end{align*}
\end{minipage}\right.$$
Let $\pspoly{W_1,W_2}{\vec s_1,\vec s_2}\subseteq \mathbb R^{|\prefspace|}$  denote the set of all points in $R^{|\prefspace|}$ that satisfy these inequalities.

\myparagraph{Analyzing likelihood in \eqref{inequ:1-theta-1}.} Because $|\top_k(\vec p)|\ge 2$, there exist $W_1,W_2\in \top_k(\vec p)$  such that $|W_1\cap W_2| = k-1$. W.l.o.g.~let $W_1= \{1,3,\ldots, k+1\}$ and $W_2= \{2,3,\ldots, k+1\}$. Let $\vec s_1 = (s_{10},\ldots,  s_{1k})$ and $\vec s_2 = (s_{20},\ldots,  s_{2k})$. It is without loss of generality to assume that $s_{10} = s_{20}=0$ and $s_{1k} = s_{2k}>0$. Because $\vec s_1\ne \vec s_2$, there exists $  i_1, i_2\in \{0,\ldots, k-1\}$ such that 
$$s_{1(i_1+1)} - s_{1 i_1 } < s_{2(i_1+1)} - s_{2 i_1 }\text{ and }s_{1(i_2+1)} - s_{1 i_2 } > s_{2(i_2+1)} - s_{2 i_2 }$$
Let $\Delta_{11} \triangleq s_{1(i_1+1)} - s_{1 i_1 }$, $\Delta_{12} \triangleq s_{1(i_2+1)} - s_{1 i_2 }$, $\Delta_{21} \triangleq s_{2(i_1+1)} - s_{2 i_1 }$, $\Delta_{22} \triangleq s_{2(i_2+1)} - s_{2 i_2 }$. Then, we have 
$$\Delta_{11} < \Delta_{21}\text{ and }\Delta_{12} > \Delta_{22}$$
Next, we   prove that there exist $a,b\in \mathbb N$ such that 
$$ a\Delta_{12} - b\Delta_{11}>0 \text{ and } a\Delta_{22} - b\Delta_{21}<0 $$
This can be proved by constructions summarized in the following table.
\renewcommand{\arraystretch}{1.2}
\begin{table}[htp]
\centering
\begin{tabular}{|c|c|c|c|}
\hline 
$\Delta_{11}$ & $\Delta_{22}$ & $a$ & $b$\\
\hline $=0 $ & $=0$ & $1$ & $1$\\
\hline $=0 $ & $>0$ & $1$ & $ \lceil \frac{\Delta_{22}+1}{\Delta_{21}}  \rceil $\\
\hline $>0 $ & $=0$ &  $ \lceil \frac{\Delta_{11}+1}{\Delta_{12}}  \rceil $& $1$\\
\hline $>0 $ & $>0$ &  $\Delta_{11}\Delta_{22}+1$& $\Delta_{12}\Delta_{22}$\\
\hline
\end{tabular}
\end{table}
\renewcommand{\arraystretch}{1}
Define 
$$P_1^* \triangleq a\times \{1,3,\ldots i_1+2\} + b\times \{1,3,\ldots i_2+2\}$$
It follows that 
\begin{equation}
\label{eq:compare-W1-W2}
\begin{minipage}{0.8\textwidth}
$$\score_{\vec s_1}(P_1^*, W_1) - \score_{\vec s_1}(P_1^*, W_2) = a\Delta_{12} - b\Delta_{11}>0\text{ and }$$
$$\score_{\vec s_2}(P_1^*, W_1) - \score_{\vec s_2}(P_1^*, W_2) = a\Delta_{22} - b\Delta_{21}<0$$
\end{minipage}
\end{equation}
Let $A^*\triangleq \{3,\ldots, k+2\}$.  Define 
$$P_{k-1}\triangleq \{1\}\cup A^* + \{2\}\cup A^*$$
Then, for every $i\le k-2$, let $\calA_{i}^*$ denote the sets of all $i$-subsets of $A^*$, i.e., $\cal A_{i}^* = \{A\subseteq A^*: |A| = i\}$. Define 
$$P_{i}\triangleq  \underbrace{\left({k-1\choose i} +1\right)\times  \cal A_{i+1}^*}_{P_i^1} + \underbrace{\{\{1\}\cup A: A\in \cal A_{i}^*\}+ \{\{2\}\cup A: A\in \cal A_{i}^*\}}_{P_i^2}$$
Clearly for any score vector $\vec s$ and  all $i\le k-1$,  $\score_{\vec s }(P_i,W_1)=\score_{\vec s }(P_i,W_2)$. Next, we prove that for any $W\in \committee k\setminus\{W_1,W_2\}$ and any $i\le k-1$, 
\begin{equation}
\label{eq:score-diff-Pi}\score_{\vec s}(P_i,W_1) - \score_{\vec s}(P_i,W)\ge s_{i+1}-s_i
\end{equation}
It is easy to verify that~\eqref{eq:score-diff-Pi} hold for $i=k+1$. For any $i\le k-2$, we prove~\eqref{eq:score-diff-Pi} in the following two cases.
\begin{itemize}
\item When $A^*\subseteq W$, we have $\score_{\vec s}(P_i^1,W_1) = \score_{\vec s}(P_i^1,W)$ and $\score_{\vec s}(P_i^2,W_1) - \score_{\vec s}(P_i^2,W) = s_{i+1} -s_i$.
\item When $A^*\not\subseteq W$, we have $\score_{\vec s}(P_i^1,W_1) - \score_{\vec s}(P_i^1,W)\ge \left(2{k-1\choose i} +1\right)(s_{i+1}-s_i)$ and $\score_{\vec s}(P_i^2,W_1) - \score_{\vec s}(P_i^2,W) \ge  -{k-1\choose i}(s_{i+1} -s_i)$. Therefore, $\score_{\vec s}(P_i ,W_1) - \score_{\vec s}(P_i,W)\ge  (s_{i+1}-s_i)$.
\end{itemize}
This proves~\eqref{eq:score-diff-Pi}. Let 
\begin{equation}
\label{eq:def-P2-star}
P_2^*\triangleq   P_1\cup\cdots\cup P_{k-1}
\end{equation}
Then, for any $\vec s$ and any $W\in\committee k\setminus\{W_1,W_2\}$, we have 
$$\score_{\vec s}(P_2^*,W_1)=\score_{\vec s}(P_2^*,W_2)>\score_{\vec s}(P_2^*,W)$$
Define  $P^*\triangleq P_1^*\cup P_2^*$. It follows from~\eqref{eq:compare-W1-W2} and~\eqref{eq:score-diff-Pi} that  for any sufficiently large $n$, $\pspolynint{W_1,W_2}{\vec s_1,\vec s_2} \ne\emptyset$. This proves that the $0$ case of~\eqref{eq:polyhedral} does not hold. Because $W_1,W_2\in \top_k(\vec p)$, they are the co-winners under $\pi_{\vec p}$ for any Thiele method. Therefor, $\pi_{\vec p}\in \pspolyz{W_1,W_2}{\vec s_1,\vec s_2}$, which means that the exponential case of~\eqref{eq:polyhedral} does not hold either. Finally, notice that the $\vec b$ vector in $\pspoly{W_1,W_2}{\vec s_1,\vec s_2}$ is negative. Therefore, $\hist(P^*)$ is an interior point of $\pspolyz{W_1,W_2}{\vec s_1,\vec s_2}$, which means that $\dim(\pspolyz{W_1,W_2}{\vec s_1,\vec s_2})=2^m$. This proves~\eqref{inequ:1-theta-1} and finishes the proof of the  $\subq{\Thiele_{\vec s_1}}{\Thiele_{\vec s_2}}$ part of the Theorem~\ref{thm:Thiele-refinement}.

The $\equalsq{\Thiele_{\vec s_1}}{\Thiele_{\vec s_2}}$ part and the $ {\Thiele_{\vec s_1}}\cap {\Thiele_{\vec s_2}}$ part of the theorem are proved similarly. Specifically, their  $\Theta(1)\wedge (1-\Theta(1))$ cases follow after~\eqref{inequ:theta-1} and~\eqref{inequ:1-theta-1}. 
\end{proof}

\subsection{Proof of Theorem~\ref{thm:core-iad}}
\label{app:proof-thm:core-iad}

\appThmnoname{thm:core-iad}{
For any fixed $m$, any fixed $k<m$, and any fixed $\vec p\in (0,1)^m$,
$$\Pr\nolimits_{P\sim(\pi_{\vec p})^n}\left(\top_k(\vec p)\subseteq \core(P)\right) = 1-\exp(-\Omega(n))$$
}
\begin{proof}
Again, we apply the polyhedral approach to prove the theorem. 

\myparagraph{Modeling.} In light of the linearity of $\core$, for any $k$-committee $W$ and any $W'\subseteq \ma$, we define LP$^{W,W'}$ to model the event that $W$ is not in $\core$ because a qualified group wants to deviate to $W'$. In fact, LP$^{W,W'}$ consists of a single inequality that corresponds to \eqref{eq:pa-inequality} in the definition of GST axioms. 
$$\text{LP}^{W,W'} = \left\{ \frac{|W'|}{k}\sum\nolimits_{A\in\prefspace} x_A- \sum\nolimits_{A\in \calM_{W,W'}} x_A  \le 0,  \right.$$
where we recall that $\calM_{W,W'} = \{A\subseteq \ma: |A\cap W'| > |A\cap W|\}$. 
Let $\ppoly{W,W'}$ denote all vectors that satisfy $\text{LP}^{W,W'}$ and  let $\calU^W = \bigcup_{W'\in\prefspace} \ppoly{W,W'}$. Clearly, $W\notin\core(P)$ if and only if $\hist(P)\in \calU^W$.

\myparagraph{Analyzing likelihood.} We now apply \eqref{eq:polyhedral} (the i.i.d.~case of~\citep[Theorem 1]{Xia2021:How-Likely}) to analyze the probability for $\hist(P)$ to be in $\ppoly{W,W'}$ under distribution $\pi_{\vec p}$. We only need to consider $W'\not \subseteq W$ with $|W'|\le k$. For each such $W'$ and every $n\in\mathbb N$, let $P=n\times \{W'\setminus W\}$. Then, $\hist(P)\in \ppoly{W,W'}$, which means that $\ppolynint{W,W'}\ne\emptyset$. Therefore, the $0$ case of \eqref{eq:polyhedral} does not hold. 

 Next, we prove that the exponential case of \eqref{eq:polyhedral} holds for all $W'\not \subseteq W$ with $|W'|\le k$. That is, $\pi_{\vec p}\notin \ppolyz{W,W'}=\ppoly{W,W'}$. In other words,  we need to prove   $\left(\vec w_{\calM_{W,W'}}-\frac {|W'|}{k}\cdot \vec 1\right)\cdot \pi_{\vec p}<0,$ 
which is equivalent to 
\begin{equation*}
\Pr\nolimits_{A\sim\pi_{\vec p}}(A\in \calM_{W,W'})< \frac {|W'|}{k}
\end{equation*}
Notice that every $A\in \calM_{W,W'}$ can be enumerated by considering the combination of following three sets $A_1$, $A_2$, and $A_3$: 
$$\underbrace{A\cap (W\setminus W')}_{A_1},\underbrace{ A\cap (W'\setminus W)}_{A_2}, \underbrace{A\cap [(W\cap W')\cup (\neg W\cap \neg W')]}_{A_3}$$
Clearly, $|A_1|<|A_2|$ and $A_3$ can be any subset of $[(W\cap W')\cup (\neg W\cap \neg W')]$. For every set $A'\subseteq \ma = [m]$, let $B_{A',\vec p}$ denote the Poisson binomial random variable that is the sum of $|A'|$ independent Bernoulli trials distributed with success probabilities $\{p_i: i\in A'\}$. Then, 
\begin{align*}
&\Pr\nolimits_{A\sim\pi_p}(A\in \calM_{W,W'}) =   \Pr(B_{W\setminus W',\vec p} < B_{W'\setminus W,\vec p})
 \end{align*}
Because $\frac{|W'\setminus W|}{|W\setminus W'|}\le \frac{|W'|}{k}$, to prove \eqref{eq:core-p}, it suffices to prove $\Pr(B_{W\setminus W',\vec p} < B_{W'\setminus W,\vec p}) <\frac{|W'\setminus W|}{|W\setminus W'|}$. 
Let $p$ denote the minimum probability in $\vec p$ indexed by $W\setminus W'$, that is, $p = \min_{i\in W\setminus W'} p_i$. This means that for all $i\in W\setminus W'$ we have $p_i\ge p$, and for every $i'\in W'\setminus W$ we have $p_{i'}\le p$ (because $W\in\top_k(\vec p)$). For any $n'\in\mathbb N$, let  $B_{n',p}$ denote the binomial random variable $(n',p)$, i.e., the sum of $n'$ independent binary  random variables, each of which takes $1$ with probability $p$. Let $k_1=|W\setminus W'| $ and $k_2 = |W'\setminus W|$.  We first prove
\begin{equation*}
\Pr(B_{k_1,p} < B_{k_2,p})\cdot \frac {k_1}{k_2} <1 
\end{equation*} 
This is proved in the following calculations. Let $q=1-p$ to simplify notation.
\begin{align*}
\Pr(B_{k_1,p} < B_{k_2,p})\cdot \frac {k_1}{k_2}  = &\sum _{s=0}^{k_2-1}\underbrace{p^s q^{k_1-s} { k_1 \choose s }}_{B_{k_1,p}=s} \left(\sum _{\ell = s+1}^{k_2} \underbrace{p^\ell  q^{k_2-\ell}{ k_2 \choose \ell }}_{B_{k_2,p} =\ell}\right) \frac {k_1}{k_2}\\
 = &\sum _{s=0}^{k_2-1} p^{s+1} q^{k_1-s} { k_1+1 \choose s+1 } \frac{s+1}{k_1+1} \left(\sum _{\ell = s+1}^{k_2} p^{\ell-1} q^{k_2-\ell}{ k_2 -1 \choose \ell -1 } \frac{k_2}{\ell}\right) \frac {k_1}{k_2}\\
  =& \sum _{s=0}^{k_2-1} p^{s+1} q^{k_1-s} { k_1+1 \choose s+1 } \frac{k_1}{k_1+1} \left(\sum _{\ell = s+1}^{k_2} p^{\ell-1} q^{k_2-\ell}{ k_2 -1 \choose \ell -1 } \frac{s+1}{\ell}\right) \\
  < &\sum _{s=0}^{k_2-1} \underbrace{p^{s+1} q^{k_1-s} { k_1+1 \choose s+1 }}_{B_{k_1+1,p} = s+1} \left(\sum _{\ell = s+1}^{k_2} \underbrace{p^{\ell-1} q^{k_2-\ell}{ k_2 -1 \choose \ell -1 }}_{B_{k_2-1,p}=\ell-1} \right) \\
 < &\Pr(B_{k_1+1,p}-1\le B_{k_2-1,p})\le 1
\end{align*}

Then, because (first-order stochastic) dominance is preserved under the addition of random variables,  $B_{W\setminus W',\vec p}$ dominates  $B_{|W\setminus W'|,p}$ and $B_{|W'\setminus W|,p}$ dominates $B_{W'\setminus W,\vec p}$. Therefore, 
\begin{align*}
&\Pr(B_{W\setminus W',\vec p} < B_{W'\setminus W,\vec p}) = \sum\nolimits_{t=1}^{|W'\setminus W|}
\Pr (B_{W\setminus W',\vec p} < t)\times \Pr(B_{W'\setminus W,\vec p}=  t)\\
\le& \sum\nolimits_{t=1}^{|W'\setminus W|}\Pr (B_{|W\setminus W'|,p}< t) \times   \Pr(B_{|W\setminus W'|,p} < B_{W'\setminus W,\vec p}) = \Pr(B_{|W\setminus W'|, p} < B_{W'\setminus W,\vec p}) \\
=& \sum\nolimits_{t=1}^{|W'\setminus W|}
\Pr (B_{|W\setminus W'|, p} = t-1)\times \Pr(B_{W'\setminus W,\vec p}) \ge   t)\\
\le & \sum\nolimits_{t=1}^{|W'\setminus W|}\Pr (B_{|W\setminus W'|, p} = t-1)\times \Pr(B_{|W'\setminus W|, p}) \ge   t) \\
=& \Pr(B_{|W\setminus W'|,p} < B_{|W'\setminus W|,p}) < \frac{|W'\setminus W|}{|W\setminus W'|} \hspace {20mm} (\textbf{from \eqref{ineq:S-p}})
\end{align*}
This proves \eqref{eq:core-p} and verifies that the exponential case of \eqref{eq:polyhedral} holds for all $W'\not \subseteq W$ with $|W'|\le k$. Therefore, for any $W\in\top_k(\pi_{\vec p})$, after applying \eqref{eq:core-p} to all $W'\not \subseteq W$ with $|W'|\le k$ and then applying the union bound, we have
$$\Pr\nolimits_{P\sim(\pi_{\vec p})^n}(\hist(P)\in \calU^W) = \exp(-\Theta(n))$$
The theorem follows after applying the union bound to all $W\in \top_k(\pi_{\vec p})$.
\end{proof}


\subsection{Proof of Proposition~\ref{prop:jr-core}}
\label{app:proof-prop:jr-core}

\appPropnoname
{prop:jr-core}
{
For every fixed $m\ge 4$ and $k\ge 3$, there exists $\vec p\in (0,1)^m$ such that for any sufficiently large $n$, 
 $\Pr\nolimits_{P\sim(\pi_{\vec p})^n}\left(\jr(P)\not\subseteq \core(P)\right) = 1-\exp(-\Omega(n))$.
}
\begin{proof}
Let $\epsilon = \frac{1}{2k^2}$ and let $\vec p = (\underbrace{\epsilon,\ldots,\epsilon}_{k-1}, \underbrace{1-\epsilon,\ldots,1-\epsilon}_{m-k+1})$. Let $W^* = \{1,\ldots, k\}$. The proposition is proved by combining the following two observation by the union bound.
\begin{equation}
\label{eq:ob1}
\Pr\nolimits_{P\sim(\pi_{\vec p})^n}\left(W^*\in \jr(P)\right) = 1-\exp(-\Omega(n))
\end{equation}
\begin{equation}
\label{eq:ob2}
\Pr\nolimits_{P\sim(\pi_{\vec p})^n}\left(W^*\notin \core(P)\right) = 1-\exp(-\Omega(n))
\end{equation}
\eqref{eq:ob1} holds because the expected number of votes that does not contain alternative $k$ is $\epsilon$, which means that with exponentially small probability the number of votes that does not contain $k$ is $2\epsilon n<\frac{1}{k}n$.

To prove \eqref{eq:ob2}. We consider the complement of ${\text{LP}}^{W,W'}$, denoted by $\overline{\text{LP}}^{W,W'}$, that represent the event that no sufficiently large group of voters want to deviate to $W'$. 
$$\overline{\text{LP}}^{W,W'} = \left\{ \sum\nolimits_{A\in \calM_{W,W'}} x_A -  \frac{|W'|}{k}\sum\nolimits_{A\in\prefspace} x_A< 0,  \right.$$
Let $\overline\calH^{W,W'}$ denote all vectors that satisfy $\overline{\text{LP}}^{W,W'}$. Let $W' = \{m-k+2,\ldots, m\}$. Notice that $|W'|=k-1$. We will prove
\begin{equation}
\label{eq:bar-h}
\Pr\nolimits_{P\sim(\pi_{\vec p})^n}\left(\hist(P) \in \overline\calH^{W^*,W'} \right) =  \exp(-\Omega(n))
\end{equation}
That is, the exponential case of \eqref{eq:polyhedral} holds. We first verify that $\pi_{\vec p}\not\in \overline\calH^{W^*,W'}_{\le 0}$. In other words, we will prove
$$\Pr\nolimits_{A\sim\pi_{\vec p}}(A\in \calM_{W^*,W'})> \frac{k-1}{k}$$
Like the proof of Theorem~\ref{thm:core-iad}, the inequality is equivalent to
$$\Pr (B_{W^*\setminus W',\vec p}< B_{W'\setminus W^*,\vec p})> \frac{k-1}{k},$$
where we recall that $B_{A',\vec p}$ is  Poisson binomial random variable that is the sum of $|A'|$ independent Bernoulli trials distributed with success probabilities $\{p_i: i\in A'\}$. Let $k_1=|W^*\setminus W'|$ and $k_2 = |W'\setminus W^*|$. We have $k_1=k_2+1$. Also notice that $B_{W^*\setminus W',\vec p} = B_{k_1,\epsilon}$ and $B_{W'\setminus W^*,\vec p}) = B_{k_2,1-\epsilon}$. Therefore, it suffices to prove that for any number $\ell\le k-1$, 
$$\Pr (B_{\ell+1,\epsilon}< B_{\ell,1-\epsilon})> \frac{k-1}{k}$$
According to the union bound, $\Pr(B_{\ell+1,\epsilon} =0)\ge 1- (\ell+1)\epsilon\ge 1-k\epsilon$ and $\Pr(B_{\ell,1-\epsilon}< 1)\le \epsilon$. Therefore, 
$$\Pr (B_{\ell+1,\epsilon}< B_{\ell,1-\epsilon})> 1-{k+1}\epsilon = 1-\frac{k+1}{2k^2} >\frac{k-1}{k},$$
which verifies the exponential case of \eqref{eq:polyhedral}  for $\overline{\text{LP}}^{W,W'}$. This proves \eqref{eq:bar-h} and concludes the proof of Proposition~\ref{prop:jr-core}.
\end{proof}
\subsection{Proof of Theorem~\ref{thm:Thiele-core}}
\label{app:proof-thm:Thiele-core}
\appThmnoname{thm:Thiele-core}
{
For any fixed $m$, any fixed $k<m$,  any fixed $\vec p\in (0,1)^m$, any  $\Thiele_{\vec s}$,
$$\Pr\nolimits_{P\sim(\pi_{\vec p})^n}\left(\Thiele_{\vec s}(P)\subseteq \core(P)\right) = 1-\exp(-\Omega(n))$$
}
\begin{proof}
Recall from Claim~\ref{claim:thiele-co-winners} that $\Thiele_{\vec s}(\pi_{\vec p}) = \top_k(\pi_{\vec p})$, which means that under $\pi_{\vec p}$, the Thiele score of any $k$-committee in $\top_k(\pi_{\vec p})$ is strictly larger than that of any other $k$-committee. Therefore, it follows after Hoeffding's inequality and union bound that with $1-\exp(-\Theta(n))$ probability, $\Thiele_{\vec s}(P) \subseteq \top_k(\pi_{\vec p})$. Recall from Theorem~\ref{thm:core-iad} that with high probability $\top_k(\pi_{\vec p})\subseteq \core(P)$. The theorem follows after applying union bound to these observations.
\end{proof}

\subsection{Remaining Proof of Theorem~\ref{thm:Thiele-GST}}
\label{app:proof-thm:Thiele-GST}
\begin{proof}
Event 1 is further divided into $|2^{2^m}|$ events, each of which is represented by an LP defined as follows. For any $G\in 2^{2^\ma}$, define
$$\text{LP}^{W_1,G} = \left\{\begin{minipage}{0.8\textwidth}
\begin{align*}
\forall W\in \committee k,  \sum\nolimits_{A\in \prefspace} (s_{|A\cap W|}-s_{|A\cap W_1|})x_A &\le  0 & \text{$W_1$ has the highest Thiele score}\\
 \frac{\tau(W_1,G)}{k}\sum\nolimits_{A\in\prefspace} x_A- \sum\nolimits_{A\in G} x_A  &\le 0 & \text{$W_1$ is disqualified by $G$} \end{align*}
\end{minipage}\right.$$
Let $\ppoly{W_1,G}$ denote the vectors that satisfy $\text{LP}^{W_1,G}$ and let   $\calU^{W_1,G}\triangleq \bigcup _{G\in 2^{2^\ma}} \ppoly{W_1,G}$. Then, Event 1 happens at profile $P$ if and only if $\hist(P)\in \calU^{W_1,G}$. Next, we use an LP$^{W_1,W_2,\tau}$ to represent Event 2. 
$$\text{LP}^{W_1,W_2,\tau} = \left\{\begin{minipage}{0.8\textwidth}
\begin{align*}
 \sum\nolimits_{A\in \prefspace} (s_{|A\cap W_2|}-s_{|A\cap W_1|})x_A &\le -1 & \\
 \forall  G \in 2^{2^\ma}, \sum\nolimits_{A\in G} x_A - \frac{\tau(W_1,G)}{k}\sum\nolimits_{A\in\prefspace} x_A &\le -1 & \end{align*}
\end{minipage}\right.$$

Let $\ppoly{W_1,W_2,\tau}$ denote the vectors that satisfy $\text{LP}^{W_1,W_2,\tau}$. Let $P_2^*$ denote the profile defined in~\eqref{eq:def-P2-star} in the proof of Theorem~\ref{thm:Thiele-refinement}. Recall that $P_2^*$ guarantees that  $\score_{\vec s}(P_2^*,W_1)=\score_{\vec s}(P_2^*,W_2)>\score_{\vec s}(P_2^*,W)$. Next, we prove that if $W_1\notin \GST_\tau(P_2^*)$ then Event 2 holds; and if $W_1\in \GST_\tau(P_2^*)$ then Event 1 holds.

{\bf\boldmath When $W_1\notin \GST_\tau(P_2^*)$,} let $G\in 2^{2^\ma}$ denote the group that disqualifies $W_1$ at $P_2^*$, which means
$$\frac{\tau(W_1,G)}{k}\cdot \vec \hist(P_2^*) - \sum\nolimits_{G\in 2^{2^\ma}} \vec w_{G}\cdot \hist(P_2^*) \le 0$$
Let $\vec x^* = \hist( P_2^*) +\vec 1+ \epsilon\hist( 2^{W_1})$ for some sufficiently small $\epsilon>0$ such that $\Thiele_{\vec s}(\vec x^*) =\{W_1\}$ and 
$$\frac{\tau(W_1,G)}{k}\cdot \vec x^* - \sum\nolimits_{G\in 2^{2^\ma}} \vec w_{G}\cdot \vec x^* <0$$
Such $\epsilon>0$ exists because $\vec 1$ does not change the Thiele score differences between any pair of $k$-committees and  $\hist( 2^{W_1})$ is used to break ties between $W_1$ and when $\epsilon$ is sufficiently small, $G$ still disqualifies $W_1$. We use~\eqref{eq:polyhedral} to prove
\begin{equation}
\label{eq:prob-W1-G}\Pr\nolimits_{P\sim(\pi_{\vec p})^n}(P\in \ppoly{W_1,G}) = \Theta(1)
\end{equation}
Notice that $\vec x^*$ is an interior point of $\ppoly{W_1,G}$ whose $\vec b$ is $\vec 0$. The $0$ case does not hold for sufficiently large $n$ because adding a small perturbation to $\vec x^*$, the resulting vector is still in $\ppoly{W_1,G}$, which means that for any sufficiently large $n^*$, any  discretization of $n^*\vec x^*$ (which is the histogram of a profile) is in $\ppoly{W_1,G}$. It is not hard to verify that the polynomial case of~\eqref{eq:polyhedral} holds and $\dim(\ppolyz{W_1,G}) = \dim(\ppoly{W_1,G}) = 2^m$ (because $\vec x^*$ is an interior point). This proves~\eqref{eq:prob-W1-G}.

{\bf\boldmath When $W_1\in \GST_\tau(P_2^*)$,} for all $G\in 2^{2^\ma}$, we have
\begin{equation}
\label{eq:inequ:G-disqualifies-W1}\frac{\tau(W_1,G)}{k}\cdot \vec \hist(P_2^*) - \sum\nolimits_{G\in 2^{2^\ma}} \vec w_{G}\cdot \hist(P_2^*) > 0
\end{equation}
Let $\vec x'  = \gamma\hist(P_2^*)+\hist(2^{W_2})$ for some sufficiently large $\gamma>0$, such that \eqref{eq:inequ:G-disqualifies-W1}
still holds for all $G\in 2^{2^\ma}$. It is not hard to verify that $\Thiele_{\vec s}(\vec x ') = \{W_2\}$ and $\vec x'$ is an interior point of $\ppoly{W_1,W_2,\tau}$, which means that it is an interior point of $\ppolyz{W_1,W_2,\tau}$ as well because the $\vec b$ vector in $\ppoly{W_1,W_2,\tau}$ is $-\vec 1$. Following a similar analysis and application of~\eqref{eq:polyhedral} as the $W_1\in \GST_\tau(P_2^*)$ case, we can prove 
$$\Pr\nolimits_{P\sim(\pi_{\vec p})^n}(P\in \ppoly{W_1,W_2,\tau}) = \Theta(1)$$
This completes the proof of Theorem~\ref{thm:Thiele-GST}.

\end{proof}
\end{document}

%% file: macro.tex
\usepackage[utf8]{inputenc}

\usepackage[ruled]{algorithm2e} 
\usepackage{algorithmic} 
\usepackage{pifont}
\usepackage{nicefrac}
\usepackage{color}
\usepackage{xcolor,colortbl}
\usepackage{wrapfig}
\usepackage{graphicx}
\usepackage{amsmath,amsthm,accents}
\usepackage{mathtools}
\usepackage{arydshln}
\usepackage{diagbox}
\usepackage[a]{esvect}
 
\usepackage{bm}
\usepackage{rotating}
\usepackage{multicol}
\usepackage{multirow}
\usepackage{latexsym}
\usepackage{enumitem}
\usepackage{hyperref}[backref=page]
\usepackage{float}
\hypersetup{
     colorlinks   = true,
     linkcolor    = red, 
     urlcolor     = blue, 
	 citecolor    = blue 
} 

\usepackage[framemethod=TikZ]{mdframed}
\newcounter{theo}[section] 

\newtheorem{thm}{Theorem}
\newtheorem{dfn}{Definition}

\newtheorem{lem}{Lemma}

\newtheorem{prop}{Proposition}
\newtheorem{coro}{Corrollary}
\newenvironment{sketch}{\noindent{\bf Proof sketch.}\rm }{\hfill $\Box$ }

\newtheorem{claim}{Claim}

\newcounter{newct}

\newcommand{\bv}{\begin{array}}

\newcommand{\appThm}[3]{\vspace{1mm} {\sc Theorem~\ref{#2}} {({\bf #1}). }{\em #3 \vspace{1mm}}}

\newcommand{\appPropnoname}[2]{\vspace{1mm}\noindent{\bf Proposition~\ref{#1}.} {\em #2}}
\newcommand{\appThmnoname}[2]{\vspace{1mm} {\sc Theorem~\ref{#1}.} {\em #2 \vspace{1mm}}}
\newcommand{\appClaimnoname}[2]{\vspace{1mm}\noindent{\bf Claim~\ref{#1}.} {\em #2 \vspace{1mm}}}

\newcommand{\myparagraph}[1]{\vspace{2mm}\noindent {\bf\boldmath #1}}

\newcommand{\cal}{\mathcal }
\newcommand{\ma}{\mathcal A}
\newcommand{\calH}{\mathcal H}
\newcommand{\calP}{\mathcal P}
\newcommand{\vH}{{\vec H}}

\newcommand{\ml}{\mathcal L}

\newcommand{\calT}{\mathcal T}

\newcommand{\calG}{\mathcal G}

\newcommand{\calS}{\mathcal S}

\newcommand{\calX}{\mathcal X}

\newcommand{\ra}{\rightarrow}

\newcommand{\rank}{\text{Rank}}
\newcommand{\sign}{\text{Sign}}
\newcommand{\signH}{\text{Sign}_{\vH}}

\newcommand{\others}{\text{others}}

\newcommand{\hist}{\text{Hist}}

\newcommand{\Omit}[1]{}
\newcommand{\calU}{\mathcal U}
\newcommand{\calM}{\mathcal M}
\newcommand{\calA}{\mathcal A}

\newcommand{\calD}{\mathcal D}

\newcommand{\ba}{{\mathbf A}}

\newcommand{\score}{\text{Score}}

\newcommand{\poly}{{\mathcal H}}

\newcommand{\polynint}{{\mathcal H}_n^{\mathbb Z}}
\newcommand{\polyz}{{\mathcal H}_{\leq 0}}

\newcommand{\ppoly}[1]{{\mathcal H}^{#1}}

\newcommand{\pspoly}[2]{{\mathcal H}^{#1}_{#2}}
\newcommand{\ppolyz}[1]{{\mathcal H}_{\leq 0}^{#1}}
\newcommand{\pspolyz}[2]{{\mathcal H}_{#2,\leq 0}^{#1}}
\newcommand{\ppolynint}[1]{{\mathcal H}_n^{#1,\mathbb Z}}
\newcommand{\pspolynint}[2]{{\mathcal H}_{#2,n}^{#1,\mathbb Z}}

\newcommand{\invert}[1]{\left(#1\right)^\top}

\newcommand{\CC}{\text{\sc CC}}

\newcommand{\prefspace}{{\cal E}}
\newcommand{\decspace}{{\cal D}}
\newcommand{\adecspace}{{\accentset{\circ}{\decspace}}}
\newcommand{\listset}[1]{{\cal L}_{#1}}
\newcommand{\committee}[1]{{\cal M}_{#1}}

\newcommand{\jr}{\text{JR}}
\newcommand{\pjr}{\text{PJR}}
\newcommand{\pjrp}{\text{PJR+}}
\newcommand{\ejr}{\text{EJR}}
\newcommand{\ejrp}{\text{EJR+}}
\newcommand{\core}{\text{CORE}}
\newcommand{\acore}{\text{aCORE}}
\newcommand{\fjr}{\text{FJR}}

\newcommand{\pav}{\text{PAV}}
\newcommand{\phrag}{\text{Phragm\'en}}
\newcommand{\coX}{{X}}

\newcommand{\IC}{\text{IC}}

\newcommand{\amap}{{f}}

\newcommand{\equalsq}[2]{#1  =_{\scriptstyle?} #2}
\newcommand{\subq}[2]{#1 \subseteq_{\scriptstyle?} #2}

\newcommand{\hypspace}{{\mathcal F}}
\newcommand{\Ndim}{\text{\sc Ndim}}
\newcommand{\GST}{\text{GST}}
\newcommand{\hypGST}{\hypspace_\text{GST}}
\newcommand{\hypsGST}{\hypspace_\text{sGST}}
\newcommand{\hypThiele}{\hypspace_\text{Thiele}}
\newcommand{\hypABCS}{\hypspace_\text{ABCS}}
\newcommand{\hypGABCS}{\hypspace_\text{GABCS}}
\newcommand{\hypOWA}{\hypspace_\text{OWA}}
\newcommand{\hypOWAr}{\hypspace_\text{oOWA}}
\newcommand{\hypPos}{\hypspace_\text{Pos}}
\newcommand{\hypSRSF}{\hypspace_\text{SRSF}}
\newcommand{\hypnSRSF}{\hypspace_\text{nSRSF}}
\newcommand{\hypCSR}{\hypspace_\text{CSR}}
\newcommand{\hyplinear}{\hypspace_{(\prefspace,\decspace,K)}}

\newcommand{\OWA}{\text{OWA}}
\newcommand{\Pos}{\text{Pos}}
\newcommand{\SRSF}{\text{SRSF}}
\newcommand{\CSR}{\text{CSR}}
\newcommand{\Thiele}{\text{Thiele}}
\newcommand{\ABCS}{\text{ABCS}}
\newcommand{\GABCS}{\text{GABCS}}

\newcommand{\PD}{\text{PD}}
\newcommand{\bw}{{\mathbf w}}

%% file: arXiv-linear-multi-winner.bbl
\begin{thebibliography}{43}
\providecommand{\natexlab}[1]{#1}
\providecommand{\url}[1]{\texttt{#1}}
\expandafter\ifx\csname urlstyle\endcsname\relax
  \providecommand{\doi}[1]{doi: #1}\else
  \providecommand{\doi}{doi: \begingroup \urlstyle{rm}\Url}\fi

\bibitem[Alon(1996)]{Alon1996:Tools}
Noga Alon.
\newblock {Tools from Higher Algebra}.
\newblock In R.~L. Graham, M.~Gr{\"o}tschel, and L.~Lov{\'a}sz, editors,
  \emph{{Handbook of combinatorics}}, volume~2, pages 1749--1783. MIT Press,
  1996.

\bibitem[Aziz and Shah(2021)]{Aziz2021:Participatory}
Haris Aziz and Nisarg Shah.
\newblock {Participatory budgeting: Models and approaches}.
\newblock In Tam{\'a}s Rudas and G{\'a}bor P{\'e}li, editors, \emph{Pathways
  Between Social Science and Computational Social Science: Theories, Methods,
  and Interpretations}, pages 215--236. Springer, 2021.

\bibitem[Aziz et~al.(2017)Aziz, Brill, Conitzer, Elkind, Freeman, and
  Walsh]{Aziz2017:Justified}
Haris Aziz, Markus Brill, Vincent Conitzer, Edith Elkind, Rupert Freeman, and
  Toby Walsh.
\newblock {Justified representation in approval-based committee voting}.
\newblock \emph{Social Choice and Welfare}, 48:\penalty0 461--485, 2017.

\bibitem[Aziz et~al.(2018)Aziz, Elkind, Huang, Lackner, Sanchez-Fernandez, and
  Skowron]{Aziz2018:On-the-Complexity}
Haris Aziz, Edith Elkind, Shenwei Huang, Martin Lackner, Luis
  Sanchez-Fernandez, and Piotr Skowron.
\newblock {On the Complexity of Extended and Proportional Justified
  Representation}.
\newblock In \emph{Proceedings of AAAI}, 2018.

\bibitem[Brandl et~al.(2018)Brandl, Brandt, Eberl, and
  Geist]{Brandl2018:Proving}
Florian Brandl, Felix Brandt, Manuel Eberl, and Christian Geist.
\newblock {Proving the Incompatibility of Efficiency and Strategyproofness via
  SMT Solving}.
\newblock \emph{Journal of the ACM}, 65\penalty0 (2):\penalty0 1--28, 2018.

\bibitem[Bredereck et~al.(2019)Bredereck, Faliszewski, Kaczmarczyk, and
  Niedermeier]{Bredereck2019:An-Experimental}
Robert Bredereck, Piotr Faliszewski, Andrzej Kaczmarczyk, and Rolf Niedermeier.
\newblock {An Experimental View on Committees Providing Justified
  Representation}.
\newblock In \emph{Proceedings of IJCAI}, 2019.

\bibitem[Brill and Peters(2023)]{Brill2023:Robust}
Markus Brill and Jannik Peters.
\newblock {Robust and Verifiable Proportionality Axioms for Multiwinner
  Voting}.
\newblock In \emph{Proceedings of ACM EC}, 2023.

\bibitem[Caragiannis and Fehrs(2022)]{Caragiannis2022:The-Complexity}
Ioannis Caragiannis and Karl Fehrs.
\newblock {The Complexity of Learning Approval-Based Multiwinner Voting Rules}.
\newblock In \emph{Proceedings of AAAI}, 2022.

\bibitem[Conitzer et~al.(2009)Conitzer, Rognlie, and
  Xia]{Conitzer09:Preference}
Vincent Conitzer, Matthew Rognlie, and Lirong Xia.
\newblock Preference functions that score rankings and maximum likelihood
  estimation.
\newblock In \emph{Proceedings of the Twenty-First International Joint
  Conference on Artificial Intelligence (IJCAI)}, pages 109--115, Pasadena, CA,
  USA, 2009.

\bibitem[Diss and Merlin(2021)]{Diss2021:Evaluating}
Mostapha Diss and Vincent Merlin, editors.
\newblock \emph{{Evaluating Voting Systems with Probability Models}}.
\newblock Studies in Choice and Welfare. Springer International Publishing,
  2021.

\bibitem[Elkind et~al.(2017)Elkind, Faliszewski, Skowron, and
  Slinko]{Elkind2017:Properties}
Edith Elkind, Piotr Faliszewski, Piotr Skowron, and Arkadii Slinko.
\newblock {Properties of multiwinner voting rules}.
\newblock \emph{Social Choice and Welfare}, 48:\penalty0 599--632, 2017.

\bibitem[Elkind et~al.(2023)Elkind, Faliszewski, Igarashi, Manurangsi,
  Schmidt-Kraepelin, and Suksompong]{Elkind2023:Justifying}
Edith Elkind, Piotr Faliszewski, Ayumi Igarashi, Pasin Manurangsi, Ulrike
  Schmidt-Kraepelin, and Warut Suksompong.
\newblock {Justifying groups in multiwinner approval voting}.
\newblock \emph{Theoretical Computer Science}, 969, 2023.

\bibitem[Fain et~al.(2018)Fain, Munagala, and Shah]{Fain2018:Fair}
Brandon Fain, Kamesh Munagala, and Nisarg Shah.
\newblock {Fair allocation of indivisible public goods}.
\newblock In \emph{Proceedings of ACM EC}, 2018.

\bibitem[Fish et~al.(2024)Fish, G{\"o}lz, Parkes, Procaccia, Rusak, Shapira,
  and W{\"u}thrich]{Fish2024:Generative}
Sara Fish, Paul G{\"o}lz, David~C. Parkes, Ariel~D. Procaccia, Gili Rusak, Itai
  Shapira, and Manuel W{\"u}thrich.
\newblock {Generative Social Choice}.
\newblock In \emph{Proceedings of ACM EC}, 2024.

\bibitem[Geist and Endriss(2011)]{Geist2011:Automated}
Christian Geist and Ulle Endriss.
\newblock {Automated Search for Impossibility Theorems in Social Choice Theory:
  Ranking Sets of Objects}.
\newblock \emph{Journal of Artificial Intelligence Research}, 40:\penalty0
  143--174, 2011.

\bibitem[Geist and Peters(2017)]{Geist17:Computer-aided}
Christian Geist and Dominik Peters.
\newblock {Computer-aided Methods for Social Choice Theory}.
\newblock In Ulle Endriss, editor, \emph{{Trends in Computational Social
  Choice}}, chapter~13. AI Access, 2017.

\bibitem[Halpern et~al.(2023)Halpern, Kehne, Procaccia, Tucker-Foltz, and
  W{\"u}thrich]{Halpern2023:Representation}
Daniel Halpern, Gregory Kehne, Ariel~D. Procaccia, Jamie Tucker-Foltz, and
  Manuel W{\"u}thrich.
\newblock {Representation with Incomplete Votes}.
\newblock In \emph{Proceedings of AAAI}, 2023.

\bibitem[Jha and Zick(2020)]{Jha2020:A-Learning}
Tushant Jha and Yair Zick.
\newblock {A Learning Framework for Distribution-Based Game-Theoretic Solution
  Concepts}.
\newblock In \emph{Proceedings of ACM EC}, 2020.

\bibitem[Jiang et~al.(2020)Jiang, Munagala, and Wang]{Jiang2020:Approximately}
Zhihao Jiang, Kamesh Munagala, and Kangning Wang.
\newblock {Approximately stable committee selection}.
\newblock In \emph{Proceedings of STOC}, 2020.

\bibitem[Lackner and Skowron(2021)]{Lackner2021:Consistent}
Martin Lackner and Piotr Skowron.
\newblock {Consistent Approval- Based Multi-Winner Rules}.
\newblock \emph{Journal of Economic Theory}, 192\penalty0 (105173), 2021.

\bibitem[Lackner and Skowron(2023)]{Lackner2023:Multi-Winner}
Martin Lackner and Piotr Skowron.
\newblock \emph{{Multi-Winner Voting with Approval Preferences}}.
\newblock Springer, 2023.

\bibitem[Mohsin et~al.(2022)Mohsin, Liu, Chen, Rossi, and
  Xia]{Mohsin2022:Learning}
Farhad Mohsin, Ao~Liu, Pin-Yu Chen, Francesca Rossi, and Lirong Xia.
\newblock {Learning to Design Fair and Private Voting Rules}.
\newblock \emph{Journal of Artificial Intelligence Research}, 75, 2022.

\bibitem[Munagala et~al.(2022)Munagala, Shen, Wang, and
  Wang]{Munagala2022:Approximate}
Kamesh Munagala, Yiheng Shen, Kangning Wang, and Zhiyi Wang.
\newblock {Approximate core for committee selection via multilinear extension
  and market clearing}.
\newblock In \emph{Proceedings of SODA}, 2022.

\bibitem[Peters(2025)]{Peters2025:The-Core}
Dominik Peters.
\newblock {The Core of Approval-Based Committee Elections with Few Candidates}.
\newblock arXiv:2501.18304, 2025.

\bibitem[Peters and Skowron(2020)]{Peters2020:Proportionality}
Dominik Peters and Piotr Skowron.
\newblock {Proportionality and the Limits of Welfarism}.
\newblock In \emph{Proceedings of ACM EC}, 2020.

\bibitem[Plott(1976)]{Plott76:Axiomatic}
Charles~R. Plott.
\newblock {Axiomatic Social Choice Theory: An Overview and Interpretation}.
\newblock \emph{American Journal of Political Science}, 20\penalty0
  (3):\penalty0 511--596, 1976.

\bibitem[Procaccia et~al.(2009)Procaccia, Zohar, Peleg, and
  Rosenschein]{Procaccia09:Learnability}
Ariel~D. Procaccia, Aviv Zohar, Yoni Peleg, and Jeffrey~S. Rosenschein.
\newblock The learnability of voting rules.
\newblock \emph{Artificial Intelligence}, 173:\penalty0 1133---1149, 2009.

\bibitem[Robbins(1955)]{Robbins1955:Remark}
Herbert Robbins.
\newblock {A Remark on Stirling's Formula}.
\newblock \emph{The American Mathematical Monthly}, 62\penalty0 (1):\penalty0
  26--29, 1955.

\bibitem[S{\'a}nchez-Fern{\'a}ndez et~al.(2017)S{\'a}nchez-Fern{\'a}ndez,
  Elkind, Lackner, Escuela, Fisteus, Val, and
  Skowron]{Sanchez-Fernandez2017:Proportional}
Luis S{\'a}nchez-Fern{\'a}ndez, Edith Elkind, Martin Lackner,
  Norberto~Fern{\'a}ndez Escuela, Jes{\'u}s Fisteus, Pablo~Basanta Val, and
  Piotr Skowron.
\newblock {Proportional Justified Representation}.
\newblock In \emph{Proceedings of AAAI}, 2017.

\bibitem[Sen(1999)]{Sen1999:The-Possibility}
Amartya Sen.
\newblock {The Possibility of Social Choice}.
\newblock \emph{American Economic Review}, 89\penalty0 (3):\penalty0 349--378,
  1999.

\bibitem[Shalev-Shwartz and Ben-David(2014)]{Shalev-Shwartz2014:Understanding}
Shai Shalev-Shwartz and Shai Ben-David.
\newblock \emph{{Understanding Machine Learning: From Theory to Algorithms}}.
\newblock Cambridge University Press, 2014.

\bibitem[Skowron et~al.(2016)Skowron, Faliszewski, and
  Lang]{Skowron2016:Finding}
Piotr Skowron, Piotr Faliszewski, and J{\'e}r{\^o}me Lang.
\newblock {Finding a collective set of items: From proportional
  multirepresentation to group recommendation}.
\newblock \emph{Artificial Intelligence}, 241:\penalty0 191--216, 2016.

\bibitem[Tang and Lin(2009)]{Tang2009:Computer-aided}
Pingzhong Tang and Fangzhen Lin.
\newblock {Computer-aided proofs of Arrow's and other impossibility theorems}.
\newblock \emph{Artificial Intelligence}, 173\penalty0 (11):\penalty0
  1041--1053, 2009.

\bibitem[Thiele(1985)]{Thiele1985:Om-Flerfoldsvalg}
Thorvald~N. Thiele.
\newblock {Om Flerfoldsvalg}.
\newblock \emph{Oversigt over det Kongelige Danske Videnskabernes Selskabs
  Forhandlinger}, 1985.

\bibitem[Utgoff(1988)]{Utgoff1988:Perceptron}
Paul~E. Utgoff.
\newblock {Perceptron trees: a case study in hybrid concept representations}.
\newblock In \emph{Proceedings of AAAI}, 1988.

\bibitem[Warren(1968)]{Warren1968:Lower}
Hugh~E. Warren.
\newblock {Lower Bounds for Approximation by Nonlinear Manifolds}.
\newblock \emph{Transactions of the American Mathematical Society},
  133\penalty0 (1):\penalty0 167--178, 1968.

\bibitem[Xia(2013)]{Xia13:Designing}
Lirong Xia.
\newblock Designing social choice mechanisms using machine learning.
\newblock In \emph{Proceedings of the 2013 International Conference on
  Autonomous Agents and Multi-Agent Systemsulti-agent systems}, pages 471--474,
  2013.

\bibitem[Xia(2015)]{Xia15:Generalized}
Lirong Xia.
\newblock {Generalized Decision Scoring Rules: Statistical, Computational, and
  Axiomatic Properties}.
\newblock In \emph{Proceedings of the Sixteenth ACM Conference on Economics and
  Computation}, pages 661--678, Portland, Oregon, USA, 2015.

\bibitem[Xia(2020)]{Xia2020:The-Smoothed}
Lirong Xia.
\newblock {The Smoothed Possibility of Social Choice}.
\newblock In \emph{Proceedings of NeurIPS}, 2020.

\bibitem[Xia(2021)]{Xia2021:How-Likely}
Lirong Xia.
\newblock {How Likely Are Large Elections Tied?}
\newblock In \emph{Proceedings of ACM EC}, 2021.

\bibitem[Xia(2022)]{Xia2022:Group}
Lirong Xia.
\newblock {Group decision making under uncertain preferences: powered by AI,
  empowered by AI}.
\newblock \emph{Annals of The New York Academy of Sciences}, 2022.

\bibitem[Xia and Conitzer(2008)]{Xia08:Generalized}
Lirong Xia and Vincent Conitzer.
\newblock Generalized scoring rules and the frequency of coalitional
  manipulability.
\newblock In \emph{Proceedings of the ACM Conference on Electronic Commerce},
  pages 109--118, 2008.

\bibitem[Zhang and Conitzer(2019)]{Zhang2019:A-PAC-Framework}
Hanrui Zhang and Vincent Conitzer.
\newblock {A PAC Framework for Aggregating Agents' Judgments}.
\newblock In \emph{Proceedings of AAAI}, 2019.

\end{thebibliography}
